\documentclass[reprint,onecolumn,superscriptaddress,showpacs,preprintnumbers,nofootinbib,aps,prd]{revtex4-1}

\usepackage{dsfont}
\usepackage[section]{placeins}
\usepackage{graphicx,color}
\usepackage{amsmath,amssymb}
\usepackage{stackrel}
\usepackage{enumitem}
\usepackage[english]{babel}
\usepackage{verbatim}
\usepackage[normalem]{ulem}
\usepackage{xcolor}
\usepackage{subcaption}
\usepackage{adjustbox}
\usepackage{multirow,float}
\usepackage{dcolumn}
\usepackage{physics}
\usepackage[bottom]{footmisc}
\usepackage[toc,page]{appendix}
\usepackage{bm}
\usepackage{slashed}
\usepackage{booktabs}
\definecolor{darkred}{rgb}{.7,.1,.1}
\usepackage{hyperref}
\hypersetup{
    pdfnewwindow=true,  
    colorlinks=true,     
    linkcolor=darkred,     
    citecolor=blue,      
    filecolor=blue,      
    urlcolor=blue        
}
 
\graphicspath{
  {./}
  {figures/}
}
\usepackage{colortbl}

\newcommand{\nn}{\nonumber}
\renewcommand{\vec}[1]{{\mathbf #1}}

\newcommand{\C}{{\cal C}}

\newcommand{\Op}{{\cal O}}		
\newcommand{\todo}[1]{{\color{red} 
\ifmmode\else[todo]\fi #1}}
\newcommand{\lrpartial}{\negthickspace\stackrel{\leftrightarrow}{\partial}\negthickspace{}}

\newcommand{\lrD}{\negthickspace\stackrel{\leftrightarrow}{D}\negthickspace{}}
\definecolor{byzantine}{rgb}{0.74, 0.2, 0.64}

\newcommand{\hq}{\hat{q}}

\newcommand{\gev}{~{\rm GeV}}
\newcommand{\mev}{~{\rm MeV}}
\newcommand{\tf}{\texorpdfstring}

\begin{document}

\title{
Comprehensive Effective Field Theory Framework for Coherent Elastic Neutrino-Nucleus Scattering
}

\author{Gang Li}
\email{ligang65@mail.sysu.edu.cn}
\affiliation{School of Physics and Astronomy, Sun Yat-sen University, Zhuhai 519082, P.R. China}
\affiliation{Guangdong Provincial Key Laboratory of Quantum Metrology and Sensing, Sun Yat-Sen University, Zhuhai 519082, P.R. China}

\author{Chuan-Qiang Song}
\email{songchuanqiang21@mails.ucas.ac.cn}
\affiliation{School of Fundamental Physics and Mathematical Sciences, Hangzhou Institute for Advanced Study, UCAS, Hangzhou 310024, China}
\affiliation{School of Physical Sciences, University of Chinese Academy of Sciences, Beijing 100049, P.R. China}
\affiliation{Institute of Theoretical Physics, Chinese Academy of Sciences, Beijing 100190, P.R. China}

\author{Feng-Jie Tang}
\email{tangfj7@mail2.sysu.edu.cn}
\affiliation{School of Physics and Astronomy, Sun Yat-sen University, Zhuhai 519082, P.R. China}

\author{Jiang-Hao Yu}
\email{jhyu@itp.ac.cn}
\affiliation{School of Fundamental Physics and Mathematical Sciences, Hangzhou Institute for Advanced Study, UCAS, Hangzhou 310024, China}
\affiliation{School of Physical Sciences, University of Chinese Academy of Sciences, Beijing 100049, P.R. China}
\affiliation{Institute of Theoretical Physics, Chinese Academy of Sciences, Beijing 100190, P.R. China}
\affiliation{International Centre for Theoretical Physics Asia-Pacific, Beijing/Hangzhou, P.R. China}

\begin{abstract}
Coherent elastic neutrino-nucleus scattering (CE$\nu$NS) stands out as a pivotal process for precision tests of the Standard Model electroweak sector, investigations of neutrino properties, and searches for new physics. Recent experimental measurements by COHERENT, CONUS+, and ton-scale xenon detectors--including PandaX-4T and XENONnT--underscore the need for a systematic theoretical framework to bridge high-energy physics scenarios with low-energy observational data. In this work, we develop a comprehensive end-to-end effective field theory (EFT) framework for CE$\nu$NS, encompassing the complete energy scale hierarchy spanning the ultraviolet regime down to the nuclear sector.
We consider the low-energy EFT (LEFT) operators up to dimension 8, incorporating their QCD renormalization group running effects, and employ the systematic spurion method to achieve matching between these operators and the chiral Lagrangian. A full power counting analysis is performed, extending to nuclear response functions, which 
evaluates contributions from LEFT operators up to dimension 8 while accounting for the nucleon number enhancement effect intrinsic to CE$\nu$NS. Moreover, we match the relevant LEFT operators for CE$\nu$NS onto operators up to dimension 8 within the Standard Model EFT. By also providing their complete tree-level ultraviolet completions, this procedure establishes a consistent top-down theoretical workflow.
Leveraging a broad suite of CE$\nu$NS experimental data, this framework enables a combined analysis to extract constraints on the scales of EFT operators and neutrino non-standard interaction parameters.
\end{abstract}

\maketitle
\emergencystretch 3em

\newpage
\section{Introduction}

Coherent elastic neutrino-nucleus scattering (CE$\nu$NS) is a process in which the neutrino scatters elastically off an entire nucleus~\cite{Freedman:1973yd}. It provides a powerful tool for precision tests of the Standard Model (SM) electroweak sector, studies of neutrino properties, nuclear form factors, and astrophysical phenomena~\cite{Coloma:2017ncl,Liao:2017uzy,Cadeddu:2017etk,Ge:2017mcq,AristizabalSierra:2017joc,Ciuffoli:2018qem,Farzan:2018gtr,Billard:2018jnl,AristizabalSierra:2018eqm,Cadeddu:2018izq,Altmannshofer:2018xyo,Miranda:2019skf,AristizabalSierra:2019ufd,Papoulias:2019txv,Giunti:2019xpr,Canas:2019fjw,Hoferichter:2020osn,Skiba:2020msb,Cadeddu:2020nbr,Du:2021rdg,Dasgupta:2021fpn,AtzoriCorona:2022qrf,DeRomeri:2022twg,Li:2024gbw,Li:2024iij,Liao:2025hcs,Denton:2018xmq,Denton:2022nol} (see also Ref.~\cite{Abdullah:2022zue}).
Since the experimental discovery of this process~\cite{COHERENT:2017ipa}, CE$\nu$NS measurements have been performed across accelerator-, solar-, and reactor-based neutrino sources and multiple target nuclei~\cite{CONNIE:2019swq,CONUS:2020skt,nGeN:2022uje,Colaresi:2022obx,Su:2023klh,Ricochet:2023yek,CONUSCollaboration:2024kvo,Yang:2024exl,RELICS:2024opj,CICENNS,COHERENT:2021xmm,COHERENT:2020iec,COHERENT:2020ybo,COHERENT:2024axu}.
In parallel, ton-scale xenon detectors have reported the first indications of solar $^8$B neutrinos via CE$\nu$NS~\cite{PandaX:2024muv,XENON:2024ijk}. The reactor experiment CONUS has set stringent constraints~\cite{CONUS:2020skt,CONUSCollaboration:2024kvo}, while recently CONUS+ achieved a direct observation of CE$\nu$NS~\cite{Ackermann:2025obx}.

In the SM, CE$\nu$NS is mediated by the exchange of the $Z$ boson, and thus provides a sensitive probe of neutral-current neutrino interactions. To fully exploit this sensitivity in precision studies, recent works have emphasized that reliable predictions for such processes require incorporating radiative and hadronic corrections~\cite{Tomalak:2020zfh,Tomalak:2025tls}. This process can also be induced by new physics (NP) beyond the SM, such as the $Z^\prime$ boson~\cite{Barranco:2005yy,Liao:2017uzy} or leptoquark~\cite{Calabrese:2022mnp,DeRomeri:2023cjt}. 
While these model-dependent scenarios have been explored, effective field theory (EFT) offers a systematic way to study CE$\nu$NS~\cite{Altmannshofer:2018xyo,Bischer:2019ttk,Breso-Pla:2023tnz,Breso-Pla:2025pds}. In the EFT approach, NP effects are encoded in a series of effective operators, for which the Wilson coefficients and NP scales are given in specific ultraviolet (UV) models. Besides,
in comparison with the studies using the weak charge supplemented by empirical nuclear form factors~\cite{Helm:1956zz,Klein:1999qj}, the EFT approach allows for power countings and controlled theoretical uncertainties, which enables {\it ab initio} calculations of the nuclear structures~\cite{Roth:2011vt,Stroberg:2019bch,Payne:2019wvy,Arthuis:2020toz,Hu:2021trw,Hu:2021awl}. 
A similar EFT framework has been used in dark-matter phenomenology~\cite{Cirigliano:2012pq,Menendez:2012tm,Klos:2013rwa,Vietze:2014vsa,Hoferichter:2015ipa,Fitzpatrick:2012ix,Anand:2013yka,Bishara:2016hek,Bishara:2017pfq} and can also be extended to CE$\nu$NS-like inelastic channels with dark-sector particles in the final state~\cite{Feng:2026jsi}.

\begin{figure}
    \centering
    \includegraphics[scale=0.3]{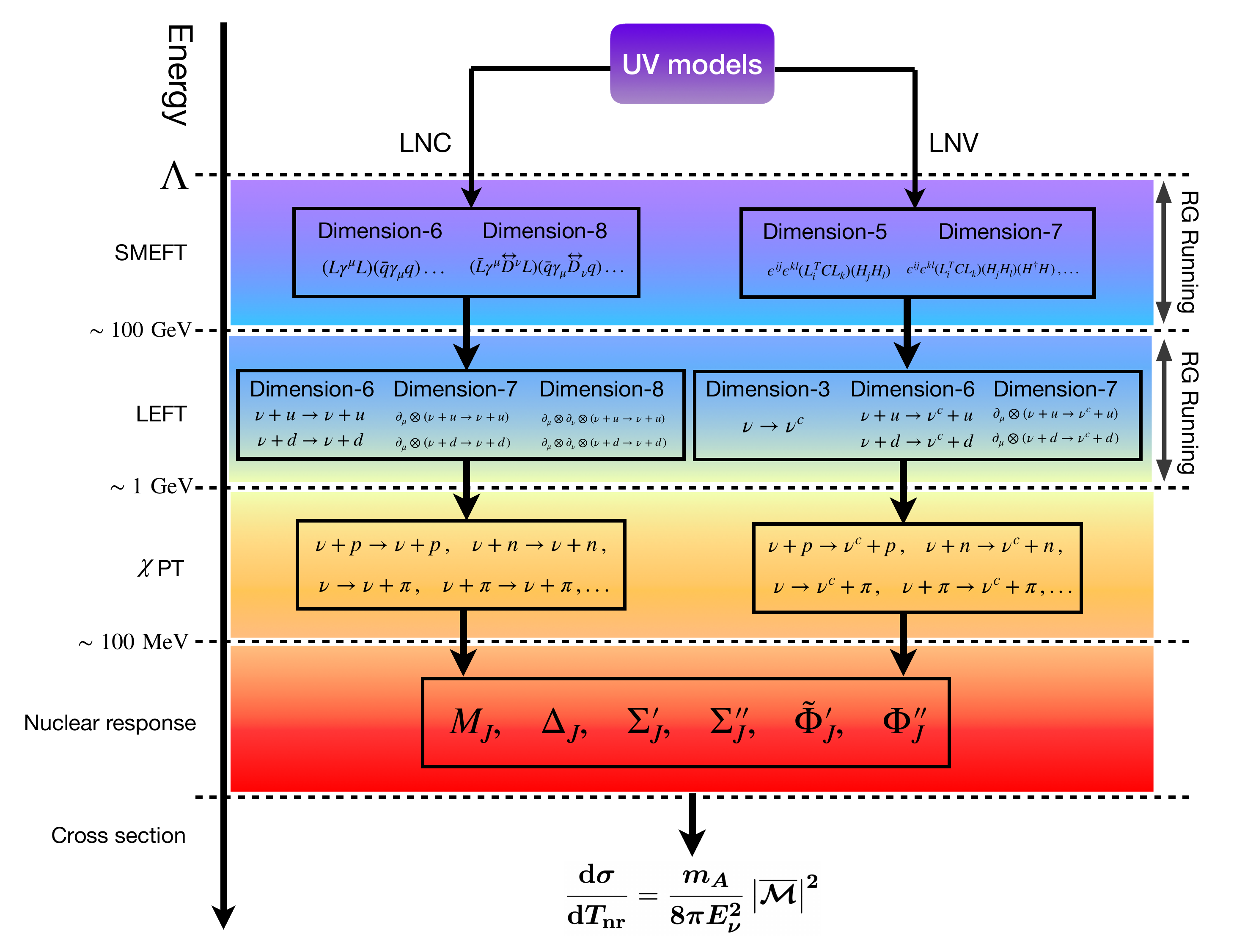}
    \caption{Comprehensive effective field theory framework for evaluating the cross section of CE$\nu$NS.}
    \label{fig:my_label}
\end{figure}

The neutrino-matter interactions in CE$\nu$NS considered are shown in Fig.~\ref{fig:my_label}. After integrating out possible new particles, which are assumed to be heavier than $\sim 100\gev$, the interactions are given by dimension-5, 6, 7 and 8 operators in the standard model EFT (SMEFT)~\cite{Weinberg:1979sa,Buchmuller:1985jz,Grzadkowski:2010es,Lehman:2014jma,Liao:2016hru,Li:2020gnx,Murphy:2020rsh}, respecting the SM gauge symmetries. At lower energy scale, CE$\nu$NS is characterized by effective interactions between neutrinos and quarks, such as $\nu + u \to \nu + u$, in the low-energy EFT (LEFT)~\cite{Jenkins:2017jig,Liao:2020zyx,Li:2021tsq} systematically, including neutrino non-standard interactions (NSIs), electromagnetic dipole operators and gluonic operators. 

At energy scales down to $\sim 1\gev$, CE$\nu$NS is described as the scattering between neutrinos and the constituent hadrons, namely pions and nucleons. Within the framework of chiral perturbation theory ($\chi$PT), the relevant weak and potential NP interactions involving the neutrino current are incorporated using the external source method~\cite{Gasser:1982ap,Gasser:1983yg,Gasser:1984gg} or, more generally, the systematic spurion method~\cite{Li:2025xmq,Song:2025snz}. The low-energy constants (LECs) that arise from non-perturbative QCD effects are determined either from experiments/lattice QCD calculations or estimated using naive dimensional analysis (NDA)~\cite{Manohar:1983md,Gavela:2016bzc,Jenkins:2013sda}.
At the scale below $\sim 100\mev$, the neutrino interacts coherently with the entire nucleus. Analogous to dark matter-nucleus scattering~\cite{Fitzpatrick:2012ix,Anand:2013yka,Vietze:2014vsa,Korber:2017ery,deVries:2023sux}, the nuclear response in CE$\nu$NS can be described by a limited number of nuclear response functions~\cite{Altmannshofer:2018xyo,Hoferichter:2020osn}.

In this work, we establish an EFT framework for the studies of CE$\nu$NS with the following key features:
\begin{itemize}
    \item We consider the LEFT operators up to dimension 8 and the corresponding chiral Lagrangian in the spurion method.
    \item A complete matching of the EFTs for CE$\nu$NS across the relevant energy scales is accomplished, and the QCD renormalization-group (RG) running effects are included.
    \item A systematic power counting at different levels is developed. At the hadronic level, it enables a clear comparison between contributions from one-body and two-body currents. At the nuclear level, the CE$\nu$NS cross section is enhanced by the nucleon number, which is incorporated via the nuclear response functions. 
    \item Full tree-level UV completions for relevant SMEFT operators are provided, relating CE$\nu$NS signals to the parameters of UV models.
    \item New constraints are derived from the latest CE$\nu$NS observation by the CONUS+ experiment.
\end{itemize}
Therefore, we aim to build a more comprehensive EFT pipeline compared to Refs.~\cite{Altmannshofer:2018xyo,Hoferichter:2020osn}.

The remainder of this paper is organized as follows. In Sec.~\ref{sec:LEFT}, we provide the LEFT operators for CE$\nu$NS up to dimension 8, and the QCD RG running or evolution (RGE) is considered. 
In Sec.~\ref{sec:ChEFT}, we perform the matching of the relevant LEFT operators onto the chiral Lagrangian using a systematic spurion method. The power counting from the electroweak scale down to the hadronic scale is then discussed using a unified NDA master formula in Sec.~\ref{sec:power}. In Sec.~\ref{sec:CS}, neutrino-nucleus scattering cross section is calculated in the approach of nuclear response function framework. In Sec.~\ref{sec:expts}, the constraints on the NP scales as well as the non-standard interaction (NSI) parameters are derived using the experimental measurements. In Sec.~\ref{sec:SMEFTUV}, we match the LEFT operators onto the SMEFT operators, and obtain various kinds of tree-level UV completions. 
We conclude in Sec.~\ref{sec:conclusion}, and provide more details of the matching, NDA, LECs and chiral expansion in the appendices.

\section{LEFT Operators}
\label{sec:LEFT}

Starting from the scale below $\sim 100\gev$, the effective Lagrangian in the LEFT can be written as
\begin{equation}
\mathcal{L}_{\text{LEFT}}=\sum_{d,a}\hat{\mathcal{C}}_a^{d}\mathcal{O}_a^{d}\,, \quad \hat{\mathcal{C}}_a^{d} \equiv \dfrac{\mathcal{C}_a^{d}}{\Lambda_{\rm EW}^{d-4}}\,,
\label{coefficient}
\end{equation}
where $\mathcal{O}_a^{d}$ denotes the $a$-th independent LEFT operator of mass dimension $d$, and $\hat{\mathcal{C}}_a^{d}$ is the corresponding Wilson coefficient with mass dimension $4-d$. Here, $\mathcal{C}_a^{d}$ is dimensionless and $\Lambda_{\rm EW}$ denotes the electroweak scale.

Compared with Refs.~\cite{Altmannshofer:2018xyo,Hoferichter:2020osn}, which included the LEFT operators up to dimension 7, we consider the lepton-number-conserving (LNC) operators up to dimension 8 as follows~\cite{Jenkins:2017jig,Liao:2020zyx,Li:2021tsq}:
\begin{align}
\label{eq:OpLEFTL0}
\Op_1^{6u/d}&=(\bar \nu_{L\alpha}\gamma^\mu \nu_{L\beta})(\bar q\gamma_\mu\tau^{u/d} q)\,,&\Op_2^{6u/d}&=(\bar \nu_{L\alpha}\gamma^\mu \nu_{L\beta})(\bar q\gamma^5\gamma_\mu\tau^{u/d} q)\,,\\
\Op_1^{7u/d}&=(\bar \nu_{L\alpha}\gamma^\mu \nu_{L\beta})(\bar q\lrD_\mu\tau^{u/d} q)\,,&\Op_2^{7u/d}&=(\bar \nu_{L\alpha}\gamma^\mu \nu_{L\beta})(\bar q\gamma^5\lrD_\mu\tau^{u/d} q)\,,\\
\Op_{1}^{8u/d} &= (\bar\nu_{L\alpha} \gamma^\mu\lrD^\nu  \nu_{L\beta})  (\bar{q} \gamma_\mu\lrD_\nu\tau^{u/d} q )\,,
&
\Op_{2}^{8u/d} &= (\bar\nu_{L\alpha} \gamma^\mu\lrD^\nu  \nu_{L\beta})  (\bar{q} \gamma^5\gamma_\mu\lrD_\nu \tau^{u/d}q)\,,
\\
\label{eq:OpLEFTL083}
\Op_3^{8u/d}&=(\bar \nu_{L\alpha}\gamma^\mu \nu_{L\beta})(\bar q\gamma^\nu T^A\tau^{u/d} q)G_{\mu\nu}^A\,,
&
\Op_4^{8u/d}&=(\bar \nu_{L\alpha}\gamma^\mu \nu_{L\beta})(\bar q\gamma^5\gamma^\nu T^A \tau^{u/d}q)G_{\mu\nu}^A\,,\\
\Op_{5}^{8u/d}&= (\bar \nu_{L\alpha}\gamma^\mu \nu_{L\beta})D^2( \bar{q} \gamma_\mu\tau^{u/d} q )\,,
&
\Op_{6}^{8u/d}&= (\bar \nu_{L\alpha}\gamma^\mu \nu_{L\beta})D^2( \bar{q} \gamma^5\gamma_\mu\tau^{u/d} q )\,,
\\
\Op_7^{8u/d}&=(\bar \nu_{L\alpha}\gamma^\mu \nu_{L\beta})(\bar q\gamma^\nu T^A\tau^{u/d} q)\Tilde{G}_{\mu\nu}^A\,,
&
\Op_8^{8u/d}&=(\bar \nu_{L\alpha}\gamma^\mu \nu_{L\beta})(\bar q\gamma^5\gamma^\nu T^A \tau^{u/d}q)\Tilde{G}_{\mu\nu}^A\,,\\
\Op_{9}^{8} &= (\bar\nu_{L\alpha} \gamma^\mu\lrD^\nu  \nu_{L\beta}) G^A_{\mu\rho}G^{A\rho}_\nu\,.
\end{align}
In the above, the isospin doublet quark field $q=(u,d)^T$, and $\nu_L$ denotes the left-handed neutrino field, with Greek indices $\alpha$ and $\beta$ labeling the neutrino flavors. The isospin projectors are defined as $\tau^{u/d} \equiv (1 \pm \tau^3)/2$, where $\tau^3$ is the third Pauli matrix. The Hermitian derivative acting on a fermion bilinear is $\bar\psi\lrD^\mu\psi \equiv \bar\psi(D^\mu\psi) - (D^\mu\bar\psi)\psi$. The covariant derivative is
\begin{eqnarray}
\label{eq:CD}
D_\mu = \partial_\mu + i g_3 T^A G_\mu^A + i e Q A_\mu\;,
\end{eqnarray}
where $A_\mu$ and $G_\mu^A$ represent the photon and gluon fields, respectively. $eQ$ is the electromagnetic charge operator, $g_3$ is the strong coupling constant, and $T^A$ ($A=1,\dots,8$) are the generators of the $SU(3)$ color group. The dual of the QCD field-strength tensor is defined as $\tilde{G}_{\mu\nu}^A = \frac{1}{2} \varepsilon_{\mu\nu}^{\quad\rho\lambda} G_{\rho\lambda}^A$, with the Levi‑Civita tensor $\varepsilon_{0123} = -1$.

The lepton-number-violating (LNV) operators up to dimension 7 are~\cite{Liao:2020zyx,Li:2021tsq}
\begin{align}
\label{eq:OpLEFTL2}
\Op_1^{5}&=\frac{e}{8\pi^2}( \nu_{L\alpha}^TC\sigma^{\mu\nu} \nu_{L\beta})F_{\mu\nu}\,,&\Op_3^{7u/d}&=m_q( \nu_{L\alpha}^TC\sigma^{\mu\nu} \nu_{L\beta})(\bar q\sigma_{\mu\nu} \tau^{u/d}q)\,,\\
    \Op_4^{7u/d}&=m_q( \nu_{L\alpha}^TC \nu_{L\beta})(\bar q \tau^{u/d}q)\,,&\Op_5^{7u/d}&=m_q( \nu_{L\alpha}^TC \nu_{L\beta})(\bar q\gamma^5\tau^{u/d}q)\,,\\
    \Op_6^{7}&=\frac{\alpha_s}{\pi}( \nu_{L\alpha}^TC \nu_{L\beta})G_{\mu\nu}^AG^{A\mu\nu}\,,&\Op_7^{7}&=\frac{\alpha_s}{\pi}( \nu_{L\alpha}^TC \nu_{L\beta})\tilde{G}_{\mu\nu}^AG^{A\mu\nu}\,,
\end{align}
Here, $F_{\mu\nu}$ denotes the electromagnetic field-strength tensor, $\alpha_s$ is the QCD coupling constant, and $C = i\gamma^{0}\gamma^{2}$ denotes the usual charge-conjugation matrix. The quark mass $m_q$ corresponds to the flavor $q = u, d$. The LNV operators $\Op_1^5$ and $\Op_3^{7u/d}$ vanish for the same neutrino flavor, $\alpha = \beta$.

The QCD RG running of the Wilson coefficients in Eq.~\eqref{coefficient} from the electroweak scale $\sim 100\gev$ to the hadronic scale $\sim 1\gev$ is given by~\cite{Jenkins:2017dyc,Naterop:2023dek} 
\begin{align} \label{eq:RGEC}
  \frac{d}{d~{\rm ln}\mu}\hat{\mathcal{C}}_3^{7u/d}=&2C_F\frac{\alpha_s}{4\pi}\hat{\mathcal{C}}_3^{7u/d}\,,\\  
  \frac{d}{d~{\rm ln}\mu}\hat{\mathcal{C}}_{4,5}^{7u/d}=&-6C_F\frac{\alpha_s}{4\pi}\hat{\mathcal{C}}_{4,5}^{7u/d}\,,\\  
  \frac{d}{d~{\rm ln}\mu}\hat{\mathcal{C}}_{1,2}^{7u/d}=&2C_F\frac{\alpha_s}{4\pi}\hat{\mathcal{C}}_{1,2}^{7u/d}\,, \\
  \frac{d}{d~\rm{ln}\mu}\left(\begin{array}{c}
         \hat{\mathcal{C}}_{1,2}^{8u/d}  \\
          \hat{\mathcal{C}}_{5,6}^{8u/d}\\
    \end{array}\right)=&\frac{\alpha_s}{4\pi}\left(\begin{array}{cc}
          -\frac{8}{3}C_F &0\\
           \frac{4}{3}C_F &0
    \end{array}\right)\left(\begin{array}{c}
         \hat{\mathcal{C}}_{1,2}^{8u/d}  \\
          \hat{\mathcal{C}}_{5,6}^{8u/d}\\
    \end{array}\right)\,,
\end{align}
where $C_F \equiv (N_c^2-1)/(2N_c)$ with $N_c$ being the color number. These RG running effects were not considered in Ref.~\cite{Altmannshofer:2018xyo}. Note that the neutrino dipole operator and operators involving vector or axial-vector currents do not run in QCD. The RG running for the dimension-8 operators involving explicit gluon field-strength tensors are not available in the literature and are therefore not considered in this work.

\section{Chiral Lagrangian}
\label{sec:ChEFT}

Having obtained the LEFT operators and considered the QCD RG running, the next step is to match the quark-level interactions onto a low-energy hadronic description at the scale $\Lambda_\chi\sim 1~\mathrm{GeV}$ in chiral perturbation theory\,\footnote{Ref.~\cite{Hoferichter:2020osn} describes neutrino-nucleon interactions in terms of nucleon matrix elements and form factors, which does not provide an order-by-order organization of the individual contributions in the chiral expansion, although some two-body current effects predicted by chiral EFT are accounted for.}. 

The spontaneous breaking of the chiral symmetry $SU(2)_L \times SU(2)_R$ down to its vector subgroup $SU(2)_V$ gives rise to three pseudo-Nambu-Goldstone bosons (pions). Within the Coleman-Callan-Wess-Zumino coset construction~\cite{Coleman:1969sm,Callan:1969sn}, these bosons are nonlinearly realized and are collectively parametrized by a unitary matrix field 
\begin{equation}
    \label{eq:CCWZ}
    u(x) = \exp\left(\frac{i\Pi(x)
    }{2f}\right)\,,
\end{equation}
where the Hermitian matrix $\Pi(x)$ contains the three pion fields

\begin{equation}
    \label{eq:NGBPi}
    \Pi(x)=\sum_{a=1}^3\phi^a(x)\tau^a = \left(\begin{array}{cc}
    \pi^0  & \sqrt{2}\pi^+  \\
    \sqrt{2}\pi^- & -\pi^0
    \end{array}\right)\,.
\end{equation}
Here $\tau^{a}$ are the Pauli matrices and $\phi^{a}$ represent the three NGB degrees of freedom. Under chiral transformations, $u(x)$ transforms as $u(x) \rightarrow Ru(x)V^{-1}=Vu(x)L^{-1}$, with $L\in SU(2)_L$, $R\in SU(2)_R$, and $V\in SU(2)_V$ being the compensating transformation that preserves the vector subgroup. The constant $f\simeq 92\ \mathrm{MeV}$ is the pion decay constant at leading order (LO) in the chiral expansion.

The matching between LEFT and $\chi$PT has been systematically performed in Ref.~\cite{Li:2025xmq}. Following that work, we define the following building blocks
\begin{align}
    \label{eq:buildblock}
     u_\mu &= i(u^\dagger D_\mu u-u D_\mu u^\dagger ) \,,\notag \\
      \chi_\pm&=u^\dagger \chi u^\dagger \pm u \chi^\dagger u\,,\notag \\
    \Sigma_{\pm} &= u^\dagger \tau u^\dagger \pm u \tau u \,,\notag\\
     Q_{\pm}&=u^\dagger\tau u\pm u\tau u^\dagger \,,\notag\\
     \nabla_\mu&=\partial_\mu+\frac{1}{2}(u^\dagger D_\mu u+u D_\mu u^\dagger )\,,
\end{align}
where the covariant derivative including the photon field is given by $D_\mu = \partial_\mu + i eQ A_\mu$, and $\tau$ is the spurion corresponding to the matrix $\tau^{u/d}$ appearing in the LEFT operators. Notice that the covariant derivative $\nabla_\mu$ and chiral vielbein $u_\mu$ do not contain the vector and axial-vector external sources. Since photon contributions from the expansion are irrelevant for our purposes, we will simply use the partial derivative $D_\mu \to \partial_\mu$ hereafter. 
We have also defined 
\begin{align}
    \chi = 2 B_0 \mathcal{M}_q \;,\quad 
    \mathcal{M}_q \equiv\left(\begin{array}{cc}
         m_u& 0 \\
        0 & m_d
    \end{array}\right)\,,
\end{align}
where $B_0\simeq 2.8~\mathrm{GeV}$ parametrizes the quark condensate and is evaluated at the scale $\mu=2~\mathrm{GeV}$. 
The $C$ and $P$ transformation properties of the building blocks are summarized in Tab.~\ref{tab:building_block}.

\begin{table}[H]
\tabcolsep=4pt
\renewcommand\arraystretch{1.5}
\begin{center}
    \begin{tabular}{c|c|c|c|c|c|c|c}
    \hline
    \hline
           & $\Sigma_+$ & $\Sigma_-$&$Q_+$& $Q_-$&$\chi_+$& $\chi_-$ & $ u_\mu$ \\
    \hline
    $P$ & $+$ & $-$ &$+$&$-$&$+$&$-$& $-$\\
    \hline
$C$  & $+$ & $+$ &$+$&$-$ &$+$&$+$ & $+$\\
    \hline
    \hline
    \end{tabular}
    \end{center}
    \caption{The $P$ and $C$ transformation properties of the chiral building blocks. 
} 
\label{tab:building_block}
\end{table}

In addition, the nucleon is included in $\chi$PT and forms a doublet $N$ in the fundamental representation of $SU(2)_V$,
\begin{equation}
    N = \left(\begin{array}{c}
p \\
 n 
    \end{array}\right)\,,
\end{equation}
which transforms as $N\rightarrow VN$ under the chiral symmetry. 

The chiral Lagrangian relies on an expansion in $ p/\Lambda_\chi$. However, each operator in the chiral Lagrangian comes with a LEC.
Fortunately, several standard LECs--including those associated with vector, axial-vector, scalar, and pseudo-scalar interactions--have been determined in hadronic physics. The unknown LECs for interactions involving higher-dimensional LEFT operators can be estimated using NDA.

\subsection{Mesonic Sector}

In the mesonic sector, we construct the chiral Lagrangian from the building blocks $(u_\mu,\Sigma_\pm,Q_\pm,\chi_\pm)$, together with covariant derivatives and neutrino fields. Moreover, LEFT operators involving gluons need to be matched onto the chiral Lagrangian. The matching of gluonic operators can be derived analogously using a systematic spurion method. 

As a concrete illustration, we consider the dimension-8 operator $\Op_3^{8u/d}$ in Eq.~\eqref{eq:OpLEFTL083}.
The gluon field-strength tensor transforms trivially under the chiral symmetry, and is $C-P+$ under $CP$ transformations.
Thus, the chiral properties of $\Op_3^{8u/d}$ are entirely determined by the quark-bilinear factor $(\bar q\gamma^\nu T^A\tau^{u/d} q)$. 
The relevant chiral building blocks are $(Q_\pm, u_\mu, \chi_\pm,\nu_L)$, which transform as $(V Q_\pm V^\dagger, Vu_\mu V^\dagger, V\chi_\pm V^\dagger,\nu_L)$ under chiral symmetry. Furthermore, the quark-gluon component of $\Op_3^{8u/d}$ is $C+P+$, and the corresponding chiral operator must likewise possess identical $CP$ transformation properties.

Chiral operators are constructed order-by-order in the chiral expansion, i.e., in powers of $p/\Lambda_\chi$. At LO, the operator of type $\nu^2_LQ_\pm u_\mu$ is constructed by the building blocks $(u_\mu,Q_\pm,\nu_L)$. Using the Young tensor technique~\cite{Li:2020gnx,Li:2020xlh,Li:2022tec}, we obtain two chiral operators
\begin{align}
    {\rm LO}:\quad (\bar \nu_{L\alpha}\gamma^\mu \nu_{L\beta})\langle u_\mu Q_+\rangle\;,\quad (\bar \nu_{L\alpha}\gamma^\mu \nu_{L\beta})\langle u_\mu Q_-\rangle\,.
\end{align}
However, the hadronic parts of these two operators are $C+P-$ and $C-P+$, respectively, and are therefore incompatible with $\Op_3^{8u/d}$. Therefore, next-to-leading order (NLO) chiral operators are required. We then construct the complete set of independent chiral operators of the types $\nu^2_LQu\chi$, $\nu^2_LQu^3$, and $\nu^2_LQu^2\nabla$ as
\begin{align}
    {\rm NLO}:\quad &(\bar \nu_{L\alpha}\gamma^\mu \nu_{L\beta})\langle u_\mu [Q_+,\chi_+]\rangle\,,\quad (\bar \nu_{L\alpha}\gamma^\mu \nu_{L\beta})\langle u_\mu [Q_+,\chi_-]\rangle\,,\quad(\bar \nu_{L\alpha}\gamma^\mu \nu_{L\beta})\langle u_\mu [Q_-,\chi_+]\rangle\,,\quad\notag\\
    &(\bar \nu_{L\alpha}\gamma^\mu \nu_{L\beta})\langle u_\mu [Q_-,\chi_-]\rangle\,,\quad(\bar \nu_{L\alpha}\gamma^\mu \nu_{L\beta})\langle u_\mu Q_+ \rangle\langle u_\nu u^\nu\rangle\,,\quad(\bar \nu_{L\alpha}\gamma^\mu \nu_{L\beta})\langle u_\mu Q_- \rangle\langle u_\nu u^\nu\rangle\,,\quad\notag\\
    &(\bar \nu_{L\alpha}\gamma^\mu \nu_{L\beta})\langle u_\nu Q_+ \rangle\langle u_\mu u^\nu\rangle\,,\quad(\bar \nu_{L\alpha}\gamma^\mu \nu_{L\beta})\langle u_\nu Q_- \rangle\langle u_\mu u^\nu\rangle\,,\quad\notag\\
    &(\bar \nu_{L\alpha}\gamma^\mu \nu_{L\beta})\langle Q_+[\nabla_\nu u_\mu,u^\nu] \rangle\,,\quad(\bar \nu_{L\alpha}\gamma^\mu \nu_{L\beta})\langle Q_-[\nabla_\nu u_\mu,u^\nu] \rangle\,.
\end{align}
Using the $CP$ properties listed in Tab.~\ref{tab:building_block}, we identify a single chiral operator with $C+P+$ transformation properties:
\begin{align}
    (\bar \nu_{L\alpha}\gamma^\mu \nu_{L\beta})\langle u_\mu [Q_-,\chi_+]\rangle\,.
\end{align}

Analogously, the corresponding chiral operators for $\Op_{3,4,5,6,7,8}^{8u/d}$ can be derived. 
It is noted that $\Op_{5,8}^{8u/d}$ and $\Op_{6,7}^{8u/d}$ share identical $CP$ properties and spurion structures with $\Op_{1,2}^{6u/d}$. 
Consequently, these higher-dimensional
operators yield the same chiral operators
and therefore contribute to the same physical processes as the dimension-6 operators.
However, the dimension-8 operators are always suppressed by a factor of order $q^2/\Lambda^2$ relative to their dimension‑6 counterparts. For this reason, the matching between LEFT and $\chi$PT for operators $\Op_{5,6,7,8}^{8u/d}$ is omitted in the present analysis.

In particular, $\Op_6^{7}$, $\Op_7^7$ and $\Op_9^{8}$ contain no quark bilinear and therefore transform in the trivial representation of $SU(2)_L\times SU(2)_R$. Accordingly, the corresponding chiral building blocks are $u_\mu$, $\chi_\pm$, and $\nu_L$, without additional spurions. 
The operator $\Op_7^{7}$ is absorbed into the QCD $\theta$ term, which is removed by an anomalous chiral rotation, and this mesonic sector is neglected in this work, as it does not contribute to CE$\nu$NS via the leading-order pion exchange. In addition, the covariant derivative $D_\mu$ in Eq.~\eqref{eq:CD} reduces to $\partial_\mu$ in the matching procedure, owing to the intrinsic properties of neutrinos.

Different chiral operators may involve different LECs. 
In Ref.~\cite{Li:2025xmq}, the spurion method has been compared to the external source method, which allows us to extract LECs for the corresponding chiral operators. 
In addition, LEFT operators of dimensions 7 and 8 are not included in the external source method; we therefore use NDA to estimate the corresponding LECs. 
Consequently, the matching results for the full set of LEFT operators in the mesonic sector can be written as
\begin{eqnarray} \label{eq:Lpion}
    \mathcal{L}_\pi&=&\frac{f^2}{2}\mathcal{\hat C}_{1,2}^{6u/d}(\bar\nu_{L\alpha}\gamma^\mu\nu_{L\beta})\langle Q_{\mp} u_\mu\rangle+\frac{f^2B_0m_q}{2} \mathcal{\hat C}_{4,5}^{7u/d}( \nu_{L\alpha}^TC \nu_{L\beta})\langle\Sigma_\pm\rangle\nn\\
    &&+\mathcal{\hat C}_{3}^{7u/d} m_q \Lambda_2( \nu_{L\alpha}^TC \sigma^{\mu\nu}\nu_{L\beta})\langle\Sigma_+[u_\mu,u_\nu]\rangle\nn\\
    &&+\mathcal{\hat C}_{1,2}^{7u/d}(\bar\nu_{L\alpha}\gamma^\mu\nu_{L\beta})\bigg\{g_1^\pi\langle \Sigma_\pm[\nabla^\nu u_\mu,u_\nu]\rangle+g_2^\pi\langle \nabla^\nu\Sigma_\pm[ u_\mu,u_\nu]\rangle\notag\\
    &&\quad+g_3^\pi\langle \Sigma_\pm[ u_\mu,\chi_-]\rangle+g_4^\pi\langle \Sigma_\mp[ u_\mu,\chi_+]\rangle+g_5^\pi\varepsilon_{\mu\nu\rho\lambda}\langle u^\nu u^\rho \nabla^\lambda \Sigma_\mp\rangle\bigg\}\nn\\
    &&+\mathcal{\hat C}_{6}^{7}( \nu_{L\alpha}^TC \nu_{L\beta})\bigg\{a_1\frac{f^2}{4}\langle u_\mu u^\mu\rangle+a_2\frac{f^2}{4}\langle\chi_+\rangle \bigg\}\nn\\
    &&+\mathcal{\hat C}_{1,2}^{8u/d}g_6^\pi(\bar \nu_{L\alpha}\gamma^\mu\lrpartial^\nu\nu_{L\beta})\langle Q_\pm\chi_+\rangle\langle u_\mu u_\nu\rangle\notag\\
    &&+\mathcal{\hat C}_{3,4}^{8u/d}g_7^\pi(\bar \nu_{L\alpha}\gamma^\mu \nu_{L\beta})\langle u_\mu [Q_\mp,\chi_+]\rangle\nn\\
    &&+\mathcal{\hat C}_9^{8} g_8^\pi(\bar \nu_{L\alpha}\gamma^\mu\lrpartial^\nu\nu_{L\beta})\langle u_\mu u_\nu\rangle+...\,,
\end{eqnarray}
where the tensor LEC is determined to be 
$\Lambda_2=6.0(3)\mev$~\cite{Baum:2011rm,Dekens:2018pbu,Hoferichter:2018zwu,Jiang:2012ir},
the LECs $a_1={4}/{27}$, $a_2={2}/{9}$~\cite{Donoghue:1990xh}, $(\nu_{L\alpha}\gamma^\mu\lrpartial^\nu\nu_{L\beta})=(\partial^\nu\nu_{L\alpha}\gamma^\mu\nu_{L\beta})-(\nu_{L\alpha}\gamma^\mu\partial^\nu\nu_{L\beta})$.
The dots denote terms that are of higher orders in the chiral expansion and are omitted.

The $u/d$ designation of the LEFT Wilson coefficients is tied to the spurion structure defined in Eq.~\eqref{eq:OpLEFTL0} via $\tau^{u/d}$. Furthermore, the subscript index of each Wilson coefficient corresponds to the $\pm$ designation of the associated spurion. For instance, $\hat{\mathcal{C}_{1}}^{6u/d}$ corresponds to the chiral operator $(\bar\nu_{L\alpha}\gamma^\mu\nu_{L\beta})\langle Q_-u_\mu\rangle$ and $\hat{\mathcal{C}_{2}}^{6u/d}$ corresponds to $(\bar\nu_{L\alpha}\gamma^\mu\nu_{L\beta})\langle Q_+u_\mu\rangle$. 
 The unknown LECs $g_1^\pi,\,g_2^\pi,\,g_3^\pi,\,g_4^\pi,\,g_5^\pi,\,g_6^\pi,\,g_7^\pi,\,g_8^\pi$ are estimated through NDA as detailed in App.~\ref{app:nda-matching} and the full results are summarized in Tab.~\ref{tab:estimate}.

\subsection{Nucleon Sector}
In the nucleon sector, the nucleon field $N$ is included as a degree of freedom in the fundamental representation of $SU(2)_V$ which is similar to Ref.~\cite{Song:2025snz}. All building blocks transform as
\begin{equation}
    \left(\begin{array}{c}
            \chi_\pm\\
            \Sigma_\pm\\
            Q_\pm\\
            N\\
            \bar N\\
            \nu_L\\
            \Bar{\nu}_L
    \end{array}\right)\rightarrow\left(\begin{array}{c}
            V \chi_\pm V^\dagger\\
            V\Sigma_\pm V^\dagger\\
            VQ_\pm V^\dagger\\ 
            VN\\
            \bar NV^\dagger\\
            \nu_L\\
            \Bar{\nu}_L
    \end{array}\right)\;,\quad V\in SU(2)_V\;,
\end{equation}
under the chiral symmetry. With the above framework established, the interactions within the nucleon sector are systematically constructed through the Young tensor technique. As a representative example, we consider the type $N^2\nu^2 Q$. The complete set of chiral operators at LO
is given by
\begin{align}
    &(\bar\nu_{L\alpha}\gamma^\mu\nu_{L\beta})(\bar N\gamma_\mu Q_\pm N)\,,\,\quad\quad(\bar\nu_{L\alpha}\gamma^\mu\nu_{L\beta})(\bar N\gamma_\mu N)\langle Q_\pm\rangle\,,\notag\\
    &(\bar\nu_{L\alpha}\gamma^\mu\nu_{L\beta})(\bar N\gamma^5\gamma_\mu Q_\pm N)\,,\,\quad(\bar\nu_{L\alpha}\gamma^\mu\nu_{L\beta})(\bar N\gamma^5\gamma_\mu N)\langle Q_\pm\rangle\,,\notag\\
    &(\nu_{L\alpha}^TC \nu_{L\beta})(\bar N Q_\pm N)\,,\,\quad\quad\quad\,\,(\nu_{L\alpha}^TC \nu_{L\beta})(\bar N  N)\langle Q_\pm\rangle\,,\notag\\
    &(\nu_{L\alpha}^TC \nu_{L\beta})(\bar N \gamma^5Q_\pm N)\,,\,\quad\,\quad\,(\nu_{L\alpha}^TC \nu_{L\beta})(\bar N \gamma^5 N)\langle Q_\pm\rangle\,,\notag\\
    &(\nu_{L\alpha}^TC \sigma^{\mu\nu}\nu_{L\beta})(\bar N \sigma_{\mu\nu}Q_\pm N)\,,\,\,\,(\nu_{L\alpha}^TC \sigma^{\mu\nu}\nu_{L\beta})(\bar N \sigma_{\mu\nu} N)\langle Q_\pm\rangle\,,
\end{align}
which correspond to some dimension-6 LEFT operators. For operator $\mathcal{O}_1^{6u/d}$, 
we construct the following LO hadronic operators with $C-P+$ symmetry that contain the neutrino current $(\bar\nu_{L\alpha}\gamma^\mu\nu_{L\beta})$:
\begin{align}
    &(\bar\nu_{L\alpha}\gamma^\mu\nu_{L\beta})(\bar N\gamma_\mu Q_+ N)\,,\quad(\bar\nu_{L\alpha}\gamma^\mu\nu_{L\beta})(\bar N\gamma_\mu N)\langle Q_+\rangle\,,\notag\\
    &(\bar\nu_{L\alpha}\gamma^\mu\nu_{L\beta})(\bar N\gamma^5\gamma_\mu Q_- N),\,(\bar\nu_{L\alpha}\gamma^\mu\nu_{L\beta})(\bar N\gamma^5\gamma_\mu N)\langle Q_-\rangle\,,
\end{align}
where only the first two terms are considered,
since the others generate a pion upon expanding $Q_-$ and are suppressed in the chiral power counting (see Sec.~\ref{sec:power}).
Thus, only the interactions without pions that contribute directly to neutrino-nucleon interactions are calculated in the framework of this analysis.

\begin{table}[H]
\begin{center}
\begin{tabular}{ccccc}
\hline\hline
LEC & value & LEC & value \\
\hline
$f$&$92.21(14){~\rm MeV}$&$g_A$&$1.2754(13)$\\
$B_0m_u$&$6200(400){~\rm MeV}^2$&$B_0m_d$&$13300(400){~\rm MeV}^2$\\
$L_1$&$1.855(14)$&$L_2$&$-1.017(15)$\\
$L_3$&$0.6377(7)$
&$L_4$&$-0.220(10)$\\
$L_5$&$1.38(17){~\rm MeV}$&$L_6$&$-0.49(8){~\rm MeV}$\\
$L_7$&$-6.76(37){~\rm MeV}$&$L_8$&$2.65(18){~\rm MeV}$\\
$L_9$&$11.52(25){~\rm MeV}$&$L_{10}$&$24.05(48){~\rm MeV}$\\
$L_{11}$&$-17.7(20){~\rm MeV}$&$L_{12}$&$-13.9(15){~\rm MeV}$\\
$L_{13}$&$-50.4(6){~\rm MeV}$&$L_{14}$&$-0.306(28) {~\rm GeV}$\\
\hline
$g_1^\pi$&$\mathcal{O}(\Lambda_\chi)$&$g_2^\pi$&$\mathcal{O}(\Lambda_\chi)$\\
$g_3^\pi$&$\mathcal{O}(\Lambda_\chi)$&$g_4^\pi$&$\mathcal{O}(\Lambda_\chi)$\\
$g_5^\pi$&$\mathcal{O}(\Lambda_\chi)$&$g_6^\pi$&$\mathcal{O}(1)$\\
$g^\pi_7$&$\mathcal{O}(\Lambda_\chi^2)$&$g_8^\pi$&$\mathcal{O}(\Lambda_\chi^2)$\\
$g_1$&$\mathcal{O}(\Lambda_\chi)$&$g_2$&$\mathcal{O}(\Lambda_\chi)$\\
$g_3$&$\mathcal{O}(\Lambda_\chi)$&$g_4$&$\mathcal{O}(\Lambda_\chi)$\\
$g_5$&$\mathcal{O}(1)$&$g_6$&$\mathcal{O}(1)$\\
$g_7$&$\mathcal{O}(1)$&$g_8$&$\mathcal{O}(1)$\\
$g_9$&$\mathcal{O}(1)$&&\\
\hline
\hline
\end{tabular}
\end{center}
\caption{Values of the LECs required for CE$\nu$NS. More details are given in App.~\ref{app:LECValues}. The unknown LECs are estimated using NDA. }
\label{tab:estimate}
\end{table}

In this work, we also include iso-singlet currents in the chiral operators; therefore the building blocks $\langle Q_\pm\rangle$ are retained. We can then obtain the chiral operators for LEFT operators without explicit pions; the corresponding Wilson coefficients and LECs are included as follows
\begin{eqnarray} \label{eq:LPionN1}
    \mathcal{L}_{\pi N}&=&\hat{\mathcal{C}}_{1}^{6u/d}(\bar\nu_{L\alpha}\gamma^\mu\nu_{L\beta})\bigg\{(\bar N\gamma_\mu Q_+ N)+(\bar N\gamma_\mu  N)\langle Q_+\rangle+\frac{L_1}{m_N}\partial^\nu(\Bar{N}\sigma_{\mu\nu} Q_+N)\nn\\
    &&\quad+\frac{L_2}{m_N}\partial^\nu(\Bar{N}\sigma_{\mu\nu} N)\langle Q_+\rangle\bigg\}+\hat{\mathcal{C}}_{2}^{6u/d}(\bar\nu_{L\alpha}\gamma^\mu\nu_{L\beta})\bigg\{L_3(\bar N\gamma^5\gamma_\mu Q_+ N)+L_4(\bar N\gamma^5\gamma_\mu N)\langle Q_+\rangle\bigg\}\nn\\
    &&+\hat{\mathcal{C}}_{3}^{7u/d}(\nu_{L\alpha}^TC \sigma^{\mu\nu}\nu_{L\beta})\bigg\{L_5(\bar N \sigma_{\mu\nu}\Sigma_+N)+L_6(\bar N \sigma_{\mu\nu}N)\langle\Sigma_+\rangle+\frac{L_7}{m_N}\partial_\nu(\bar N \gamma_{\mu}\Sigma_+N)\nn\\
    &&\quad+\frac{L_8}{m_N}\partial_\nu(\bar N \gamma_{\mu}N)\langle\Sigma_+\rangle\bigg\}+\hat{\mathcal{C}}_{4}^{7u/d}(\nu_{L\alpha}^TC \nu_{L\beta})\bigg\{L_9(\bar N \Sigma_+ N)+L_{10}(\bar N N)\langle\Sigma_+\rangle\bigg\}\nn\\
    &&+\hat{\mathcal{C}}_{5}^{7u/d}(\nu_{L\alpha}^TC \nu_{L\beta})\bigg\{L_{11}(\bar N \gamma^5 \Sigma_+N)+L_{12}(\bar N \gamma^5 N)\langle\Sigma_+\rangle\bigg\}\nn\\
    &&+\hat{\mathcal{C}}_{6}^{7}L_{13}(\nu_{L\alpha}^TC \nu_{L\beta})(\bar NN)+\hat{\mathcal{C}}_{7}^{7}L_{14}(\nu_{L\alpha}^TC \nu_{L\beta})(\bar N\gamma^5N)\nn\\
    &&+\hat{\mathcal{C}}_{1}^{7u/d}(\bar\nu_{L\alpha}\gamma^\mu\nu_{L\beta})\bigg\{g_1(\bar N\gamma_\mu\Sigma_+ N)+g_{2}(\bar N \gamma_\mu N)\langle\Sigma_+\rangle\bigg\}\nn\\
    &&+\hat{\mathcal{C}}_{2}^{7u/d}(\bar\nu_{L\alpha}\gamma^\mu\nu_{L\beta})\bigg\{\frac{g_{3}}{m_N}\partial^\nu(\bar N i\gamma^5\sigma_{\mu\nu}\Sigma_+N)+\frac{g_{4}}{m_N}\partial^\nu(\bar N i\gamma^5\sigma_{\mu\nu}N)\langle\Sigma_+\rangle\bigg\}\nn\\
    &&+\hat{\mathcal{C}}_{1}^{8u/d}(\bar \nu_{L\alpha}\gamma^\mu\lrpartial^\nu\nu_{L\beta})\bigg\{g_5(\bar N \gamma_\mu \lrpartial_\nu Q_+ N)+g_6(\bar N \gamma^\mu\lrpartial_\nu  N)\langle Q_+\rangle\bigg\}\nn\\
    &&+\hat{\mathcal{C}}_{2}^{8u/d}(\bar \nu_{L\alpha}\gamma^\mu\lrpartial^\nu\nu_{L\beta})\bigg\{g_7(\bar N \gamma^5\gamma_\mu \lrpartial_\nu Q_+N)+g_8(\bar N \gamma^5\gamma_\mu \lrpartial_\nu N)\langle Q_+\rangle\bigg\}\nn\\
    &&+\hat{\mathcal{C}}_{9}^{8}g_{9}(\bar \nu_{L\alpha}\gamma^\mu\lrpartial^\nu\nu_{L\beta})(\bar N\gamma_\mu \lrpartial_\nu N)+...\,,
\end{eqnarray}
where the LECs $L_{1\sim 14}$ can be calculated from experiment and lattice QCD results, and the details of these LECs are discussed in App.~\ref{app:LECValues}. The other LECs $g_{1\sim 9}$, which involve derivative dimension-7 and dimension-8 LEFT operators, can be estimated by NDA. All the relevant LECs are listed in Tab.~\ref{tab:estimate}.
It should be noted that the chiral operators we consider are independent, but their associated LECs may be related to other LECs. This follows the same methodology as Ref.~\cite{Fettes:2000gb}, which provides the complete list of chiral operators for the pion-nucleon Lagrangian at $O(p^4)$ for the QCD interactions.
Moreover, the chiral operators of $\mathcal{O}_{3,4}^{8u/d}$ are not presented in Eq.~\eqref{eq:LPionN1}, because the chiral operators of $\mathcal{O}_{3,4}^{8u/d}$ can not contribute to the neutrino-nucleon interactions directly and are highly suppressed.

Moreover, the chiral expansion $p/\Lambda_\chi$ is invalid due to the heavy nucleon mass $m_N \sim \Lambda_\chi$ of the nucleon field. The heavy baryon form~\cite{Krause:1990xc,Jenkins:1990jv} is needed for the nucleon field. According to the discussion above, the baryon is so heavy that it is non-relativistic when interacting with other degrees of freedom. It is therefore  natural to expand the operators containing baryons in terms of the inverse of their mass $1/m_N$. For any time-like vector $v^\mu$ satisfying $v^2 = 1$, we can define the projection operators as $P_v^{\pm} = (1 \pm v\!\!\!\slash)/2$, and the Dirac spinors are expressed as
\begin{equation}
    \Psi(x) = e^{-imv\cdot x}\left[\underbrace{e^{imv\cdot x} P_v^+\Psi(x)}_{\equiv N_v(x)} + \underbrace{e^{imv\cdot x}P_v^- \Psi(x)}_{\equiv \mathcal{H}_v(x)}\right]\,.
\end{equation}
Then a general pion-nucleon interaction is expanded as
\begin{equation}
\mathcal{L}_{ \pi N} = \bar{N_v}AN_v+\bar{\mathcal{H}_v}BN_v+\bar{N_v}\gamma^0B^\dagger\gamma^0\mathcal{H}_v -\bar{\mathcal{H}_v}C\mathcal{H}_v\,,
\end{equation}
where $A$, $B$, $C$ are composed of the mesonic fields. Integrating out the heavy component $\mathcal{H}_v$, and expanding $C^{-1}$ in inverse power of the nucleon mass, the Lagrangian is expanded in terms of $N_v$~\cite{Ecker:1995rk,Fettes:2000gb}, 
\begin{align}
    \mathcal{L}_{\pi N} \rightarrow &\bar{N_v} (A+\gamma^0B^\dagger\gamma^0C^{-1} B) N_v \notag\\
    &=\bar{N_v}(A^{(1)}+A^{(2)}+\frac{1}{2m_N}\gamma^0B^{\dagger{(1)}}\gamma^0B^{(1)}+...)N_v\,.
\end{align}
Then power counting in chiral Lagrangian for pion-nucleon interactions is consistent with Weinberg power counting~\cite{Weinberg:1978kz,Weinberg:1990rz,Weinberg:1991um}. In the heavy baryon framework, the nucleon bilinear can be expanded as
\begin{align}
\label{heavy baryon}
    \Bar{N} 1 N &= \Bar{N_v} 1 N_v +...\,, \notag \\
    \Bar{N} \gamma^5 N &= \frac{1}{m_N}\partial_\mu(\Bar{N_v} S^\mu N_v)+... \,, \notag \\
    \Bar{N} \gamma^\mu N &= \Bar{N_v} v^\mu N_v +\frac{1}{2m_N}\Bar{N_v}(i\overleftrightarrow\partial^\mu-v^\mu\slashed{\lrpartial}) N_v-\frac{1}{m_N}\varepsilon^{\mu\nu\rho\lambda}\partial_\nu(\Bar{N_v}v_\rho S_\lambda N_v)+...\,, \notag \\
    \Bar{N} \gamma^5\gamma^\mu N &= 2\Bar{N_v} S^\mu N_v -\frac{i}{m_N}v^\mu\Bar{N_v} S^\nu\overleftrightarrow\partial_\nu N_v...\,, \notag \\
    \Bar{N} \sigma^{\mu\nu} N &= -2\varepsilon^{\mu\nu\rho\lambda}\Bar{N_v} v_\rho S_\lambda N_v -\frac{1}{2m_N}(2i\varepsilon^{\mu\rho\lambda\sigma}\Bar{N_v} v_\sigma [S_\lambda,v^\nu]\overleftrightarrow\partial_\rho uN_v+[v^\mu ,\partial^\nu](\Bar{N_v} N_v))...\,, \notag \\
    \Bar{N} i\overleftrightarrow{D}^\mu N &= \Bar{N_v} v^\mu N_v +\frac{1}{2m_N}\Bar{N_v}i\overleftrightarrow\partial^\mu N_v-\frac{1}{m_N}\varepsilon^{\mu\nu\rho\lambda}\partial_\nu(\Bar{N_v}v_\rho S_\lambda N_v) +...\,,  
\end{align}
where $v_\mu$ is the four-velocity of the nucleon $N_v$, $S_\mu$ is the spin-operator of the nucleon, and the dots denote terms that are of higher order in the chiral expansion. The four-velocity $v_\mu$ and spin-operator satisfy the relations
\begin{equation}
    S^\mu=\frac{i}{2}\gamma^5\sigma^{\mu\nu}v^\nu\,,\quad S\cdot v=0\,,\quad\{S^\mu,S^\nu\}=\frac{1}{2}\left(v^\mu v^\nu-g^{\mu\nu}\right)\,,\quad\left[S^\mu,S^\nu\right]=i\varepsilon^{\mu\nu\rho\lambda}v_\rho S_\lambda\,.
\end{equation}

Thus, we consider the non-relativistic pion-nucleon interactions in the chiral expansion framework. Here, we expand the pion-nucleon interactions to $1/m_N$ order in the heavy baryon approach. Thus, the involved chiral operators can be selected as
\begin{eqnarray} \label{eq:LPionN}
    \mathcal{L}_{\pi N}^{\rm HB}&=&\hat{\mathcal{C}}_{1}^{6u/d}(\bar\nu_{L\alpha}\gamma^\mu\nu_{L\beta})\left\{(\bar N_vv_\mu Q_+ N_v)+\frac{1}{2m_N}(\bar N_v i\lrpartial^\mu Q_+ N_v)-\frac{1+2L_1}{m_N}\varepsilon^{\mu\nu\rho\lambda}\partial_\nu(\Bar{N_v}v_\rho S_\lambda Q_+N_v)\right\}\nn\\
    &&+\hat{\mathcal{C}}_{1}^{6u/d}(\bar\nu_{L\alpha}\gamma^\mu\nu_{L\beta})\left\{(\bar N_vv_\mu N_v)\langle  Q_+\rangle+\frac{1}{2m_N}(\bar N_v i\lrpartial^\mu N_v)\langle  Q_+\rangle-\frac{1+2L_2}{m_N}\varepsilon^{\mu\nu\rho\lambda}\partial_\nu(\Bar{N_v}v_\rho S_\lambda N_v)\langle Q_+\rangle\right\}\nn\\
    &&+\hat{\mathcal{C}}_{2}^{6u/d}(\bar\nu_{L\alpha}\gamma^\mu\nu_{L\beta})\bigg\{2L_3(\bar N_vS_\mu Q_+ N_v)-\frac{L_3}{m_N}(\Bar{N_v} v^\mu S^\nu i\lrpartial_\nu Q_+N_v)\bigg\}\nn\\
    &&+\hat{\mathcal{C}}_{2}^{6u/d}(\bar\nu_{L\alpha}\gamma^\mu\nu_{L\beta})\bigg\{2L_4(\bar N_vS_\mu N_v)\langle Q_+\rangle-\frac{L_4}{m_N}(\Bar{N_v} v^\mu S^\nu i\lrpartial_\nu N_v)\langle Q_+\rangle\bigg\}\nn\\
    &&-\hat{\mathcal{C}}_{3}^{7u/d}(\nu_{L\alpha}^TC \sigma^{\mu\nu}\nu_{L\beta})\bigg\{2L_5(\bar N_v v_\mu S_\nu\Sigma_+N_v)+\frac{L_5}{m_N}\varepsilon^{\mu\rho\lambda\sigma}\Bar{N_v} v_\sigma S_\lambda v^\nu i\lrpartial_\rho \Sigma_+N_v\nn\\
    &&\quad\quad+\frac{L_5+L_7}{m_N}\partial_\mu(\bar N_vv_\nu\Sigma_+N_v)\bigg\}-\hat{\mathcal{C}}_{3}^{7u/d} (\nu_{L\alpha}^TC \sigma^{\mu\nu}\nu_{L\beta})\bigg\{2L_6(\bar N_v v_\mu S_\nu N_v)\langle\Sigma_+\rangle\nn\\
    &&\quad\quad\quad\quad\quad+\frac{L_6}{m_N}\varepsilon^{\mu\rho\lambda\sigma}(\Bar{N_v} v_\sigma S_\lambda v^\nu i\lrpartial_\rho N_v)\langle\Sigma_+\rangle+\frac{L_6+L_8}{m_N}\partial_\mu(\bar N_vv_\nu N_v)\langle\Sigma_+\rangle\bigg\}\nn\\
    &&+\hat{\mathcal{C}}_{4}^{7u/d}(\nu_{L\alpha}^TC \nu_{L\beta})\bigg\{L_9(\bar N_v \Sigma_+ N_v)+L_{10}(\bar N_v N_v)\langle\Sigma_+\rangle\bigg\}\nn\\
    &&+\hat{\mathcal{C}}_{5}^{7u/d}(\nu_{L\alpha}^TC \nu_{L\beta})\bigg\{\frac{L_{11}}{m_N}\partial^\mu(\bar N_v S_\mu \Sigma_+N_v)+\frac{L_{12}}{m_N}\partial^\mu(\bar N_v S_\mu N_v)\langle\Sigma_+\rangle\bigg\}\nn\\
    &&+\hat{\mathcal{C}}_{6}^{7}L_{13}(\nu_{L\alpha}^TC \nu_{L\beta})(\bar N_vN_v)+\hat{\mathcal{C}}_{7}^{7}\frac{L_{14}}{m_N}(\nu_{L\alpha}^TC \nu_{L\beta})\partial^\mu(\bar N_vS_\mu N_v)\nn\\
    &&+\hat{\mathcal{C}}_{1}^{7u/d}(\bar\nu_{L\alpha}\gamma^\mu\nu_{L\beta})\bigg\{g_1(\bar N_vv_\mu\Sigma_+ N_v)+g_2(\bar N_v v_\mu N_v)\langle\Sigma_+\rangle\bigg\}\nn\\
    &&+\hat{\mathcal{C}}_{2}^{7u/d}(\bar\nu_{L\alpha}\gamma^\mu\nu_{L\beta})\left\{\frac{2g_3}{m_N}\partial^\nu(\bar N_vS_{[\nu} v_{\mu]} \Sigma_+N_v)+\frac{2g_4}{m_N}\partial^\nu(\bar N_vS_{[\nu} v_{\mu]} N_v)\langle\Sigma_+\rangle\right\}\nn\\
    &&+\hat{\mathcal{C}}_{1}^{8u/d}(\bar \nu_{L\alpha}\gamma^\mu\lrpartial^\nu\nu_{L\beta})\bigg\{g_5(\bar N_v v_\mu v_\nu Q_+ N_v)+g_6(\bar N_v v_\mu v_\nu  N_v)\langle Q_+\rangle\bigg\}\nn\\
    &&+\hat{\mathcal{C}}_{2}^{8u/d}(\bar \nu_{L\alpha}\gamma^\mu\lrpartial^\nu\nu_{L\beta})\bigg\{2g_7(\bar N_v S_\mu v_\nu Q_+N_v)+2g_8(\bar N_v S_\mu v_\nu N_v)\langle Q_+\rangle\bigg\}\nn\\
    &&+\hat{\mathcal{C}}_{9}^{8}g_9(\bar \nu_{L\alpha}\gamma^\mu\lrpartial^\nu\nu_{L\beta})(\bar N_vv_\mu v_\nu N_v)+...\,,
\end{eqnarray}
where the dots denote terms that are of higher order in the chiral expansion and are omitted.

In addition, the standard pion-nucleon Lagrangian has been constructed and we consider the LO for the pion-nucleon interaction
\begin{equation}
\label{eq:pn}
    \mathcal{L}_{\pi N}=(\bar N_v iv_\mu \nabla^\mu +g_AS_\mu u^\mu N_v)\,.
\end{equation}
There is also the photon-hadron interaction generated from the QED covariant derivative. These interactions are given by
\begin{align}
\label{AQED}
    \mathcal{L}_{\pi N}^{\rm QED}=A^\mu(\bar N_v v_\mu Q_+^A N_v)+\frac{\varepsilon_{\mu\nu\rho\lambda}}{m_N}\partial^\nu A^\mu\left[g_\mu(\bar N_v v^\rho S^\lambda Q_+^AN_v)+g_\mu^\prime(\bar N_v v^\rho S^\lambda N_v)\langle Q_+^A\rangle\right]+...\,,
\end{align}
where $A_\mu$ is the photon field, and $Q_+^A$ is the spurion.
The dots denote terms that are of higher orders in the chiral expansion and are omitted.
Here, interactions containing two spurion insertions are not considered, since they contribute only at higher orders. Moreover, the last two terms can contribute to the proton and neutron magnetic moments $(\mu^p/m_N, \mu^n/m_N)$. Thus, the values of the coefficients $g_\mu$ and $g_\mu^\prime$ should be fixed to $g_\mu=\mu^p-\mu^n$ and $g_\mu^\prime=\mu^p+2\mu^n$.

\section{Power Counting in LEFT and \tf{$\chi$PT}{ChPT}}
\label{sec:power}

At low energies, neutrino-nucleus scattering is described within $\chi$PT. 
Ref.~\cite{Altmannshofer:2018xyo} considered only the chirally leading contribution for each LEFT operator. Here, we adopt Weinberg power counting~\cite{Weinberg:1978kz,Weinberg:1990rz,Weinberg:1991um,Weinberg:1992yk,Cirigliano:2012pq,Hoferichter:2015ipa,Klos:2018idj}, in which amplitudes are organized as an expansion in a soft momentum scale $q$ relative to the chiral scale $\Lambda_\chi$, and use this counting to estimate the size of different contributions to CE$\nu$NS. The procedure consists of two steps. First, we determine the chiral order of each vertex by expanding the chiral building blocks that appear in the operators. 
Second, we determine the scaling of the amplitude by combining the chiral orders of the vertices, propagators, and loop factors for a given diagram topology.

For each LEFT operator in Eqs.~\eqref{eq:OpLEFTL0}, \eqref{eq:OpLEFTL2}, and \eqref{eq:LPionN1}, the momentum dependence of the associated vertices follows from the $\chi$PT expansion of the building blocks appearing in the corresponding chiral operators. The standard building blocks ($u_\mu$, $\Sigma_\pm$, $Q_\pm$, $\chi_\pm$, $\nabla_\mu$) have been introduced above, and their explicit expansions are summarized in App.~\ref{app:chiral-expansion}. At leading chiral order, $u_\mu$, $\Sigma_{-}$, $Q_{-}$, and $\chi_{-}$ start at $\mathcal{O}(\pi)$ and therefore contain an odd number of pion fields, whereas $\nabla_\mu$, $\Sigma_{+}$, $Q_{+}$, and $\chi_{+}$ contain an even number of pion fields and include a nonvanishing pion-independent term. Each pion field attached to a vertex contributes a factor $1/(2f)$, so that a term with $n_\pi$ pion fields carries an additional factor $[1/(2f)]^{n_\pi}$. The overall normalization of the induced chiral operators is fixed using the NDA matching rules~\cite{Gavela:2016bzc,Jenkins:2013sda,Song:2025snz} given in App.~\ref{app:nda-matching}; representative estimates are given in Tab.~\ref{tab:estimate}. In addition, insertions proportional to the light-quark mass $m_q$ count as $\mathcal{O}(q^2)$.

As an explicit example of vertex counting, consider the chiral operator
\begin{align}
\label{eq:ex-op}
  \mathcal{\hat C}_{1}^{6u/d}\,(\bar\nu_{L\alpha}\gamma^\mu\nu_{L\beta})\,\langle Q_{-} u_\mu\rangle \, .
\end{align}
According to the NDA matching rule in Eq.~\eqref{eq:NDA_2}, its overall normalization scales as $(\Lambda_\chi/\Lambda_{\rm EW})^2$ where $\Lambda_{\rm EW}$ denotes the scale corresponding to LEFT operators. The leading nonvanishing contribution arises from the two-pion term with $n_\pi=2$, which supplies a factor $[1/(2f)]^{2}$. Moreover, $u_\mu$ contains one derivative and therefore contributes a factor $\mathcal{O}(q)$. 
The leading chiral order of the two-pion vertex is 
\begin{align}
\label{eq:ex-scaling}
  \left(\frac{\Lambda_\chi}{\Lambda_{\rm EW}}\right)^2
  \left(\frac{q}{\Lambda_\chi}\right)
  \left(\frac{1}{2f}\right)^{2}\, .
\end{align}

For the full scattering process, power counting requires accounting not only for the chiral orders of the vertices, but also for the orders associated with propagators, loops, and the diagram topology.
In multi-nucleon systems, intermediate states can include purely nucleonic configurations in which nucleons interact through pion exchange and short-range contact interactions. Analogous two-body current effects have been discussed in the EFT literature on dark-matter-nucleus scattering~\cite{Cirigliano:2012pq,Korber:2017ery,deVries:2023sux}.
In Weinberg power counting, the pion mass is counted as soft, $m_\pi\sim \mathcal{O}(q)$~\cite{Weinberg:1990rz,Weinberg:1991um,Weinberg:1992yk,Klos:2018idj}. For irreducible diagrams, i.e.\ diagrams that do not contain purely nucleonic intermediate states that can go on shell, pion and nucleon propagators scale as $1/q^{2}$ and $1/q$, respectively, and each loop contributes a factor $q^{4}/(16\pi^{2})$. For reducible diagrams, which do contain purely nucleonic intermediate states, the non-relativistic nucleon propagator scales as $m_{N}/q^{2}$, each loop contributes $q^{5}/(16\pi^{2} m_{N})$, and pion propagators still scale as $1/q^{2}$~\cite{Cirigliano:2012pq}.

\begin{figure}[t]
    \centering
    \subfloat[]{\includegraphics[scale=0.8]{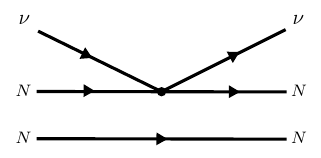}}
    \subfloat[]{\includegraphics[scale=0.8]{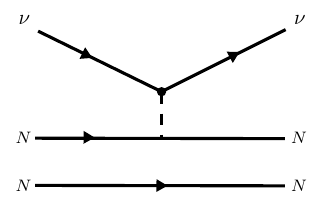}} 
    \subfloat[]{\includegraphics[scale=0.8]{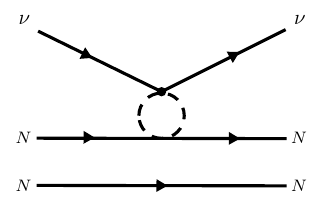}} \\
    \subfloat[]{\includegraphics[scale=0.8]{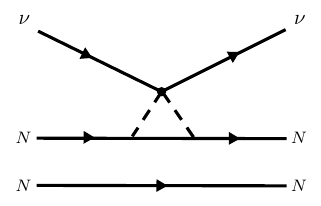}}
    \subfloat[]{\includegraphics[scale=0.8]{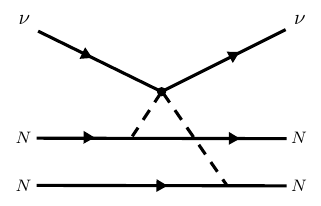}}
    \caption{
    Diagrams (a) and (b) show the leading-order one-body current contributions. Diagrams (c) and (d) show loop corrections to the one-body currents. Diagram (e) shows the two-body current contribution.}
        \label{fig:OneTwoBodyDiag}
\end{figure}

Different chiral operators contribute to different topologies in Fig.~\ref{fig:OneTwoBodyDiag} because the expansions of the operators contain different numbers of pion fields. Operators whose leading term involves no pion fields generate one-body $\nu N$ contact currents in diagram~(a), whereas operators whose leading term contains a single pion field induce long-range contributions mediated by single-pion exchange in diagram~(b).
In the following, we work at leading order with the mesonic Lagrangian in Eq.~\eqref{eq:Lpion} and the pion-nucleon Lagrangian in Eq.~\eqref{eq:LPionN}. NP effects enter only through semileptonic LEFT operators acting on the leptonic current, while all $\pi N$ vertices are from the SM.

Separating the operator normalization from the chiral order generated by vertices, propagators, and loops, we parametrize the scaling factor of the contributions to the amplitude as
\begin{align}
\label{eq:CountingFormula}
  \left(\frac{\Lambda_{\chi}}{\Lambda_{\rm EW}}\right)^{d-4}
  \left(\frac{q}{\Lambda_{\chi}}\right)^{n}
  \left(\frac{q}{m_N}\right)^{m}
  \sim \mathcal{O}(\Lambda_{\rm EW}^{4-d}) \cdot \mathcal{O}(\Lambda_{\chi}^{-n}) \cdot \mathcal{O}(m_N^{-m})\,,
\end{align}
for an amplitude induced by a dimension-$d$ LEFT operator, where the overall factor $(\Lambda_{\chi}/\Lambda_{\rm EW})^{d-4}$ is fixed by the operator dimension and follows from the NDA matching rules. The integer $n$ collects the chiral suppression in powers of $q/\Lambda_\chi$ arising from derivatives, insertions of $m_\pi$, and the counting of propagators and loops. For bookkeeping, the integer $m$ tracks explicit $1/m_N$ effects from the heavy baryon $\chi$PT Lagrangian, written as a separate factor $(q/m_N)^m$.

Applying this counting to the diagrams in Fig.~\ref{fig:OneTwoBodyDiag}, we focus on the leading contributions induced by dimension-6 LEFT operators, which carry the electroweak suppression $\mathcal{O}(\Lambda_{\rm EW}^{-2})$. We further retain only the leading term in the $1/m_N$ expansion, i.e.\ $\mathcal{O}(m_N^{0})$. Under these assumptions, the contributions to the amplitude from the tree-level one-body currents in Fig.~\ref{fig:OneTwoBodyDiag} (a) and (b) scale as
\begin{equation}
   \mathcal{O}(\Lambda_{\rm EW}^{-2}) \cdot
   \mathcal{O}(\Lambda_{\chi}^{0}) \cdot
   \mathcal{O}(m_N^{0}) \,,
\end{equation}
while one-loop one-body current contributions in Fig.~\ref{fig:OneTwoBodyDiag}~(c) and (d) scale as
\begin{equation}
    \mathcal{O}(\Lambda_{\rm EW}^{-2}) \cdot
    \mathcal{O}(\Lambda_{\chi}^{-2}) \cdot
    \mathcal{O}(m_N^{0}) \, .
\end{equation}
Moreover, the two-body current diagram in Fig.~\ref{fig:OneTwoBodyDiag}~(e) is irreducible and shares the same scaling,
\begin{equation}
   \mathcal{O}(\Lambda_{\rm EW}^{-2}) \cdot
   \mathcal{O}(\Lambda_{\chi}^{- 2}) \cdot
   \mathcal{O}(m_N^{0}) \, .
\end{equation}

Thus, diagrams (a) and (b) contribute to the amplitude at the same chiral order, while the one-loop one-body currents in diagrams (c) and (d) and the tree-level two-body current in diagram (e) are suppressed by two additional powers of $q/\Lambda_\chi$ with respect to the leading tree-level one-body amplitude. Diagrams with three-body currents or with two loops are further suppressed in our power counting and are not considered. The corresponding chiral orders for one- and two-body neutrino-nucleus currents induced by different LEFT operators can be expressed in terms of $\varepsilon_\chi \equiv q/\Lambda_\chi$ and $\varepsilon_N \equiv q/m_N$.

\begin{table}[h]
\centering
\begin{tabular}{c c c c c}
\hline \hline
LEFT operator & One-body (a)  &  One-body (b)  & One-body (c, d) & Two-body (e)  \\
\hline
$\mathcal{O}^{6u/d}_{1}$
& $1$ 
& $-$
& $\varepsilon_\chi^2$
& $\varepsilon_\chi^2$
\\
\hline
$\mathcal{O}^{6u/d}_{2}$
& $1$
& $1$
& $-$
& $-$
\\
\hline
\hline
$\mathcal{O}^{7u/d}_{1}$
& $1$
& $-$
& $\varepsilon_\chi^2$
& $\varepsilon_\chi^2$
\\
\hline
$\mathcal{O}^{7u/d}_{2}$
& $\varepsilon_N$
& $-$
& $\varepsilon_\chi^2$
& $\varepsilon_\chi^2$
\\ 
\hline
$\mathcal{O}^{7u/d}_{3}$
& $\varepsilon_\chi^2$
& $-$
& $\varepsilon_\chi^4$
& $\varepsilon_\chi^4$
\\
\hline
$\mathcal{O}^{7u/d}_{4}$
& $\varepsilon_\chi^2$
& $-$
& $\varepsilon_\chi^4$
& $\varepsilon_\chi^4$
\\
\hline
$\mathcal{O}^{7u/d}_{5}$
& $\varepsilon_\chi^2 \varepsilon_N$
& $\varepsilon_\chi$
& $-$
& $-$
\\
\hline
$\mathcal{O}^{7}_{6}$
& $\varepsilon_\chi^2$
& $-$
& $\varepsilon_\chi^4$
& $\varepsilon_\chi^4$
\\
\hline
$\mathcal{O}^{7}_{7}$
& $\varepsilon_\chi^2 \varepsilon_N$
& $-$
& $-$
& $-$
\\
\hline
\hline
$\mathcal{O}^{8u/d}_{1}$
& $1$
& $-$
& $\varepsilon_\chi^2$
& $\varepsilon_\chi^2$
\\
\hline
$\mathcal{O}^{8u/d}_{2}$
& $1$
& $-$
& $-$
& $-$
\\
\hline
$\mathcal{O}^{8}_{9}$
& $1$
& $-$
& $\varepsilon_\chi^2$
& $\varepsilon_\chi^2$
\\
\hline
\hline
\end{tabular}
\caption{Chiral power counting for one- and two-body neutrino-nucleus currents induced by LEFT operators. Entries indicate the leading suppression in powers of $q/\Lambda_\chi$ and $q/m_N$ in the amplitude, as in Eq.~\eqref{eq:CountingFormula}; we define $\varepsilon_\chi \equiv q/\Lambda_\chi$ and $\varepsilon_N \equiv q/m_N$. 
The tree-level one-body currents contribute at the leading chiral order, while one-loop one-body currents and two-body currents appear at higher orders.}
\label{tab:chiral-counting}
\end{table}

Tab.~\ref{tab:chiral-counting} shows that all dimension-6 operators contributing to tree-level one-body currents appear at the same chiral order as diagrams (a) and (b), while one-loop one-body currents and tree-level two-body currents are suppressed by 
$\varepsilon_\chi^2$. 
Dimension-7 and dimension-8 operators can generate chiral structures analogous to those from dimension-6 operators; despite the additional $(\Lambda_\chi/\Lambda_{\rm EW})$ suppression, their ultraviolet origin may differ, so their contributions are not necessarily negligible in well-motivated scenarios.

In summary, the power counting implies that contributions from one-body currents dominate CE$\nu$NS. We therefore focus on tree-level one-body contributions hereafter, leaving the inclusion of one-loop one-body currents and tree-level two-body currents to future work. 
In addition to the one-body currents at leading  chiral order, which are depicted in Fig.~\ref{fig:OneTwoBodyDiag}~(a) and (b), 
we incorporate contributions that are subleading in chiral expansion into the calculations of the CE$\nu$NS cross section. It should be noted that Ref.~\cite{Altmannshofer:2018xyo} omits the mesonic chiral Lagrangian; in that approach, the chirally leading pion-exchange contributions illustrated in Fig.~\ref{fig:OneTwoBodyDiag}~(b) are incorporated through the form factors, which will be discussed in the next section.

\enlargethispage{1.2\baselineskip}
\vspace{-0.8\baselineskip}
\pagebreak[0]

\section{Neutrino-Nucleus Scattering Cross Section}
\label{sec:CS}

Having established the power counting down to the hadronic level for CE$\nu$NS, we will derive the leading one-body non-relativistic neutrino-nucleon interactions from the chiral Lagrangian, and express the scattering amplitude in terms of nuclear response functions using a multipole expansion~\cite{Serot:1978vj,Haxton:2008zza} to obtain the differential cross section.

\subsection{Interactions of Neutrinos with Non-relativistic Nucleons}

From diagrams (a) and (b) in Fig.~\ref{fig:OneTwoBodyDiag}, the neutrino interaction with non-relativistic nucleons is described by the effective Lagrangian
\begin{align}
    \mathcal{L}_{\rm NR}=c_{i,N} \, \mathcal{O}_{i,N} \,,
\end{align}
where $N=p,n$ labels the nucleon species. 
The momenta of the incoming and outgoing neutrinos (nucleons) are denoted by $p_1$ and $p_2$ ($k_1$ and $k_2$), and we define
\begin{equation}
    q = p_{1} - p_{2}\,, \qquad
    P = p_{1} + p_{2}\,, \qquad
    K = k_{1} + k_{2}\,.
\end{equation}
The nuclear recoil energy is defined as $T_{\rm nr}=|q^{0}|$, where $q^0$ is the zero component of the momentum transfer $q$.
The incoming neutrino energy is $E_{\nu}=p_{1}^{0}$.

The non-relativistic neutrino-nucleon interactions are obtained by evaluating the Feynman diagrams (a) and (b) in Fig.~\ref{fig:OneTwoBodyDiag} using the chiral Lagrangian introduced in Sec.~\ref{sec:ChEFT}. The neutrino-nucleon interaction terms in Eq.~\eqref{eq:LPionN} contribute directly to diagram (a). In addition, diagram (b) receives contributions from the neutrino-pion interactions in Eq.~\eqref{eq:Lpion} together with the pion-nucleon interactions in Eq.~\eqref{eq:pn}. Moreover, the dipole operator $\mathcal{O}_1^5$ induces neutrino-nucleon interactions via the photon-nucleon couplings in Eq.~\eqref{AQED}. Consequently, the corresponding NR operators are given by
\begin{align} \label{eq:NROperators}
    \mathcal{O}_{1,p}=&
    (\bar{\nu}_{L\alpha}\gamma^\mu\nu_{L\beta})v_\mu(\bar p_vp_v)\,,&\mathcal{O}_{2,p}=&(\bar{\nu}_{L\alpha}\gamma^\mu\nu_{L\beta})(\bar p_vS_{\mu}p_v)\,, \\
    \mathcal{O}_{3,p}=&(\nu_{L\alpha}^TC\nu_{L\beta})(\bar p_vp_v)\,,&\mathcal{O}_{4,p}=&(\nu_{L\alpha}^TC\sigma^{\mu\nu}\nu_{L\beta})v_\nu(\bar p_vS_{\mu}p_v)\,,\\
    \mathcal{O}_{5,p}=&(\bar{\nu}_{L\alpha}\gamma^\mu\nu_{L\beta})(\bar p_v\frac{K^\mu}{2m_N}p_v)\,,&\mathcal{O}_{6,p}=&\varepsilon_{\mu\nu\rho\lambda}(\bar{\nu}_{L\alpha}\gamma^\mu\nu_{L\beta})v^\rho(\bar p_v\frac{q^\nu S^\lambda}{m_N}p_v)\,,\\
    \mathcal{O}_{7,p}=&(\bar{\nu}_{L\alpha}\gamma^\mu\nu_{L\beta})v_\mu(\bar p_v\frac{K_\nu S^\nu}{2m_N}p_v)\,,&\mathcal{O}_{8,p}=&\varepsilon_{\mu\rho\lambda\sigma}(\nu_{L\alpha}^TC\sigma^{\mu\nu}\nu_{L\beta})v_\nu v_\sigma(\bar p_v\frac{K^\rho S^\lambda}{m_N}p_v)\,,\\
    \mathcal{O}_{9,p}=&(\nu_{L\alpha}^TC\sigma^{\mu\nu}\nu_{L\beta})v_\nu(\bar p_v \frac{q_\mu}{m_N} p_v)\,,&\mathcal{O}_{10,p}=&(\nu_{L\alpha}^TC\nu_{L\beta})(\bar p_v\frac{S^\mu q_\mu}{m_N} p_v)\,,\\
    \mathcal{O}_{11,p}=&(\bar{\nu}_{L\alpha}\gamma^\mu\nu_{L\beta})v_\mu(\bar p_v\frac{q^\nu S_\nu}{m_N}p_v)\,,&\mathcal{O}_{12,p}=&\varepsilon_{\mu\nu\rho\lambda}(\nu_{L\alpha}^TC\nu_{L\beta})\frac{P^\mu q^\nu}{m_N^2} v^\rho(\bar p_v S^\lambda p_v)\,,\\
    \mathcal{O}_{13,p}=&(\bar{\nu}_{L\alpha}\gamma^\mu\nu_{L\beta})P^\nu v_\mu v_\nu(\bar p_vp_v)\,,&\mathcal{O}_{14,p}=&(\bar{\nu}_{L\alpha}\gamma^\mu\nu_{L\beta})P^\nu v_\nu(\bar p_vS_\mu p_v)\,.
\end{align}

According to the chiral operators in Sec.~\ref{sec:ChEFT}, the corresponding coefficients are
\begin{align}
    c_{1,p}&=2\hat{\mathcal{C}}_1^{6u}+\hat{\mathcal{C}}_1^{6d}+(g_1+g_2)\hat{\mathcal{C}}_1^{7u}+g_2\hat{\mathcal{C}}_1^{7d}\,,\\
    c_{2,p}&=2(L_3+L_4) \hat{\mathcal{C}}_2^{6u}+2L_4 \hat{\mathcal{C}}_2^{6d}\,,\\
    c_{3,p}&=\sigma^p_u\hat{\mathcal{C}}_4^{7u}+\sigma^p_d\hat{\mathcal{C}}_4^{7d}+L_{13}\hat{\mathcal{C}}_{6}^{7}-\frac{2\alpha \bar E_\nu}{\pi q^2}\hat{\mathcal{C}}_1^5\,,\\
    c_{4,p}&=2(L_5+L_6)\hat{\mathcal{C}}_3^{7u}+2L_6\hat{\mathcal{C}}_3^{7d}\,,\\
    c_{5,p}&=2\hat{\mathcal{C}}_1^{6u}+\hat{\mathcal{C}}_1^{6d}\,,\\
    c_{6,p}&=(2+2L_1+2L_2)\hat{\mathcal{C}}_1^{6u}+(1+2L_2)\hat{\mathcal{C}}_1^{6d}\,,\\
    c_{7,p}&=2(L_3+L_4)\hat{\mathcal{C}}_2^{6u}+2L_4 \hat{\mathcal{C}}_2^{6d}\,,\\
    c_{8,p}&=(L_5+L_6)\hat{\mathcal{C}}_3^{7u}+L_6\hat{\mathcal{C}}_3^{7d}\,,\\
    c_{9,p}&=(L_5+L_6+L_7+L_8)\hat{\mathcal{C}}_3^{7u}+(L_6+L_8)\hat{\mathcal{C}}_3^{7d}\,,\\
    c_{10,p}&=\frac{B_0g_A m_N}{m_\pi^2-q^2}(m_u\hat{\mathcal{C}}_5^{7u}-m_d\hat{\mathcal{C}}_5^{7d})+(L_{11}+L_{12})\hat{\mathcal{C}}_5^{7u}+L_{12}\hat{\mathcal{C}}_5^{7d}+L_{14}\hat{\mathcal{C}}_{7}^{7}\,,\\
c_{11,p}&=4(g_3+g_4)\hat{\mathcal{C}}_2^{7u}+4g_4\hat{\mathcal{C}}_2^{7d}\,,\\
    c_{12,p}&=\frac{\alpha m_N}{3\pi q^2}(1+2L_1)\hat{\mathcal{C}}^{5}_1\,,\\
    c_{13,p}&=(g_5+g_6)\hat{\mathcal{C}}_1^{8u}+g_6\hat{\mathcal{C}}_1^{8d}+g_9\hat{\mathcal{C}}_9^{8}\,,\\
    c_{14,p}&=2(g_7+g_8)\hat{\mathcal{C}}_{2}^{8u}+2g_8\hat{\mathcal{C}}_{2}^{8d}\,,
\end{align}
where $\bar E_\nu$ is the average energy of the neutrino before and after scattering. The analogous NR operators and corresponding coefficients for neutrons are obtained from the above expressions with $p \to n$. The matching relation between the scalar quark current and the nucleon current has been described by the nucleon sigma terms $\sigma_{q}^N$, which can be evaluated through the LECs. The values of sigma terms are given by~\cite{Hoferichter:2023ptl}
\begin{align}
\label{eq:sigmaterm}
    &\sigma^p_u=18.5(2.5) {~\rm MeV}\,,\quad \sigma^p_d=36.0(2.5){~\rm MeV}\,,\nn\\
    &\sigma^n_u=16.7(1.2){~\rm MeV}\,,\quad \sigma^n_d=39.7(2.4){~\rm MeV}\,.
\end{align}
More details of the relation between LECs and sigma terms are given in App.~\ref{app:LECValues}. The operators $\mathcal{O}_{3,p}$ and $\mathcal{O}_{12,p}$ can be generated via photon exchange induced by the QED dipole operator $\mathcal{O}_1^5$. The SM photon-nucleon interactions are given in Eq.~\eqref{AQED}. Notably, for neutron scattering, the contribution mediated by $\mathcal{O}_1^5$ does not generate $\mathcal{O}_{3,n}$ because the neutron carries zero electric charge. In addition, the pseudo-scalar current can contribute to the neutrino-nucleon interactions via the pion exchange, while its direct contact contributions are at higher chiral order relative to the pion-exchange channel. Both types of contributions are included in our analysis. Specifically, the operator $\mathcal{O}_5^{7u/d}$ contributes to the one-body pion-exchange current shown in Fig.~\ref{fig:OneTwoBodyDiag}~(b), which is included in the coefficient $c_{10,p}$. This effect corresponds to the pion-pole contribution encoded in the pseudo-scalar form factor $F_P^{q/N}$ in the approach of Ref.~\cite{Altmannshofer:2018xyo}. By contrast, the axial-vector current contributions arising from pion exchange are proportional to the neutrino masses and are therefore neglected. Consequently, the contribution from $\mathcal{O}_2^{6u/d}$ via pion exchange is not included in the coefficients $c_{i,N}$ above, in agreement with Ref.~\cite{Altmannshofer:2018xyo}.

\subsection{Nuclear Response}
With the one-body neutrino-nucleon interaction specified, we now incorporate nuclear-structure effects by performing a multipole expansion of the nuclear current. The effective operators can be written as matrix elements between external neutrino and nuclear states. Since the neutrino is ultra-relativistic while nucleons inside the nucleus are well described in the non-relativistic limit, we factorize the matrix element into leptonic and hadronic pieces,
\begin{equation}
    \langle f|\mathcal{O}|i\rangle \sim
    \langle \nu_{L\alpha}' |\mathcal{O}_{\nu}| \nu_{L\beta}\rangle
    \langle N' |\mathcal{O}_{N}| N \rangle \,,
\end{equation}
where $\langle f|$ and $|i\rangle$ denote the final and initial states, respectively, and $|N\rangle=|p\rangle,|n\rangle$ denotes a proton or neutron state.

We formulate the nuclear dynamics in terms of non-relativistic three-vector components. In the laboratory frame, the nucleon four-velocity is $v^{\mu}=(1,0,0,0)$, which contracts with the neutrino Dirac structures. The spin four-vector $S_\mu$ then reduces to the three-component nucleon spin vector $\vec{S}_N$. The momentum transfer $q_\mu$ can be identified with its spatial component $\vec{q}$, while $K_\mu$ corresponds to the nucleon three-momentum $\vec{K}_N$. 
The NR operators are classified into the charge ($\ell_{0}$), axial charge ($\ell_0^A$), axial vector ($\bm{\ell}_5$), vector magnetic ($\bm{\ell}_M$) and vector electric ($\bm{\ell}_E$) ~\cite{Fitzpatrick:2012ix,Anand:2013yka}, where the $\ell$ presents the neutrino bilinears and collects the corresponding coefficients:
\begin{align}
\label{eq:LDEFT}
\ell_0 \cdot 1_N + \bm{\ell}_5 \cdot (2\vec{S}_{N}) + \ell_{0,A} \cdot \left( {\vec{K} \cdot \vec{S}_{N} \over m_N}\right)   +  \bm{\ell}_M \cdot \left( {\vec{K}_N  \over 2m_N} \right)+  \bm{\ell}_E \cdot \left(-i~ {\vec{K} \times \vec{S}_{N} \over m_N} \right) \,,
\end{align}
where $1_{N}=\bar{N}N$, $ \vec{S}_N = \bar N \vec{S} N $, $ \vec{K}_N = \bar N \vec{K} N $. The operators containing nucleus information ($\vec{K}$, $\vec{S}$) can be separated as follows:
\begin{equation} \label{eq:non-reO}
    \Big\{1_N, 2\vec{S}_{N}, {\vec{K} \cdot \vec{S}_{N} \over m_N}, {\vec{K}  \over 2m_N} , -i~ {\vec{K}_{N} \times \vec{S}_{N} \over m_N} \Big\}  \,.
\end{equation}
The remaining coefficients and the neutrino bilinears do not carry any
information about the nuclear structure and are therefore grouped into the
leptonic structure introduced above. In our framework, the operators $1_N$ and $2\vec S_N$ arise already at leading order in the chiral expansion, which have been considered in Ref.~\cite{Altmannshofer:2018xyo}, while the operators with the momentum-dependent structures $\vec K \cdot \vec S_N/m_N$, $\vec K/(2m_N)$ and $-i\,\vec{K}_{N} \times \vec S_N/m_N$ are suppressed by one power of $1/m_N$.

The nuclear operators are local density operators in coordinate space and the matrix element can be written as a Fourier transform,
\begin{equation} \label{eq:FourierMatrixElements}
     \langle f|\mathcal{O}|i\rangle \rightarrow
    \langle \nu_{L\alpha}' |\mathcal{O}_{\nu}| \nu_{L\beta}\rangle
    \int d \vec{x} e^{-i \vec{q} \cdot \vec{x}}
    \langle N' |\mathcal{O}_{N} (\vec{x})| N \rangle \,.
\end{equation}

In neutrino-nucleus scattering, assuming that the nucleus remains
unexcited and in an eigenstate of total angular momentum $|j_N\rangle$, the nuclear matrix element $e^{-i\vec q\cdot\vec x}\langle j_{N'}|\mathcal{O}_N(\vec x)|j_N\rangle$ can be expanded in multipoles.
The multipole operators can be expressed in terms of eleven single-particle operators ~\cite{OConnell:1972edu,Serot:1978vj,Anand:2013yka,Fitzpatrick:2012ix,DelNobile:2021wmp} 
\begin{equation}
M_{J}, \tilde{\Omega}_{J}, \Delta_{J}, \Delta_{J}', \tilde{\Delta}_{J}'', \Sigma_{J}, \Sigma_{J}', \Sigma_{J}'', \tilde{\Phi}_{J}, \tilde{\Phi}_{J}', \Phi_{J}'' \,.
\end{equation}

If we restrict the nuclear ground state to have good parity and $CP$ symmetry, the matrix elements $\langle j_{N'}|O|j_N\rangle$ vanish for
$O = \tilde{\Omega}_{J},\, \Sigma_{J},\, \Delta_{J}',\, \tilde{\Delta}_{J}'',\, \tilde{\Phi}_{J}$.
These operators therefore do not contribute to the scattering amplitude.
The non-vanishing single-particle operators are~\cite{Fitzpatrick:2012ix,Anand:2013yka}
\begin{equation} \label{eq:SixNuclearResonses}
M_{J}, \Delta_{J}, \Sigma_{J}', \Sigma_{J}'', \tilde{\Phi}_{J}', \Phi_{J}''\,.
\end{equation}
The operator $M_{J}$ is related to the total number (or charge) of nucleons in the nucleus, while the operators $\Sigma_{J}'$ and $\Sigma_{J}''$ are associated with the nuclear spin. These three responses receive contributions from leading-order chiral operators and are not suppressed in the chiral expansion. In contrast, the operators
$\Delta_{J}$, $\tilde{\Phi}_{J}'$ and $\Phi_{J}''$ arise from NLO chiral interactions and are suppressed by ${\cal O}(1/m_N)$.

The many-body nuclear matrix elements can be expressed in terms of one-body density matrices\,\footnote{Details of the one-body density matrices are given in App.~\ref{app:NuclearResponse}.}. They are parametrized through nuclear response functions,
\begin{equation}
    W_{O O^{\prime}}= \sum_{J}    \langle j_{N'} ||~ O_{J} ~ || j_N \rangle 
\langle j_{N'} ||~ O_J^{\prime} ~ || j_N \rangle \,.
\end{equation}

In terms of these response functions, the spin-averaged squared amplitude can be written as
\begin{align}
\label{eq:AmplitudeSquare2}
    |\overline{{\cal M}}|^{2} = \frac{4 \pi}{2 J_{A}+1}\sum_{\tau,\tau'}\Bigg\{&\Big[R_{MM}^{\tau\tau'} W_{MM}^{\tau\tau'}(y)+R_{\Sigma''\Sigma''}^{\tau\tau'} W_{\Sigma''\Sigma''}^{\tau\tau'}(y)+R_{\Sigma'\Sigma'}^{\tau\tau'} W_{\Sigma'\Sigma'}^{\tau\tau'}(y)\Big] 
    \notag \\
    + \frac{|\vec{q}|^2}{m_N^2}&\Big[R_{\Phi''\Phi''}^{\tau\tau'}W_{\Phi''\Phi''}^{\tau\tau'}(y)+R_{\tilde{\Phi}'\tilde{\Phi}'}^{\tau\tau'}W_{\tilde{\Phi}'\tilde{\Phi}'}^{\tau\tau'}(y)+R_{\Delta\Delta}^{\tau\tau'}W_{\Delta\Delta}^{\tau\tau'}(y)\Big] 
    \notag \\
    - \frac{2 |\vec{q}|}{m_N}&\Big[ R_{\Phi''M}^{\tau\tau'}W_{\Phi''M}^{\tau\tau'}(y) + R_{\Delta\Sigma'}^{\tau\tau'}W_{\Delta\Sigma'}^{\tau\tau'}(y) \Big]
    \Bigg\},
\end{align}
where $y = (|\vec{q}| b/2)^2$ with $b$ the harmonic-oscillator parameter, and $\hat{q} = \vec{q}/|\vec{q}|$. The indices $\tau,\tau' = 0,1$ label isoscalar and isovector components, and $J_A$ denotes the spin of the target nucleus. Notice that factors $|\vec{q}|^2/m_N^2$ and $|\vec{q}|/m_N$ arise from chiral expansion, and terms accompanied by these factors are not retained in Ref.~\cite{Altmannshofer:2018xyo}.

In this context, the amplitude squared can be separated into the kinematic terms ($R^{\tau \tau'}$) and the nuclear response functions ($W^{\tau \tau'}$)\,\footnote{We closely follow the approach of Ref.~\cite{Altmannshofer:2018xyo}, based on the framework proposed for dark matter-nucleus scattering~\cite{Anand:2013yka}.
A different decomposition of the amplitude squared is employed in Ref.~\cite{Hoferichter:2020osn}.}. The expressions of $R$ are
\begin{align} \label{eq:KineticItems}
R_{M M}^{\tau \tau^\prime}&=  l_0^\tau   l_0^{\tau^\prime {\dagger}} \,,
\nonumber \\
R_{\Sigma^{\prime \prime}\Sigma^{\prime \prime}}^{\tau \tau^\prime} &= \left( \hat{q} \cdot  \vec{l}_5^\tau \right) ~ \left( \hat{q} \cdot  \vec{l}_5^{\tau^\prime {\dagger}} \right) \,, 
\nonumber \\
R_{\Sigma^\prime \Sigma^\prime}^{\tau \tau^\prime}  &= \frac{1}{2} \Big(  \vec{l}_5^\tau  \cdot  \vec{l}_5^{\tau^\prime{\dagger}} - \hat{q} \cdot  \vec{l}_5^\tau  ~\hat{q} \cdot  \vec{l}_5^{\tau^\prime {\dagger}}  \Big) \,,
\nonumber \\ 
R_{\Phi^{\prime \prime} \Phi^{\prime \prime}}^{\tau \tau^\prime} &= \left( \hat{q} \cdot  \vec{l}_E^\tau \right)  ~  \left( \hat{q} \cdot  \vec{l}_E^{\tau^\prime {\dagger}}  \right) \,,
\nonumber \\
R_{\tilde{\Phi}^\prime \tilde{\Phi}^\prime}^{\tau \tau^\prime}  &=  {1 \over 2} \big(  \vec{l}_E^\tau  \cdot  \vec{l}_E^{\tau^\prime {\dagger}} - \hat{q} \cdot  \vec{l}_E^\tau ~ \hat{q} \cdot  \vec{l}_E^{\tau^\prime {\dagger}} \big) \,,
\nonumber \\
R_{\Delta \Delta}^{\tau \tau^\prime}  &= {1 \over 2} \big(  \vec{l}_M^\tau  \cdot  \vec{l}_M^{\tau^\prime {\dagger}} - \hat{q} \cdot  \vec{l}_M^\tau  ~ \hat{q} \cdot  \vec{l}_M^{\tau^\prime {\dagger}} \big) \,,
\nonumber \\
R_{\Phi^{\prime \prime}M}^{\tau \tau^\prime} &= \hq \cdot \mathrm{Re} \left[   \vec{l}_E^\tau ~  l_0^{\tau^\prime {\dagger}}  \right] \,,
\nonumber \\
R_{\Delta \Sigma^\prime}^{\tau \tau^\prime}  &= 2 \hq \cdot  \mathrm{Re} \left[ i  \vec{l}_M^\tau  \times  \vec{l}_5^{\tau^\prime {\dagger}}  \right] \,,
\end{align}
where $l_0^\tau$, $\vec{l}_5^\tau$, $\vec{l}_M^\tau$ and $\vec{l}_E^\tau$ denote the neutrino matrix elements (including the corresponding Wilson coefficients) of the leptonic structures introduced in Eq.~\eqref{eq:LDEFT}.

In the long-wavelength limit, the nuclear response functions $W_{MM}$, $W_{\Phi^{\prime \prime}\Phi^{\prime \prime}}$ and $W_{\Phi^{\prime \prime}M}$ receive coherent enhancement proportional to the nucleon number (nuclear number enhancement). In contrast, the responses $W_{\Sigma^{\prime \prime}\Sigma^{\prime \prime}}$, $W_{\Sigma^\prime \Sigma^\prime}$, $W_{\tilde{\Phi}^\prime\tilde{\Phi}^\prime}$, $W_{\Delta \Delta}$ and $W_{\Delta \Sigma^\prime}$ do not exhibit such enhancement and become phenomenologically important in regimes where the spin-independent contribution is suppressed.

\begin{table}[h]
\centering
\begin{tabular}{c c c c}
\hline \hline
\multicolumn{2}{c}{LEFT operator} & Response function &  Overall scaling\\
\hline
\multirow{4}{*}{$\mathcal{O}^{6u/d}_{1}$}
&
\multirow{4}{*}{$(\bar \nu_{L\alpha}\gamma^\mu \nu_{L\beta})(\bar q\gamma_\mu\tau^{u/d} q)$}
&$W_{MM}$ 
&$ A $ 
\\
&
&$W_{\Sigma' \Sigma'} $ 
&$ 1 $ 
\\
&
&$W_{\Delta \Delta}$ 
&$\varepsilon \cdot 1 $ 
\\
&
&$W_{\Delta \Sigma'}$ 
&$\varepsilon \cdot 1 $ 
\\
\hline
$\mathcal{O}^{6u/d}_{2}$
&
$(\bar \nu_{L\alpha}\gamma^\mu \nu_{L\beta})(\bar q\gamma^5\gamma_\mu\tau^{u/d} q)$
&$W_{\Sigma' \Sigma'}$
&$  1 $ 
\\
\hline
\hline
$\mathcal{O}^{7u/d}_{1}$
&
$(\bar \nu_{L\alpha}\gamma^\mu \nu_{L\beta})(\bar q\lrpartial_\mu\tau^{u/d} q)$ 
&$W_{MM}$
&$  A $ 
\\
\hline
$\mathcal{O}^{7u/d}_{2}$
&
{$(\bar \nu_{L\alpha}\gamma^\mu \nu_{L\beta})(\bar q\gamma^5\lrpartial_\mu\tau^{u/d} q)$}
&$W_{\Sigma'' \Sigma''}$
&{$ \varepsilon  \cdot 1 $} 
\\
\hline
\multirow{5}{*}{$\mathcal{O}^{7u/d}_{3}$} 
&
\multirow{5}{*}{$m_q (\nu_{L\alpha}^TC\sigma^{\mu\nu} \nu_{L\beta})(\bar q\sigma_{\mu\nu} \tau^{u/d}q)$} 
& $W_{MM}$ 
& $ \varepsilon^3 \cdot A $ 
\\
&
&$W_{\Sigma' \Sigma'}, W_{\Sigma'' \Sigma''}$
&$  \varepsilon^2  \cdot 1 $ 
\\
&
&$W_{\Phi^{\prime \prime}\Phi^{\prime \prime}}$
&$ \varepsilon^3 \cdot A $ 
\\
&
&$W_{\tilde{\Phi}^\prime\tilde{\Phi}^\prime}$
&$ \varepsilon^3 \cdot 1 $ 
\\
&
&$W_{{\Phi}^{\prime \prime}M}$
&$ \varepsilon^3 \cdot A $ 
\\
\hline
$\mathcal{O}^{7u/d}_{4}$
&
$ m_q ( \nu_{L\alpha}^TC \nu_{L\beta})(\bar q \tau^{u/d}q)$
&$W_{MM}$ 
&$ \varepsilon^2  \cdot A $
\\
\hline
$\mathcal{O}^{7u/d}_{5}$
&
{$m_q ( \nu_{L\alpha}^TC \nu_{L\beta})(\bar q\gamma^5\tau^{u/d}q)$}
&$W_{\Sigma'' \Sigma''}$
& \multirow{1}{*}{$  \varepsilon^3 \cdot 1$} 
\\
\hline
$\mathcal{O}^{7}_{6}$
&
$  ( \nu_{L\alpha}^TC \nu_{L\beta})G_{\mu\nu}^AG^{A\mu\nu}$
&$W_{MM}$ 
&$  \varepsilon^2  \cdot A$
\\
\hline
$\mathcal{O}^{7}_{7}$
&
{$( \nu_{L\alpha}^TC \nu_{L\beta})\tilde{G}_{\mu\nu}^AG^{A\mu\nu}$}
&$ W_{\Sigma'' \Sigma''}$
& \multirow{1}{*}{$  \varepsilon^3 \cdot 1$} 
\\
\hline
\hline
$\mathcal{O}^{8u/d}_{1}$
&
$(\bar\nu_{L\alpha} \gamma^\mu\lrpartial^\nu  \nu_{L\beta})  (\bar{q} \gamma_\mu\lrpartial_\nu\tau^{u/d} q )$ 
&$W_{MM}$
&$ A $ 
\\
\hline
$\mathcal{O}^{8u/d}_{2}$
&
$(\bar\nu_{L\alpha} \gamma^\mu\lrpartial^\nu  \nu_{L\beta})  (\bar{q} \gamma^5\gamma_\mu\lrpartial_\nu \tau^{u/d}q)$ 
&$W_{\Sigma' \Sigma'}$
&$ 1 $ 
\\
\hline
$\mathcal{O}^{8}_{9}$
&
$(\bar\nu_{L\alpha} \gamma^\mu\lrpartial^\nu  \nu_{L\beta}) G^A_{\mu\rho}G^{A\rho}_\nu$ 
&$W_{MM}$
&$ A $ 
\\
\hline \hline
\end{tabular}
\caption{
Response functions and the corresponding overall scaling of amplitude associated with different LEFT operators for the tree-level diagram (a) in Fig.~\ref{fig:OneTwoBodyDiag}. In the last column, we have incorporated the factors in chiral power counting. Here, $A$ denotes the nucleon number, and we adopt a common expansion parameter $\varepsilon$ with $\varepsilon \sim \varepsilon_\chi \sim \varepsilon_N$.
}
\label{tab:ResponseEnhance}
\end{table}

Collecting the above results, we can summarize the relative sizes of the nuclear responses as follows. The response $W_{MM}$ receives a fully coherent enhancement and scales as $W_{MM} \sim A^{2}$, where $A$ represents the number of nucleons. The response $W_{\Phi^{\prime\prime}\Phi^{\prime\prime}}$ is also coherently enhanced, but the corresponding term $(q^2/m_N^2)W_{\Phi^{\prime\prime}\Phi^{\prime\prime}}$ 
in Eq.~\eqref{eq:AmplitudeSquare2} scales as $\mathcal{O}(\varepsilon_N^2)\cdot A^2$. The additional factor of $q^2/m_N^2$ originates from the subleading terms of the chiral Lagrangian in Eq.~\eqref{eq:LPionN}. 
Likewise, the interference term behaves as $(q/m_N) W_{\Phi'' M} \sim \mathcal{O}(\varepsilon_N) \cdot A^2$.
In contrast, the responses $W_{\Sigma^\prime \Sigma^\prime}$ and $W_{\Sigma^{\prime\prime}\Sigma^{\prime\prime}}$ do not receive coherent enhancement and are of order ${\cal O}(1)$.

Notice that for heavy nuclei,
the scaling ${\cal O}(\varepsilon_N^{2})\cdot A^{2}$ can become comparable to, or even larger than ${\cal O}(1)$. We thus assume that ${\cal O}(\varepsilon_N^2)\cdot A^{2}\sim {\cal O}(1)$,
the hierarchy of responses then can be written schematically as
\begin{align}
    W_{MM}  
    \gg \frac{q}{m_N} W_{\Phi'' M}  
    &\gg  \left\{W_{\Sigma^\prime \Sigma^\prime},\,W_{\Sigma^{\prime \prime}\Sigma^{\prime \prime}},\,  \frac{q^2}{m_N^2}W_{MM},\, \frac{q^2}{m_N^2} W_{\Phi^{\prime \prime}\Phi^{\prime \prime}}\right\} \notag \\
    &\gg \left\{\frac{q}{m_N} W_{\Delta \Sigma'},\frac{q^2}{m_N^2}W_{\Delta \Delta},\, \frac{q^2}{m_N^2} W_{\tilde{\Phi}^\prime\tilde{\Phi}^\prime}\right\}   \,.  
\end{align}

Incorporating these response-function enhancements allows us to estimate the overall scaling of the amplitude associated with different LEFT operators for tree-level diagram (a)\,\footnote{For diagram (b), the nucleon current is proportional to $\bar N S^{\mu} N$ and therefore does not give rise to coherent enhancement. Its contribution is purely subleading and mainly provides a controlled correction to the leading-order diagram (a).} in Fig.~\ref{fig:OneTwoBodyDiag}, as summarized in Tab.~\ref{tab:ResponseEnhance}.
In the last column, we have included the factors in Tab.~\ref{tab:chiral-counting} for the contributions at leading chiral order. 
For those at higher chiral orders (e.g., the $W_{\Delta\Delta}$ term associated with $\mathcal{O}_1^{6u/d}$), the relevant chiral factors can be read off from Eq.~\eqref{eq:AmplitudeSquare2} and are likewise included. As Tab.~\ref{tab:ResponseEnhance} shows, the vector
LEFT operators $\mathcal{O}_1^{6u/d}$ induce coherently enhanced contributions already at leading chiral order. Although the dimension-7 and dimension-8 operators are suppressed by powers of the electroweak scale, it is possible that new physics contributes only to these higher-dimensional LEFT operators. The scalar operators $\mathcal{O}_4^{7u/d}$ are also coherently enhanced at leading chiral order given $m_q \sim \mathcal{O}(\varepsilon^2)$. In contrast, the tensor operators $\mathcal{O}_3^{7u/d}$ do not receive coherent enhancement at leading chiral order for $W_{\Sigma' \Sigma'}, W_{\Sigma'' \Sigma''}$. However, NLO chiral operators generate coherently enhanced contributions scaling as $\mathcal{O} (\varepsilon^3) \cdot A $ that can even dominate over the leading chiral order terms, which is consistent with the finding reported in Ref.~\cite{Liao:2025hcs}. Such coherently enhanced contributions lie beyond the treatment of Ref.~\cite{Altmannshofer:2018xyo}, where only the chirally leading contribution is retained for each LEFT operator.
Besides, we find that the operators $\mathcal{O}_1^{7u/d}$, $\mathcal{O}_6^{7}$, $\mathcal{O}_1^{8u/d}$, and $\mathcal{O}_9^{8}$ are coherently enhanced.

\subsection{Cross Section for Neutrino-Nucleus Scattering}
In a typical neutrino detection experiment the target nuclei are at rest in the laboratory frame. The differential scattering cross section with respect to the nuclear recoil energy $T_{\rm nr}$ is then given by
\begin{equation}
    \frac{{\rm d}\sigma}{{\rm d} T_{\rm nr}}
    = \frac{m_A}{8\pi E_{\nu}^{2}}\,
    \bigl|\overline{\mathcal M}\bigr|^{2} \,,
\end{equation}
where $m_{A}$ is the mass of the target nucleus. 
Here $|\overline{\mathcal M}|^{2}$ stands for the squared scattering amplitude, summed over final spins and averaged over the initial nuclear spin projections.

According to Eq.~\eqref{eq:LDEFT}  the neutrino-nucleon interaction can be organized in terms of the five leptonic Dirac structures
% $l$ ($l_{0}$, $\vec{l}_5$, $l_0^A$,  $\vec{l}_M$, $\vec{l}_E$)
$\ell$ ($\ell_{0}$, $\bm{\ell}_5$, $\ell_0^A$, $\bm{\ell}_M$, $\bm{\ell}_E$)
each accompanied by appropriate Wilson coefficients for the specific process under consideration. 
The leading one-body contribution to the scattering amplitude from diagrams (a) and (b) in Fig.~\ref{fig:OneTwoBodyDiag} can be written as
\begin{align}
    \mathcal{M}
    &=
    \ell_0 \cdot \bar{N} \, N
    + \bm{\ell}_5 \cdot \left( \bar N \, 2\vec{S} \, N \right) 
    + \bm{\ell}_M \cdot \left( \frac{1}{2m_N} \bar{N} \vec{K} N \right)
    + \bm{\ell}_E \cdot \left(- \frac{i}{m_N} \bar{N} ({\vec{K} \times \vec{S}}) N \right) \,,
\end{align}
where $N = p,n$.

These different leptonic structures project onto different kinematic response functions $R^{\tau\tau'}_X$ introduced in Eq.~\eqref{eq:KineticItems}. 
In particular,
$\ell_0$ contributes to $R^{\tau\tau'}_{MM}$ and $R^{\tau\tau'}_{\Phi'' M}$; $\bm{\ell}_5$ contributes to $R^{\tau\tau'}_{\Sigma'\Sigma'}$, $R^{\tau\tau'}_{\Sigma''\Sigma''}$ and $R^{\tau\tau'}_{\Delta\Sigma'}$,
while $\ell_0^A$ does not contribute at this order.
The magnetic structure $\bm{\ell}_M$ enters only the subleading responses $R^{\tau\tau'}_{\Delta\Delta}$ and $R^{\tau\tau'}_{\Delta\Sigma'}$,
and the electric structure $\bm{\ell}_E$ contributes to
$R^{\tau\tau'}_{\Phi'\Phi'}$, $R^{\tau\tau'}_{\Phi''\Phi''}$ and $R^{\tau\tau'}_{\Phi'' M}$.

The leptonic structure $\ell_0$, $\bm{\ell}_5$, $\ell_0^A$, $\bm{\ell}_M$ and $\bm{\ell}_E$ can be expressed in terms of the underlying LEFT Wilson coefficients as
\begin{align} \label{eq:LeptonStructure}
\ell_{0, p} 
&= c_{1,p}\,\left(\bar{\nu}_{L \alpha}\,\gamma^\mu \,\nu_{L \beta}\right) v_\mu
 + c_{13,p}\,\left(\bar{\nu}_{L \alpha}\,\gamma^\mu \,\nu_{L \beta} \right)  v_\mu v_\nu P^\nu 
 \notag \\
& \quad
 + c_{3,p}\,\left(\nu_{L \alpha}^T C\,\nu_{L \beta} \right) 
 + c_{9,p}\, \frac{q_\mu }{m_N} \,\left(\nu_{L \alpha}^T C\,\sigma^{\mu \nu} \,\nu_{L \beta} \right) v_\nu \,,
 \notag \\
\ell_{5, p}^{\mu} 
&=  \frac{1}{2} c_{2,p}\,\left(\bar{\nu}_{L \alpha}\, \gamma^\mu  \,\nu_{L \beta}\right)
 + c_{6,p}\,\frac{q_\rho}{2m_N} \left(\bar{\nu}_{L \alpha}\,  \gamma_\nu  \,\nu_{L \beta}\right) \varepsilon^{\mu \nu \rho \lambda} v_\lambda
 \notag \\
& \quad
 + c_{11,p} \, \frac{q^\mu}{2 m_N} \left(\bar{\nu}_{L \alpha}\, \gamma^\nu  \,\nu_{L \beta}\right) v_\nu
 + \frac{1}{2} c_{14,p} \,  \,\left(\bar{\nu}_{L \alpha}\, \gamma^\mu \,\nu_{L \beta}\right) P^\nu v_\nu
\notag \\
&\quad
 + \frac{1}{2} c_{4,p}\,\left(\nu_{L \alpha}^T C\, \sigma^{\mu\nu}\,   \nu_{L \beta} \right) v_{\nu} 
 + c_{10,p} \, \frac{q^\mu}{2 m_N}\,\left(\nu_{L \alpha}^T C\, \nu_{L \beta} \right)
 \notag \\ 
 &\quad
 + c_{12,p} \left(\nu_{L \alpha}^T C\, \nu_{L \beta} \right) P^{\mu} q^{\nu} v_{\nu} \notag
\,,
\notag \\
\ell^{\mu}_{M,p}
&= c_{5,p} \left(\bar \nu_{L \alpha}\, \gamma^\mu  \nu_{L \beta} \right) \,,
 \notag \\
\ell^{\mu}_{E,p} 
&= i c_{8,p} \left(\nu_{L \alpha}^T C\, \sigma^{\mu \nu} \, \nu_{L \beta} \right)  v_\nu \,,
\end{align}
where the subscript $N=p,n$. Eq.~\eqref{eq:LeptonStructure} contains both diagram (a) and (b).

There are two classes of neutrino bilinears in our framework, corresponding to lepton-number-conserving and lepton-number-violating structures,
\begin{align}
    &\text{LNC:}\quad \bar{\nu }_{L \alpha} \gamma^{\mu} \nu_{L \beta} \,, \notag \\
    &\text{LNV:}\quad \nu_{L \alpha}^{T} C \nu_{L \beta}, \, \nu_{L\alpha}^{T} C \sigma^{\mu \nu}  \nu_{L\beta} \,,
\end{align}
where $\nu_{L}$ denotes the left-handed neutrino field and $C$ is the charge-conjugation matrix.  Throughout this work, we neglect interference between the LNC and LNV amplitudes and treat the LNC and LNV contributions separately in the rate.

After response expansion, the one-body kinematic terms can be written (expanded to the $\vec{q}/m_A$ and $(\vec{q}/m_N)^2$ order) as
\begin{align}
R_{MM}^{\tau\tau'} &=
\left(4E_{\nu}^2-\vec{q}^{\,2}-\frac{2E_{\nu}\vec{q}^{\,2}}{m_A}\right)
\left(c_{1}^{\tau\dagger}c_{1}^{\tau'}\right)
+\left(16E_{\nu}^4-4E_{\nu}^2\vec{q}^{\,2}\right)
\left(c_{13}^{\tau\dagger}c_{13}^{\tau'}\right)
\notag\\
&\quad+
\left(-\frac{16E_{\nu}^3\vec{q}^{\,2}}{m_A}+\frac{2E_{\nu}\vec{q}^{\,4}}{m_A}\right)
\left(c_{13}^{\tau\dagger}c_{13}^{\tau'}\right)
+\vec{q}^{\,2}\,\big(c_{3}^{\tau\dagger}c_{3}^{\tau'}\big)
\notag\\
&\quad+
\left(8E_{\nu}^3-2E_{\nu}\vec{q}^{\,2}-\frac{6E_{\nu}^2\vec{q}^{\,2}}{m_A}
+\frac{\vec{q}^{\,4}}{2m_A}\right)
\Big[\left(c_{1}^{\tau\dagger}c_{13}^{\tau'}\right)
+\left(c_{13}^{\tau\dagger}c_{1}^{\tau'}\right)\Big]
\notag\\
&\quad+
\left(\frac{2 i E_{\nu}\vec{q}^{\,2}}{m_N}-\frac{i\vec{q}^{\,4}}{2m_A m_N}\right)
\Big[\big(c_{9}^{\tau\dagger}c_{3}^{\tau'}\big)
-\big(c_{3}^{\tau\dagger}c_{9}^{\tau'}\big)\Big]
+\left(\frac{4E_{\nu}^2\vec{q}^{\,2}}{m_N^2}-\frac{2E_{\nu}\vec{q}^{\,4}}{m_A m_N^2}\right)
\big(c_{9}^{\tau\dagger}c_{9}^{\tau'}\big)
\,.
\notag
\end{align}

\vspace{-6pt}
\begin{align}
R_{\Sigma^{\prime\prime}\Sigma^{\prime\prime}}^{\tau\tau'} &=
\left(\frac{E_{\nu}^2\vec{q}^{\,2}}{m_N^2}-\frac{E_{\nu}\vec{q}^{\,4}}{2m_A m_N^2}
-\frac{\vec{q}^{\,4}}{4m_N^2}\right)\big(c_{11}^{\tau\dagger}c_{11}^{\tau'}\big)
+\frac{\vec{q}^{\,4}}{4m_N^2}\,\big(c_{10}^{\tau\dagger}c_{10}^{\tau'}\big)
\notag\\
&\quad+
\left(E_{\nu}^2-\frac{E_{\nu}\vec{q}^{\,2}}{2m_A}\right)\big(c_{4}^{\tau\dagger}c_{4}^{\tau'}\big)
+\left(\frac{iE_{\nu}\vec{q}^{\,2}}{2m_N}-\frac{i\vec{q}^{\,4}}{8m_A m_N}\right)
\Big[\big(c_{10}^{\tau\dagger}c_{4}^{\tau'}\big)-\big(c_{4}^{\tau\dagger}c_{10}^{\tau'}\big)\Big]
\notag\\
&\quad+
\left(\frac{E_{\nu}^3\vec{q}^{\,2}}{m_A m_N}-\frac{E_{\nu}\vec{q}^{\,4}}{4m_A m_N}\right)
\Big[\big(c_{11}^{\tau\dagger}c_{14}^{\tau'}\big)+\big(c_{14}^{\tau\dagger}c_{11}^{\tau'}\big)\Big]
\notag\\
&\quad+
\left(\frac{E_{\nu}^2\vec{q}^{\,2}}{2m_A m_N}-\frac{\vec{q}^{\,4}}{8m_A m_N}\right)
\Big[\big(c_{2}^{\tau\dagger}c_{11}^{\tau'}\big)+\big(c_{11}^{\tau\dagger}c_{2}^{\tau'}\big)\Big]
\,.
\notag
\end{align}

\vspace{-6pt}
\begin{align}
R_{\Sigma^\prime\Sigma^\prime}^{\tau\tau'} &=
\left(2E_{\nu}^4+\frac{1}{2}E_{\nu}^2\vec{q}^{\,2}
-\frac{2E_{\nu}^3\vec{q}^{\,2}}{m_A}-\frac{E_{\nu}\vec{q}^{\,4}}{4m_A}\right)
\big(c_{14}^{\tau\dagger}c_{14}^{\tau'}\big)
\notag\\
&\quad+
\left(E_{\nu}^3+\frac{1}{4}E_{\nu}\vec{q}^{\,2}-\frac{3E_{\nu}^2\vec{q}^{\,2}}{4m_A}
-\frac{\vec{q}^{\,4}}{16m_A}\right)
\Big[\big(c_{2}^{\tau\dagger}c_{14}^{\tau'}\big)+\big(c_{14}^{\tau\dagger}c_{2}^{\tau'}\big)\Big]
\notag\\
&\quad+
\left(\frac{2 i E_{\nu}^2\vec{q}^{\,2}}{m_N}-\frac{iE_{\nu}\vec{q}^{\,4}}{m_A m_N}\right)
\Big[\big(c_{6}^{\tau\dagger}c_{14}^{\tau'}\big)-\big(c_{14}^{\tau\dagger}c_{6}^{\tau'}\big)\Big]
\notag\\
&\quad+
\left(\frac{1}{2}E_{\nu}^2+\frac{1}{8}\vec{q}^{\,2}-\frac{E_{\nu}\vec{q}^{\,2}}{4m_A}\right)
\big(c_{2}^{\tau\dagger}c_{2}^{\tau'}\big)
+\left(\frac{iE_{\nu}\vec{q}^{\,2}}{m_N}-\frac{i\vec{q}^{\,4}}{4m_A m_N}\right)
\Big[\big(c_{6}^{\tau\dagger}c_{2}^{\tau'}\big)-\big(c_{2}^{\tau\dagger}c_{6}^{\tau'}\big)\Big]
\notag\\
&\quad+
\left(\frac{2E_{\nu}^2\vec{q}^{\,2}}{m_N^2}+\frac{\vec{q}^{\,4}}{2m_N^2}-\frac{E_{\nu}\vec{q}^{\,4}}{m_A m_N^2}\right)
\big(c_{6}^{\tau\dagger}c_{6}^{\tau'}\big)
+\left(\frac{1}{2}E_{\nu}^2-\frac{E_{\nu}\vec{q}^{\,2}}{4m_A}-\frac{1}{8}\vec{q}^{\,2}\right)
\big(c_{4}^{\tau\dagger}c_{4}^{\tau'}\big)
\,.
\notag
\end{align}

\vspace{-6pt}
\begin{align}
R_{\Phi^{\prime\prime}\Phi^{\prime\prime}}^{\tau\tau'} &=
\left(4E_{\nu}^2-\frac{2E_{\nu}\vec{q}^{\,2}}{m_A}\right)\big(c_{8}^{\tau\dagger}c_{8}^{\tau'}\big)
\,,
\notag\\
R_{\tilde{\Phi}^\prime\tilde{\Phi}^\prime}^{\tau\tau'} &=
\left(2E_{\nu}^2-\frac{1}{2}\vec{q}^{\,2}-\frac{E_{\nu}\vec{q}^{\,2}}{m_A}\right)\big(c_{8}^{\tau\dagger}c_{8}^{\tau'}\big)
\,,
\notag\\
R_{\Delta\Delta}^{\tau\tau'} &=
\left(2E_{\nu}^2+\frac{1}{2}\vec{q}^{\,2}-\frac{E_{\nu}\vec{q}^{\,2}}{m_A}\right)\big(c_{5}^{\tau\dagger}c_{5}^{\tau'}\big)
\,,
\notag\\
R_{\Phi^{\prime\prime}M}^{\tau\tau'} &=
\left(-2E_{\nu}|\vec{q}|+\frac{|\vec{q}|^3}{2m_A}\right)\big(c_{3}^{\tau\dagger}c_{8}^{\tau'}\big)
+\left(-\frac{4iE_{\nu}^2|\vec{q}|}{m_N}+\frac{2iE_{\nu}|\vec{q}|^3}{m_A m_N}\right)\big(c_{9}^{\tau\dagger}c_{8}^{\tau'}\big)
\,.
\end{align}
where $m_N$ is the mass of the nucleon, the nuclear recoil energy $T_{\rm nr}=\vec{q}^2/(2m_A)$.
The isospin $\tau, \tau' = 0,1$,
\begin{equation}
    c^{0}_{i} = \frac{1}{2} (c_{i,p} + c_{i,n}) \,, \qquad 
    c^{1}_{i} = \frac{1}{2} (c_{i,p} - c_{i,n}) \,.
\end{equation}
These kinematic factors, together with the nuclear response functions discussed above, are used to constrain the LEFT Wilson coefficients and the associated new-physics scales.
At chirally leading order, our expressions for 
$R_{MM}^{\tau\tau'}$, $R_{\Sigma^{\prime\prime}\Sigma^{\prime\prime}}^{\tau\tau'}$, and 
$R_{\Sigma^\prime\Sigma^\prime}^{\tau\tau'}$ agree with Ref.~\cite{Altmannshofer:2018xyo}. 
At subleading orders, additional LEFT operators contribute to CE$\nu$NS. 
For instance, the coefficients of dimension-7 operators $\mathcal{O}_3^{7u/d}$ enter 
$R_{MM}^{\tau\tau'}$ through the coefficient $c_9^\tau$.

\section{Experimental Constraints}
\label{sec:expts}

Having derived the CE$\nu$NS differential cross section via the multipole expansion and expressed the neutrino-nucleus amplitude in terms of nuclear response functions, the next step is to confront these predictions with data and extract constraints on new physics. The amplitude is applied to the experimental analyses by providing the expected event yields for the different measurements and performing the corresponding $\chi^2$ fits. The experimental data are used to set lower bounds on the effective
scales and to constrain the electron-flavor NSI parameters.

\subsection{Experiments}
The experimental inputs employed in this analysis originate from accelerator-, solar-, and reactor-based CE$\nu$NS measurements, which differ both in their neutrino sources and in the detector-level observables used to report the data. Specifically, the COHERENT experiment detects neutrinos from the Spallation Neutron Source (SNS) using a CsI[Na] scintillator~\cite{COHERENT:2017ipa,COHERENT:2021xmm} and Ar detector~\cite{COHERENT:2020iec,COHERENT:2020ybo}.
PandaX-4T and XENONnT probe CE$\nu$NS from solar $^8$B neutrinos with liquid xenon ~\cite{PandaX:2024muv,XENON:2024ijk}, while CONUS+ measures reactor $\bar\nu_e$-induced CE$\nu$NS using high-purity germanium detectors ~\cite{CONUS:2020skt,CONUSCollaboration:2024kvo,Ackermann:2025obx}.
To treat these different data sets within a common statistical framework, we first define a universal recoil-energy rate in terms of the nuclear recoil $T_{\rm nr}$, and subsequently specify, for each experiment, the appropriate detector response, binning prescription, and $\chi^2$ function adopted in the fit, following the conventions of Ref.~\cite{Li:2024iij}.

For an incident (anti)neutrino flux $\Phi_{\nu_\alpha}(E_\nu)$ at the detector, the differential event rate per target nucleus is
\begin{equation}
\frac{{\rm d}R_{\nu_\alpha}}{{\rm d}T_{\rm nr}}
=
\int_{E_{\nu,\min}(T_{\rm nr})}^{E_{\nu,\max}}
{\rm d}E_\nu\,
\Phi_{\nu_\alpha}(E_\nu)\,
\frac{{\rm d}\sigma}{{\rm d}T_{\rm nr}}(E_\nu,T_{\rm nr})\;,
\label{eq:master_dRdT}
\end{equation}
where $T_{\rm nr}$ is the nuclear recoil energy, and $E_{\nu,\min}(T_{\rm nr})$ and $E_{\nu,\max}$ are fixed by the source and kinematics. The exposure amounts of different experiments are included in flux $\Phi_{\nu_\alpha}(E_\nu)$. In the following, we apply this common differential event rate definition to the three experiments considered here, specifying in each case the corresponding detector response and statistical treatment.

For the COHERENT (SNS) experiment, we consider the CsI[Na]~\cite{COHERENT:2021xmm} and Ar~\cite{COHERENT:2020iec} detectors.
The predicted CE$\nu$NS counts in the $i$-th experimental bin are obtained by folding
${\rm d}R_{\nu_\alpha}/{\rm d}T_{\rm nr}$ with the efficiency and the energy resolution function,
\begin{equation}
N^{\,i}_{\rm CE\nu NS}
=
n_N \sum_{\nu_\alpha=\nu_e,\nu_\mu,\bar\nu_\mu}
\int_{i}{\rm d}n_{\rm PE}\;
\epsilon(n_{\rm PE})
\int {\rm d}T_{\rm nr}\;
P(n_{\rm PE}\,|\,T_{\rm nr})\;
\frac{{\rm d}R_{\nu_\alpha}}{{\rm d}T_{\rm nr}} \,.
\label{eq:coherent_bins_compact}
\end{equation}
Here, $n_N$ denotes the number of target nuclei, $n_{\rm PE}$ denotes the photoelectron variable,  $\epsilon(n_{\rm PE})$ denotes the detection efficiency, and $P(n_{\rm PE}|T_{\rm nr})$ denotes the energy resolution function (including quenching and energy smearing).
The SNS flux components and the COHERENT energy resolution function inputs are taken from Refs.~\cite{Coloma:2017egw,Liao:2017uzy,COHERENT:2020ybo,COHERENT:2021xmm}.
The impact of systematic uncertainties related to the quenching factor on the CE$\nu$NS cross section measurement was recently investigated in Ref.~\cite{Li:2025pfw}. 

We adopt a binned least-squares function with correlated normalization nuisances for signal and beam-related background~\cite{Fogli:2002pt,AristizabalSierra:2018eqm},
\begin{equation}
\chi^2_{\rm COHERENT}
=
\sum_{i=1}^{N_{\rm bin}}
\frac{\left[N^{\,i}_{\rm meas}-(1+\alpha)\,N^{\,i}_{\rm CE\nu NS}-(1+\beta)\,N^{\,i}_{\rm bkg}\right]^2}
{\left(\sigma_i\right)^2}
+
\left(\frac{\alpha}{\sigma_\alpha}\right)^2
+
\left(\frac{\beta}{\sigma_\beta}\right)^2 ,
\label{eq:chi2_coherent_compact}
\end{equation}
where $N^{\,i}_{\rm meas}$ are the measured counts in the $i$-th bin, $N^{\,i}_{\rm CE\nu NS}$ and $N^{\,i}_{\rm bkg}$ are the theoretical predictions for the CE$\nu$NS signal and beam-related background, respectively, and $\sigma_i$ denotes the uncertainty. 
The numerical values of the systematic uncertainty $\sigma_\alpha = 0.125$ and the background uncertainty $\sigma_\beta = 0.35$ for CsI~\cite{COHERENT:2021xmm,DeRomeri:2022twg,Li:2024iij},  and $\sigma_\alpha = 0.12,\sigma_\beta = 0.146$ for Ar~\cite{Cadeddu:2020lky, Miranda:2020tif}.

For solar-$^8$B CE$\nu$NS in liquid xenon (PandaX-4T + XENONnT), we include $\nu_\alpha\in\{\nu_e,\nu_\mu,\nu_\tau\}$ and use Eq.~\eqref{eq:master_dRdT}
with the solar-$^8$B flux and oscillation-averaged flavor composition as in Ref.~\cite{Li:2024iij}.
Given the publicly available information for the solar-$^8$B analyses, we model both PandaX-4T (paired + US2) and XENONnT as single-bin counting measurements in reconstructed recoil energy, following the treatment in Refs.~\cite{Liao:2025hcs,Li:2024iij}.
\begin{equation}
N^{(k)}_{\rm CE\nu NS}
=
n_N^{(k)}\sum_{\alpha=e,\mu,\tau}
\int_{T_{\rm nr,min}^{(k)}}^{T_{\rm nr,max}^{(k)}}{\rm d}T_{\rm nr}\;
\epsilon^{(k)}(T_{\rm nr})\;
\frac{{\rm d}R_{\nu_\alpha}}{{\rm d}T_{\rm nr}}\;,
\qquad
k\in\{\rm paired, US2,XENONnT\}\;,
\label{eq:solar_xe_counts_compact}
\end{equation}
where the $\epsilon(T_{\rm nr})$ is the detection efficiency\,\footnote{
The Migdal effect can also affect the spectral interpretations~\cite{Herrera:2023xun,Blanco-Mas:2024ale}, which is not considered in this analysis.}. 
The combined $\chi^2$ is a sum of three independent single-bin contributions,
\begin{equation}
\chi^2_{{\rm Solar} ^8{\rm B}}= \sum_{k} \chi^2_{k}\;,
\qquad
\chi^2_{k}
=
\frac{\left[N^{(k)}_{\rm meas}-(1+\alpha_k)\,N^{(k)}_{\rm CE\nu NS}\right]^2}{(\sigma^{(k)})^2}
+
\left(\frac{\alpha_k}{\sigma_{\alpha_k}}\right)^2\;,
\label{eq:chi2_solar_xe_compact}
\end{equation}
where $k$ labels the PandaX-4T paired and US2 data sets, and the XENONnT data set. The $N_{\rm meas}$ is the measured count, $N_{\rm CE\nu NS}$ is the theoretical prediction. $\alpha_k$ parametrizes the normalization uncertainty for data set $k$.  The systematic uncertainty is $\sigma_{\alpha_k}=0.12$~\cite{Vitagliano:2019yzm} and the statistical uncertainties for the US2, paired, and XENONnT data sets are taken to be $37.3\%$, $37.1\%$, and $36.9\%$ of the corresponding measured counts, respectively~\cite{Liao:2025hcs,Li:2024iij}, 
with experimental details in Refs.~\cite{PandaX:2024muv,XENON:2024ijk}. 

For CONUS+ (reactor $\bar\nu_e$, Ge), we use Eq.~\eqref{eq:master_dRdT} with $\nu_\alpha=\bar\nu_e$~\cite{CONUS:2020skt,CONUSCollaboration:2024kvo} and the reactor flux model
(see Refs.~\cite{Mueller:2011nm,Kopeikin:2012zz,DeRomeri:2025csu}).
The comparison is performed in reconstructed electron-equivalent energy $E_{\rm ee}^{\rm reco}$; the predicted binned counts are obtained by folding the recoil spectrum with quenching and resolution,
\begin{equation}
N_i^{\rm th}
=
\int_{E^{i}_{\rm ee}}^{E^{i+1}_{\rm ee}}{\rm d}E_{\rm ee}^{\rm reco}\;
\left[
n_N
\int {\rm d}T_{\rm nr}\;
G\!\left(E_{\rm ee}^{\rm reco},E_{\rm ee}(T_{\rm nr})\right)\;
\frac{{\rm d}R}{{\rm d}T_{\rm nr}}
\right]\,,
\label{eq:conus_bins_compact}
\end{equation}
where the $E_{\rm ee}(T_{\rm nr})$ is defined through the adopted quenching model. The $G (E_{\rm ee}^{\rm reco},E_{\rm ee}(T_{\rm nr}))$ is the energy resolution function.
All detector-response inputs (threshold, resolution, quenching) are taken from the CONUS+ analyses in Refs.~\cite{Ackermann:2025obx,DeRomeri:2025csu}.

For binned excess-count data, we adopt a profiled least-squares function with a global normalization nuisance parameter $\alpha$,
\begin{equation}
\chi^2_{\rm CONUS+}
=
\sum_{i=1}^{N_{\rm bin}}
\frac{\left[N_{\rm meas}^i-(1+\alpha)\,N_{\rm CE\nu NS}^i\right]^2}{\sigma_i^2}
+
\left(\frac{\alpha}{\sigma_\alpha}\right)^2 \,,
\label{eq:chi2_conus_compact}
\end{equation}
where $N_{\rm meas}^i$ and $\sigma_i$ are the reported excess counts and uncertainties~\cite{Ackermann:2025obx}, 
and the relative systematic uncertainty is taken to be $\sigma_\alpha = 0.169$.

For the binned analyses, the reported best-fit CE$\nu$NS signal yields for COHERENT CsI, COHERENT Ar, and CONUS+ are $306\pm20$~\cite{COHERENT:2021xmm}, $159\pm43$~\cite{COHERENT:2020iec}, and $395\pm106$~\cite{Ackermann:2025obx}, from which we can estimate the relative statistical uncertainties as $6.54\%$, $27.0\%$, and $26.8\%$, respectively. Together with the relative statistical uncertainties of the PandaX-4T and XENONnT measurements discussed above, this indicates that the current sensitivities of PandaX-4T, XENONnT, COHERENT Ar, and CONUS+ are limited by statistics, whereas the COHERENT CsI measurement has reached a regime in which systematic uncertainties dominate.

For the CsI data set, the relative systematic uncertainty $\sigma_\alpha = 0.125$~\cite{Li:2024iij,COHERENT:2021xmm} is obtained by combining in quadrature the contributions from the quenching factor ($3.8\%$), detector acceptance ($4.1\%$), neutrino-flux prediction ($10\%$), and nuclear response modeling ($5\%$). The first two are detector-related systematics, whereas the latter two are theoretical/modeling uncertainties. As mentioned in Ref.~\cite{Li:2024iij}, we assume a universal uncertainty of $5\%$ for the nuclear response, comparable to that used in the approach of nuclear form factor~\cite{DeRomeri:2022twg}. However, the uncertainties associated with the nuclear response functions may be even larger~\cite{Altmannshofer:2018xyo}. This highlights the importance of the systematic chiral expansion and power counting, which provide a controlled way to assess order-by-order truncation effects and uncertainties associated with the nuclear response. Thus, for future precision CE$\nu$NS measurements, reducing both detector-related and theoretical/modeling uncertainties is increasingly important as statistical uncertainties are reduced.

\subsection{Constraints on NP scales and NSI parameters}
We now recast the CE$\nu$NS measurements into constraints on the LEFT parameter space. 
For convenience, the dimensionful Wilson coefficients in Eq.~\eqref{coefficient} are decomposed into possible SM and NSI contributions as
\begin{equation}
  \hat{\mathcal{C}}_a^{d} = \left.\hat{\mathcal{C}}_a^{d}\right|_{\rm SM} + \left.\hat{\mathcal{C}}_a^{d}\right|_{\rm NSI}\,,\quad \left.\hat{\mathcal{C}}_a^{d}\right|_{\rm NSI} \equiv \frac{\mathcal{C}_{a, {\rm NSI}}^{d}}{\Lambda_{a, {\rm NSI}}^{\,d-4}} \,,
\end{equation}
where the NSI contribution is parametrized in terms of a dimensionless coupling $\mathcal{C}_{a,{\rm NSI}}^{d}$ and the corresponding effective NP scale $\Lambda_{a,{\rm NSI}}$.
Setting $|\mathcal{C}_{a,{\rm NSI}}^{d}|=1$, the experimental bound on Wilson coefficients is converted into a limit on $\Lambda_{a,{\rm NSI}}$. 
In this convention, a larger lower bound on $\Lambda_{a,{\rm NSI}}$ corresponds to a stronger suppression of the corresponding NP interaction and therefore a more stringent constraint from CE$\nu$NS data.

To derive limits, we adopt the single-operator approach: for a given operator $\mathcal O_a^{d}$ we switch on only $\hat{\mathcal{C}}_a^{d}$ while setting all other coefficients to zero, and we perform a one-parameter fit to the combined CE$\nu$NS data. The $90\%$ C.L.\ lower bound on $\Lambda_{a,{\rm NSI}}$ is then obtained by requiring the appropriate one-dimensional likelihood criterion,
\begin{equation}
  \Delta \chi^2(\Lambda_{a,{\rm NSI}})\equiv \chi^2(\Lambda_{a,{\rm NSI}})-\chi^2_{\rm min} \le 2.71\,,
\end{equation}
where $\chi^2_{\rm min}$ denotes the minimum of the profiled $\chi^2$ for that operator.

\begin{figure}[t]
    \centering
    {\includegraphics[scale=0.6]{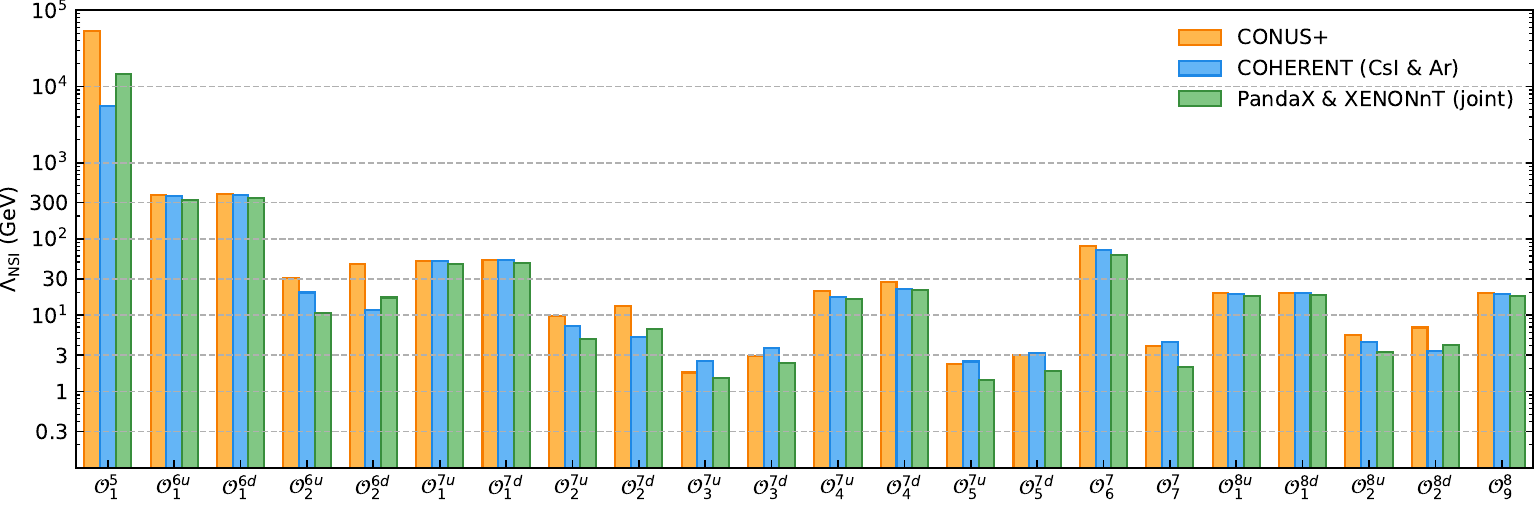}}
    \caption{
      For CE$\nu$NS processes initiated by an electron-flavor neutrino, we present lower bounds at $90\%$ C.L. on the effective scale $\Lambda_{\rm NSI}$ (in GeV) associated with individual LEFT operators $\mathcal{O}_1^{5}$, $\mathcal{O}_a^{6u}$,
      $\mathcal{O}_a^{6d}$, $\mathcal{O}_a^{7u}$, $\mathcal{O}_a^{7d}$, $\mathcal{O}_6^{7}$, $\mathcal{O}_7^{7}$, $\mathcal{O}_a^{8u}$, $\mathcal{O}_a^{8d}$ and $\mathcal{O}_9^{8}$.
      The limits are obtained by turning on one operator at a time.
      The orange, blue and green bars correspond to constraints from CONUS+,
      COHERENT (CsI and Ar combination), and the PandaX-4T and XENONnT combination, respectively.
    }
    \label{fig:Lambda_coefficients}
\end{figure}

The resulting bounds are summarized in Fig.~\ref{fig:Lambda_coefficients}, where we compare the constraints from the CONUS+, the COHERENT CsI and Ar measurements, and solar $^8$B CE$\nu$NS in liquid xenon (PandaX-4T and XENONnT). The CONUS+ source is an electron antineutrino flux. In what follows, we use the notation
$\nu_e$ to denote electron-flavor (anti)neutrinos and do not distinguish $\nu_e$ from
$\bar{\nu}_e$ in our rate-level constraints.  Presenting the results in terms of $\Lambda_{a,{\rm NSI}}$ provides an intuitive measure of the reach of each experiment in the LEFT framework and highlights the complementarity between reactor, accelerator-based and solar-neutrino CE$\nu$NS data sets across different operator structures. 

In general, operators that contribute to the coherently enhanced channel and interfere with the SM amplitude lead to more stringent bounds, whereas operators that are dominated by spin-dependent responses or lack coherent enhancement are constrained more weakly. This behavior can be understood in terms of nuclear response scaling in the long-wavelength limit:
the coherent (spin-independent) response scales as $W_M\sim \mathcal{O}(A^2)$ due to the in-phase addition of nucleon amplitudes, while spin-dependent responses scale only as $W_{\Sigma}\sim \mathcal{O}(1)$ and therefore do not benefit from the same parametric enhancement, as discussed in Tab.~\ref{tab:ResponseEnhance}. 
For instance, the vector-current dimension-6 operator $\mathcal{O}^{6u/d}_{1}$ couples to the coherent charge response and thus scales as $A^{2}$, whereas the axial operator $\mathcal{O}^{6u/d}_{2}$ is spin-dependent and incoherent and lacks the $A^{2}$ enhancement.

The comparison among experiments further highlights their complementarity. COHERENT provides spectral information in multiple bins and thus yields robust constraints on a broad set of coefficients. The solar-$^8$B liquid-xenon measurements (PandaX-4T and XENONnT) probe a different neutrino source and recoil-energy regime,
and their combined analysis offers competitive sensitivity for operators that primarily affect low-energy nuclear recoils. For instance, dipole-like contributions can be enhanced at low momentum transfer, so low-threshold recoil measurements can become particularly sensitive to such contributions. 
For higher-dimensional operators (dimension-7 and dimension-8), the overall reach in $\Lambda$ is generically reduced, as additional powers of $E_\nu/\Lambda$
or $q/\Lambda$ suppress their impact in MeV-scale CE$\nu$NS, requiring higher statistics and tighter systematics to achieve comparable sensitivity.

We also perform an analysis in terms of neutrino NSIs. At the quark level, the neutral-current NSIs~\cite{Wolfenstein:1977ue,Scholberg:2005qs,Barranco:2005yy,Davidson:2003ha,Du:2021rdg} can be parametrized as
\begin{align}
\label{eq:NSI-epsilon}
\mathcal{L}_{\mathrm{NC}}
&\supset
-2\sqrt{2}\,G_F
\Big[
\epsilon_{\alpha\beta}^{qL}
\left(\bar{\nu}_{\alpha}\gamma^{\mu}P_{L}\nu_{\beta}\right)
\left(\bar{q}\gamma_{\mu}P_{L}q\right)
+
\epsilon_{\alpha\beta}^{qR}
\left(\bar{\nu}_{\alpha}\gamma^{\mu}P_{L}\nu_{\beta}\right)
\left(\bar{q}\gamma_{\mu}P_{R}q\right)
\Big]\;,
\end{align}
where $P_{L/R}=(1\mp\gamma_{5})/2$, $\alpha,\beta$ denote neutrino flavors,
$q\in\{u,d\}$, and $G_F$ is the Fermi constant.

In the LEFT notation adopted in Eq.~\eqref{eq:OpLEFTL0}, the NSI parameters
$\epsilon_{\alpha\beta}^{qL(R)}$ are related to the dimension-6 Wilson
coefficients $\hat{\mathcal{C}}_{1,q}^{(6)}$ and $\hat{\mathcal{C}}_{2,q}^{(6)}$ via
\begin{align}
\label{eq:epsLR}
\epsilon_{\alpha\beta}^{qL/R}
=
-\frac{1}{2\sqrt{2}\,G_F}
\left(
\hat{\mathcal{C}}_{1,q}^{(6)} \big|_{\rm NSI}
\mp
\hat{\mathcal{C}}_{2,q}^{(6)} \big|_{\rm NSI}
\right)\;.
\end{align}

For comparison with existing CE$\nu$NS NSI studies, we focus on the vector
combination $\epsilon_{ee}^{qV} \equiv \epsilon_{ee}^{qL} + \epsilon_{ee}^{qR}$, which is related to the Wilson coefficient of LEFT operator as~\cite{Abdullah:2022zue}
\begin{align}
\label{eq:epsee}
\epsilon_{ee}^{qV}
=
-\frac{1}{\sqrt{2}\,G_F}
\left.\hat{\mathcal{C}}_{1,q}^{(6)}\right|_{\rm NSI}\;.
\end{align}

\begin{figure}[t]
    \centering
    {\includegraphics[scale=0.6]{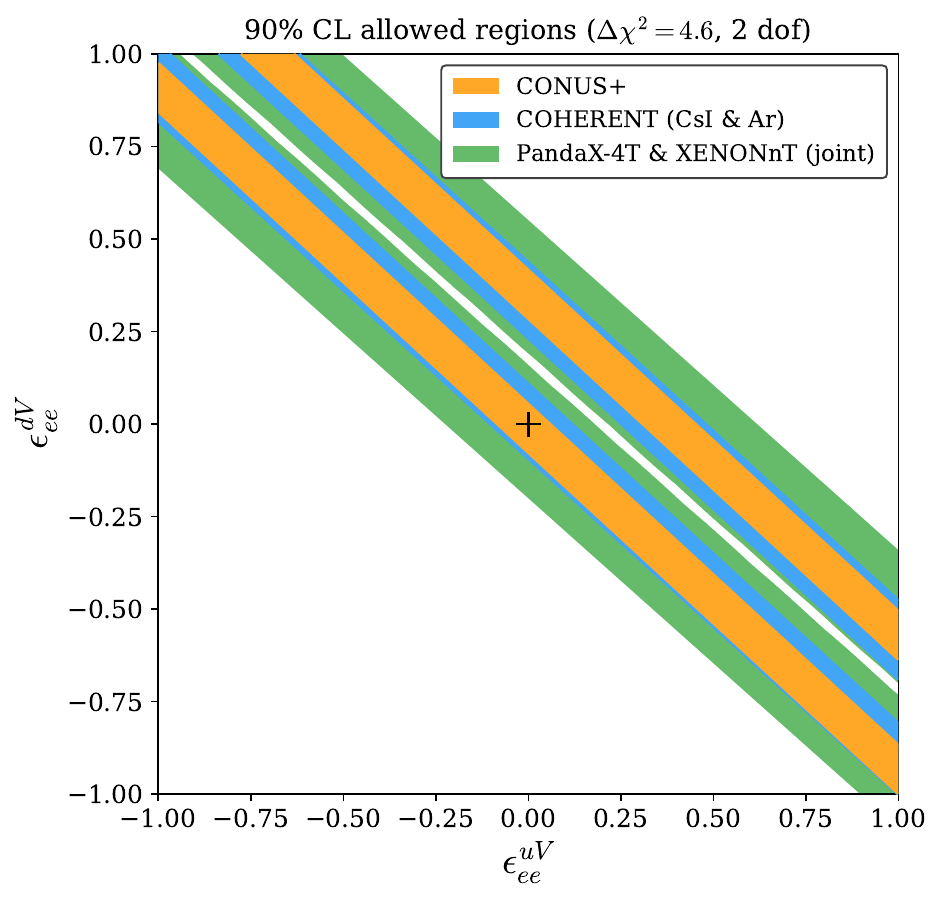}}
    \caption{The $90\%$ C.L. allowed regions in the $(\epsilon_{ee}^{uV},\,\epsilon_{ee}^{dV})$ plane are derived from the CONUS+ (orange) and COHERENT (blue) CE$\nu$NS measurements, together with the combined constraint from the PandaX-4T + XENONnT (green) solar-$^8$B analysis.}
    \label{fig:chiSquare8BCsI}
\end{figure}

Fig.~\ref{fig:chiSquare8BCsI} shows the $90\%$ C.L.\ allowed regions in the
$(\epsilon_{ee}^{uV},\,\epsilon_{ee}^{dV})$ plane derived from COHERENT (CsI and Ar),
CONUS+, and the combined solar-$^8$B (PandaX-4T$+$XENONnT) analysis. The contours are
obtained using the two-parameter criterion $\Delta\chi^{2}\le 4.61$ (2 degrees of
freedom). The resulting regions exhibit the characteristic elongated, band-like structure, reflecting the strong correlation between $\epsilon_{ee}^{uV}$ and $\epsilon_{ee}^{dV}$ in CE$\nu$NS. 
For a given nuclear target $(Z,N)$, the event
rate is primarily sensitive to an approximately linear combination of the up- and down-quark vector NSIs entering the response function $W_{MM}$, while the orthogonal combination remains only weakly constrained. Consequently, the allowed parameter space extends along a diagonal direction in the
$(\epsilon_{ee}^{uV},\,\epsilon_{ee}^{dV})$ plane rather than forming a compact elliptical region.

When we compare the widths of the allowed bands, CONUS+ yields the tightest constraint, followed by COHERENT, while the solar-$^8$B liquid-xenon combination is comparatively weaker in the present treatment.  This pattern is driven by the relative fluxes, thresholds, and dominant systematics: CONUS+ benefits from the intense reactor flux and low recoil thresholds, COHERENT is limited by SNS statistics and flux/background uncertainties, and the solar-$^8$B liquid-xenon analyses are further weakened by the broad neutrino-energy spectrum and detector-response systematics~\cite{DeRomeri:2025csu,Alpizar-Venegas:2025wor}.

\section{SMEFT Operators and UV Completions}
\label{sec:SMEFTUV}

Having established the low-energy description of CE$\nu$NS in the LEFT and carried it through
$\chi$PT to the nuclear response and differential cross section (see Fig.~\ref{fig:my_label}),
we now turn to the ultraviolet side of the same EFT chain. In particular, when the NP
scale $\Lambda$ lies above the electroweak scale, the appropriate starting point is the SMEFT,
defined between $\Lambda$ and $\Lambda_{\rm EW}$ with $SU(3)_C\times SU(2)_L\times U(1)_Y$
gauge invariance. The SMEFT operators are first evolved and matched at $\Lambda_{\rm EW}$ onto
LEFT operators, which then are incorporated into the hadronic and nuclear matching steps already developed
in the previous sections. This section therefore completes the top part of Fig.~\ref{fig:my_label}
by organizing the relevant SMEFT operator basis (up to dimension 8) and illustrating
representative UV completions that generate them.

\subsection{SMEFT Operators}
The LEFT describes the theory below the electroweak scale with the gauge group of $SU(3)_C\times U(1)_{\rm em}$. When the NP is heavier than the electroweak scale and contains the properties of the SM group $SU(3)_C\times SU(2)_L\times U(1)_Y$, the SMEFT comes into play to characterize the EFT valid between the NP scale $\Lambda$ and the electroweak scale $\Lambda_{\rm EW}$. 

The SMEFT Lagrangian comprises the SM Lagrangian of dimension 4, together with a complete set of independent higher-dimensional operators. 
\begin{equation}
    \mathcal L_{\text{SMEFT}}=\mathcal L_{\text{SM}}+\sum_{D\geq 5} \, \sum_{a}\hat{C}_a^D\,\mathcal O_a^D ~\,, \quad \hat{C}_a^D \equiv \dfrac{C_a^D} {\Lambda^{D-4}}\,,
\end{equation}
where $D$ denotes the mass dimension for the $a$-th operator $\mathcal{O}_a^D$. The Wilson coefficient $\hat{C}_a^D$ is dimensionful, while $C_a^D$ is dimensionless.

The SM Lagrangian in unbroken phase is
\begin{eqnarray}
\mathcal L_{\text{SM}} &=& \, -\dfrac14 G^A_{\mu\nu} G^{A\mu\nu} - \dfrac14 W^I_{\mu\nu} W^{I\mu\nu}
- \dfrac14 B_{\mu\nu} B^{\mu\nu} \nn\\
&& +\sum_{\psi=Q,u_R,d_R,\ell,e_R} \overline\psi i\slashed D\psi + (D_\mu H)^\dagger (D^\mu H)
- \lambda \left(H^\dagger H-\dfrac12v^2\right)^2 \nn\\
&& -\left[\overline \ell e_R(Y_e)H+\overline Qu_R(Y_u)\tilde H+\overline Qd_R(Y_d)H+\text{h.c.}\right]
\nn\\
&& +~\dfrac{\theta_3{g_s^2}}{32\pi^2}G^A_{\mu\nu}\tilde G^{A\mu\nu}
+\dfrac{\theta_2g^2}{32\pi^2}W^I_{\mu\nu}\tilde W^{I\mu\nu}+\dfrac{\theta_1{g'}^2}{32\pi^2}B_{\mu\nu}\tilde B^{\mu\nu} ~.
\label{SMLagrangian}
\end{eqnarray}
The gauge covariant derivative is $D_\mu=\partial_\mu+ig_1YB_\mu+ig_2t^IW_\mu^I+ig_3T^AG_\mu^A$. Here, $T^A=\lambda^A/2$ and $t^I=\tau^I/2$ are the $SU(3)_C$ and $SU(2)_L$ generators while $\lambda^A$ and $\tau^I$ are the Gell-Mann and Pauli matrices, and $Y$ is the $U(1)_Y$ generator. The complex conjugate of the Higgs field can be written as $H^\dagger$ or $\tilde{H}$. The $\tilde{H}$ is defined by $\tilde{H}_i=\epsilon_{ij}H^{\dagger j}$ where the $\epsilon_{ij}$ is totally antisymmetric with $\epsilon_{12}=1$. Fermion fields $Q$ and $\ell$ are left-handed doublet fields $(u_L,d_L)^T$ and $(\nu_L,e_L)^T$. The right-handed fermion fields are $u_R\,,d_R$ and $e_R$. In addition, the fermion fields have weak-eigenstate indices $r=1,2,3$ and the Yukawa couplings $Y_e$, $Y_u$ and $Y_d$ are $3\times 3$ matrices. However, only the $u$ and $d$ quarks are considered in the LEFT operators, such that quark flavor degrees of freedom are not taken into account; by contrast, the three flavor indices associated with neutrinos are considered in the analysis.

The relevant SMEFT operators up to dimension 8 are
listed in Tab.~\ref{tab:SMEFTop}. For convenience, the index $D$ is omitted.  The dimension-5 Weinberg operator $\mathcal{O}_5$ can contribute to the LNV process and has been constrained by the mass of neutrino. Thus, we neglect the contribution of $\mathcal{O}_5$ in this work. For the LNC dimension-6 and dimension-7 LEFT operators, we consider only contributions from dimension-6 SMEFT operators, as those from dimension-8 SMEFT operators are highly suppressed. The latter contributions, arising from dimension-8 SMEFT operators with two Higgs insertions, were considered in Ref.~\cite{Altmannshofer:2018xyo}. Separately, dimension-8 LEFT operators receive contributions only from dimension-8 SMEFT operators at tree-level matching; we therefore incorporate dimension-8 SMEFT operators in our analysis. For LNV LEFT operators, by contrast, we restrict our consideration to SMEFT operators up to dimension 7. The full set of matching conditions is presented in App.~\ref{app:MatchingSM2L}.

After the $SU(2)_L$ symmetry breaking, we can obtain the matching from SMEFT operators to LEFT operators as the Ref~\cite{Jenkins:2017jig}.
Similarly, the QCD RGEs of the SMEFT operators are also included; these were not considered in Ref.~\cite{Altmannshofer:2018xyo}. Only the Wilson coefficients $\hat C_{uB}\,,\hat C_{uW}\,,\hat C_{dB}\,,\hat C_{dW}\,,\hat C_{QuLLH}\,,\hat C_{dLQLH1}\,,\hat C_{dLQLH2}$ need to be considered. The RGEs of the SMEFT operators are given by~\cite{Jenkins:2013zja,Jenkins:2013wua,Alonso:2013hga,Bakshi:2024wzz,Boughezal:2024zqa}
\begin{align}
    \frac{d}{d~\rm{ln}\mu}\hat C_{u(d)B(W)}=&2C_F\frac{\alpha_s}{4\pi}\hat C_{u(d)B(W)}\,,\\
    \frac{d}{d~\rm{ln}\mu}\hat C_{QuLLH}=&-6C_F\frac{\alpha_s}{4\pi}\hat C_{QuLLH}\,,\\
    \frac{d}{d~\rm{ln}~\mu}\left(\begin{array}{c}
         \hat C_{dLQLH1}  \\
          \hat C_{dLQLH2}\\
    \end{array}\right)=&\frac{\alpha_s}{4\pi}\left(\begin{array}{cc}
        -6C_F & 0 \\
         4C_F& -6C_F
    \end{array}\right)\left(\begin{array}{c}
         \hat C_{dLQLH1}  \\
          \hat C_{dLQLH2}\\
    \end{array}\right)\,,\\
    \frac{d}{d~\rm{ln}\mu}\hat C_{\ell^2q^2D^2}^{(2)(4)}=&C_F\frac{\alpha_s}{3\pi}\hat C_{\ell^2q^2D^2}^{(2)(4)}\,,\\
    \frac{d}{d~\rm{ln}\mu}\hat C_{\ell^2[u(d)]^2D^2}=&C_F\frac{\alpha_s}{3\pi}\hat C_{\ell^2[u(d)]^2D^2}\,.
\end{align}

\begin{table}[h]
  \centering
  \resizebox{0.6\textwidth}{!}{  
\begin{tabular}{|c|c|c|c|}
\hline
\multicolumn{2}{|c|}{Dimension-6}&\multicolumn{2}{c|}{Dimension-5}\\
\hline
${\cal O}_H$ & $\left(H^\dagger H\right)^3$ & ${\cal O}_{5}$ & $\epsilon^{ik}\epsilon^{jl}(\ell^T_{i} C \ell_{j}) H_k H_l$  \\
\cline{3-4}
${\cal O}_{HW}$ & $H^\dagger H W^I_{\mu\nu} W^{I\mu\nu}$& \multicolumn{2}{c|}{Dimension-7}  \\
\cline{3-4}
${\cal O}_{HB}$ & $H^\dagger H B_{\mu\nu} B^{\mu\nu}$  & ${\cal O}_{QuLLH}$ & $\epsilon^{ij} (\overline{Q}^{k} u_{R}) (\ell^T_{k} C \ell_i) H_j $   \\
${\cal O}_{HWB}$ & $H^\dagger \tau^I H W^I_{\mu\nu} B^{\mu\nu}$ &  ${\cal O}_{dLQLH1}$ & $\epsilon^{ij}\epsilon^{kl} (\overline{d}_R\ell_{i}) (Q^T_{j} C \ell_k) H_l $    \\
${\cal O}_{HD}$ & $\left(H^\dagger D^\mu H\right)^* \left(H^\dagger D^\mu H\right) $& ${\cal O}_{dLQLH2}$ & $\epsilon^{ik}\epsilon^{jl} (\overline{d}_R\ell_{i}) (Q^T_{j} C \ell_k) H_l $\\
\cline{3-4}
${\cal O}_{H\ell}^{(1)}$ & $(H^\dagger i \overset{\leftrightarrow}{D}{}_\mu H)(\bar{\ell}\gamma^\mu \ell)$&\multicolumn{2}{c|}{Dimension-8}\\
\cline{3-4}
${\cal O}_{H\ell}^{(3)}$ & $(H^\dagger i \overset{\leftrightarrow}{D}{}_\mu^I H)(\bar{\ell}\tau^I\gamma^\mu \ell)$&$\Op_{l^2G^2D}$  &  $ (\bar \ell \gamma^\mu i\overleftrightarrow{D}^\nu \ell) G_{\mu\rho}^A G_{\nu}^{A\rho}$ \\
${\cal O}_{Hq}^{(1)}$ & $(H^\dagger i \overset{\leftrightarrow}{D}{}_\mu H)(\bar Q\gamma^\mu Q)$&$\Op_{l^2q^2G}^{(1)}$  & $(\bar \ell \gamma^\mu \ell) (\bar Q \gamma^\nu T^A Q) G^A_{\mu\nu}$\\
${\cal O}_{Hq}^{(3)}$ & $(H^\dagger i \overset{\leftrightarrow}{D}{}_\mu^I H)(\bar Q\tau^I\gamma^\mu Q)$&$\Op_{l^2q^2G}^{(2)}$  & $(\bar \ell \gamma^\mu \ell) (\bar Q \gamma^\nu T^A Q) \widetilde G^A_{\mu\nu}$\\
${\cal O}_{Hu}$ & $(H^\dagger i \overset{\leftrightarrow}{D}{}_\mu H)(\bar{u}_R\gamma^\mu u_R)$&$\Op_{l^2q^2G}^{(3)}$  & $(\bar \ell \gamma^\mu \tau^I \ell) (\bar Q \gamma^\nu T^A \tau^I Q) G^A_{\mu\nu}$\\
${\cal O}_{Hd}$ & $(H^\dagger i \overset{\leftrightarrow}{D}{}_\mu H)(\bar{d}_R\gamma^\mu d_R)$&$\Op_{l^2q^2G}^{(4)}$  & $(\bar \ell \gamma^\mu \tau^I \ell) (\bar Q \gamma^\nu T^A \tau^I Q) \widetilde G^A_{\mu\nu}$\\
 ${\cal O}_{uB}$ & $(\bar Q\sigma^{\mu\nu} u_R) \tilde{H} B_{\mu\nu}$&$\Op_{l^2u^2G}^{(1)}$  &  $(\bar \ell \gamma^\mu \ell) (\bar u_R \gamma^\nu T^A u_R) G^A_{\mu\nu}$\\
${\cal O}_{uW}$ & $(\bar Q\sigma^{\mu\nu} u_R)\tau^I \tilde{H} W^I_{\mu\nu}$&$\Op_{l^2u^2G}^{(2)}$  &  $(\bar \ell \gamma^\mu \ell) (\bar u_R \gamma^\nu T^A u_R) \widetilde G^A_{\mu\nu}$\\
${\cal O}_{dB}$ & $(\bar Q\sigma^{\mu\nu} d_R) H B_{\mu\nu}$&$\Op_{l^2d^2G}^{(1)}$  &  $(\bar \ell \gamma^\mu \ell) (\bar d_R \gamma^\nu T^A d_R) G^A_{\mu\nu}$ \\
${\cal O}_{dW}$ & $(\bar Q\sigma^{\mu\nu} d_R)\tau^I H W^I_{\mu\nu}$&$\Op_{l^2d^2G}^{(2)}$  &  $(\bar \ell \gamma^\mu \ell) (\bar d_R \gamma^\nu T^A d_R) \widetilde G^A_{\mu\nu}$\\
${\cal O}_{\ell q}^{(1)}$ & $(\bar{\ell} \gamma^\mu \ell)(\bar Q \gamma_\mu Q)$&$\Op_{l^2q^2D^2}^{(2)}$  &  $(\bar \ell \gamma^\mu \overleftrightarrow{D}^\nu \ell) (\bar Q \gamma_\mu \overleftrightarrow{D}_\nu Q)$\\
${\cal O}_{\ell q}^{(3)}$ & $(\bar{\ell} \gamma^\mu \tau^I \ell)(\bar Q \gamma_\mu \tau^I Q)$&$\Op_{l^2q^2D^2}^{(4)}$  &  $(\bar \ell \gamma^\mu \overleftrightarrow{D}^{I\nu} \ell) (\bar Q \gamma_\mu \overleftrightarrow{D}^I_\nu Q)$\\
${\cal O}_{\ell u}$ & $(\bar{\ell} \gamma^\mu \ell)(\bar{u}_R \gamma_\mu u_R)$&$\Op_{l^2u^2D^2}^{(2)}$  &  $(\bar \ell \gamma^\mu \overleftrightarrow{D}^\nu \ell) (\bar u_R \gamma_\mu \overleftrightarrow{D}_\nu u_R)$\\
${\cal O}_{\ell d}$ & $(\bar{\ell} \gamma^\mu \ell)(\bar{d}_R \gamma_\mu d_R)$&$\Op_{l^2d^2D^2}^{(2)}$  &  $(\bar \ell \gamma^\mu \overleftrightarrow{D}^\nu \ell) (\bar d_R \gamma_\mu \overleftrightarrow{D}_\nu d_R)$ \\
\hline
\end{tabular}
}
\caption{The SMEFT operators for CE$\nu$NS process.}
\label{tab:SMEFTop}
\end{table}

\subsection{UV Completions}
There are a considerable number of the UV models that can contribute to the SMEFT operators. We  derive for the first time the complete tree-level UV models and give the minimal Lagrangian of these models for CE$\nu$NS. In the notation adopted in Refs.~\cite{Li:2023cwy,Li:2023pfw}, the $(S\,,~F\,,~V\,,~R\,,~T)$ correspond to the scalar particles, spin 1/2 fermion particles, vector particles, spin 3/2 fermion particles and spin 2 tensor particles. In addition, the representation of the SM group is given as $(SU(3)_C,SU(2)_L)_{U(1)_Y}$. 

\subsection*{Dimension-6 UV Completions}
The UV completions of dimension-6 SMEFT operators are listed in Tab.~\ref{tab:UV6}. The corresponding gauge invariant UV Lagrangian for these particles can be obtained and the kinematic terms of the particles are
\begin{align}
    \mathcal{L}_{\rm UV}^{\rm kin}=&-S_{1}^{\dagger}(D^2+M_{S_{1}}^2)S_{1}-S_{4}^{\dagger i}(D^2+M_{S_{4}}^2)S_{4 i}-S_{5}^{\dagger I}(D^2+M_{S_{5}}^2)S_{5 I}-S_{6}^{\dagger I}(D^2+M_{S_{6}}^2)S_{6 I}\notag\\
    &~-S_{7}^{\dagger \rm i}(D^2+M_{S_{7}}^2)S_{7}^{\rm i}-S_{8}^{\dagger \rm i}(D^2+M_{S_{8}}^2)S_{8}^{\rm i}-S_{10}^{\dagger a}(D^2+M_{S_{10}}^2)S_{10 a}\notag\\
    &~-S_{12}^{\dagger a i}(D^2+M_{S_{12}}^2)S_{12 ai}-S_{13}^{\dagger a i}(D^2+M_{S_{13}}^2)S_{13 ai}-S_{14}^{\dagger a I}(D^2+M_{S_{14}}^2)S_{14 a I}\notag\\
    &~+\bar F_1(i{\slashed D}-M_{F_1})F_{1}+\bar F_2(i{\slashed D}-M_{F_2})F_{2}+\bar F_5^I(i{\slashed D}-M_{F_5})F_{5 I}+\bar F_6^I(i{\slashed D}-M_{F_6})F_{6 I}\notag\\
    &~+\bar F_8^a(i{\slashed D}-M_{F_8})F_{8 a}+\bar F_9^a(i{\slashed D}-M_{F_9})F_{9 a}+\bar F_{10}^{ai}(i{\slashed D}-M_{F_{10}})F_{10 ai}+\bar F_{11}^{ai}(i{\slashed D}-M_{F_{11}})F_{11 ai}\notag\\
    &~+\bar F_{12}^{ai}(i{\slashed D}-M_{F_{12}})F_{12 ai}+\bar F_{13}^{aI}(i{\slashed D}-M_{F_{13}})F_{13 aI}
    +\bar F_{14}^{aI} \left(i\slashed D-M_{F_{14}}\right)F_{14\,aI}\notag\\
    &~+V_1^{\dagger\mu}(g_{\mu\nu}D^2-D_\nu D_\mu+g_{\mu\nu} M_{V_1}^2)V_1^\nu+V_2^{\dagger\mu}(g_{\mu\nu}D^2-D_\nu D_\mu+g_{\mu\nu} M_{V_2}^2)V_2^\nu\notag\\
    &~+V_4^{\dagger\mu}(g_{\mu\nu}D^2-D_\nu D_\mu+g_{\mu\nu} M_{V_4}^2)V_4^\nu+V_5^{\dagger\mu}(g_{\mu\nu}D^2-D_\nu D_\mu+g_{\mu\nu} M_{V_5}^2)V_5^\nu\notag\\
    &~+V_7^{\dagger\mu}(g_{\mu\nu}D^2-D_\nu D_\mu+g_{\mu\nu} M_{V_7}^2)V_7^\nu+V_8^{\dagger\mu}(g_{\mu\nu}D^2-D_\nu D_\mu+g_{\mu\nu} M_{V_8}^2)V_8^\nu\notag\\
    &~+V_9^{\dagger\mu}(g_{\mu\nu}D^2-D_\nu D_\mu+g_{\mu\nu} M_{V_9}^2)V_9^\nu\,,
    \end{align}
where $g_{\mu\nu}$ is the metric tensor, $a$ is the $SU(3)_C$ triplet index, and $i$, $I$, and ${\rm i}$ are the $SU(2)_L$ doublet, triplet, and quartet indices, respectively.
Moreover, the interactions of these particles with the SM field are
\begin{align}
\label{Inter6}
    \mathcal{L}_{\rm UV}^{\rm int}=&\mathcal{C}_{S_1H^2}(H^\dagger H)S_1+\mathcal{C}_{S_1^3}(S_1S_1S_1)+\mathcal{D}_{S_1^2H^2}(H^\dagger H)S_1S_1+\mathcal{D}_{S_4 H^3}(H^\dagger H)(H^{\dagger i} S_{4i})\notag\\
    &+\mathcal{C}_{S_5H^2}(H\tau^I H^\dagger)S_5^I+\mathcal{D}_{S_5^2H^21}(H^\dagger H)S_{5I} S_5^I+\mathcal{D}_{S_5^2H^22}(\tau^I)^i_k(\tau^J)^k_jH^{\dagger j} H_iS_{5}^I S_5^J\notag\\
    &+\mathcal{C}_{S_6H^2}\epsilon^{kj}(\tau^I)^i_kH_iH_jS_6^{\dagger I}+\mathcal{D}_{S_6^2H^21}(H^\dagger H)S_6^IS_6^{\dagger I}+\mathcal{D}_{S_6^2H^22}(\tau^I)^i_k(\tau^J)^k_jH^{\dagger j} H_iS_{6}^I S_6^{\dagger J}\notag\\
    &+\mathcal{D}_{S_7H^3}\epsilon_{kl}C_{\rm i}^{ijl}H_iH_jH^{\dagger k}S_7^{\dagger \rm i}+\mathcal{D}_{S_8H^3}C_{\rm i}^{ijk}H_iH_jH_kS_8^{\dagger \rm i}+\mathcal{D}_{S_{10}\ell Q}\epsilon^{ij}(\ell _i^TCQ_{aj})S_{10}^{\dagger a}\notag\\
    &+\mathcal{D}_{S_{12}\ell d}\epsilon^{ij}(\Bar{d}^a_R\ell_i){S_{12}}_{aj}+\mathcal{D}_{S_{13}\ell u}\epsilon^{ij}(\Bar{u}^a_R\ell_i){S_{13}}_{aj}+\mathcal{D}_{S_{14}LQ}\epsilon^{jk}(\tau)^i_k(\ell^T_iCQ_{aj})S_{14}^{\dagger aI}\notag\\
    &+\mathcal{D}_{F_1H\ell}\epsilon^{ij}(F^T_1C\ell_i)H_j+\mathcal{D}_{F_2H\ell}(F_2^TC\ell_i)H^{\dagger i}+\mathcal{D}_{F_5H\ell}\epsilon^{kj}(\tau^I)^i_k(F_5^{IT}C\ell_i)H_j\notag\\
    &+\mathcal{D}_{F_6H\ell}(\tau^I)^i_j(F_6^{IT}C\ell_i)H^{\dagger j}+\mathcal{D}_{F_8HQ}(\bar F_8^aQ_{ai})H^{\dagger i}+\mathcal{D}_{F_9HQ}\epsilon^{ij}(\bar F_9^aQ_{ai})H_j\notag\\
    &+\mathcal{D}_{F_{10}Hd}\epsilon^{ij}(\bar d_R^aF_{10ai})H_j+\mathcal{D}_{F_{11}Hd}(\bar d_R^aF_{11ai})H^{\dagger i}+\mathcal{D}_{F_{11}Hu}\epsilon^{ij}(\bar u_R^aF_{11ai})H_j\notag\\
    &+\mathcal{D}_{F_{12}Hu}(\bar u_R^a F_{12ai})H^{\dagger i}+\mathcal{D}_{F_{13}HQ}(\tau^I)^i_j(\bar F_{13}^{aI} Q_{ai})H^{\dagger j}+\mathcal{D}_{F_{14}HQ}\epsilon^{kj}(\tau^I)^i_k(\bar F_{14}^{aI}Q_{ai})H_j\notag\\
    &+\mathcal{D}_{V_1DH}(H^\dagger D_\mu H)V^\mu_1+\mathcal{D}_{V_1\ell^2}(\bar\ell\gamma_\mu\ell)V_1^\mu+\mathcal{D}_{V_1Q^2}(\bar Q\gamma_\mu Q)V_1^\mu+\mathcal{D}_{V_1u^2}(\bar u_R\gamma_\mu u_R)V_1^\mu\notag\\
    &+\mathcal{D}_{V_1d^2}(\bar d_R\gamma_\mu d_R)V_1^\mu+\mathcal{D}_{V_2DH^2}\epsilon^{ij}(D_\mu H_i) H_j V_2^{\dagger \mu}+\mathcal{D}_{V_4DH^2}(\tau^I)^i_j(\tau^I)H^{\dagger j}V_4^{I\mu}\notag\\
    &+\mathcal{D}_{V_4L^2}(\bar\ell\tau^I\ell)V_4^{I\mu}+\mathcal{D}_{V_4Q^2}(\bar Q\tau^IQ)V_4^{I\mu}+\mathcal{D}_{V_5\ell Q}(\ell\gamma_\mu Q_a)V_5^{\dagger a\mu}\notag\\
    &+\mathcal{D}_{V_7\ell d}(\bar d_R^a\gamma_\mu C{\Bar{\ell^{iT}}})V_{7ai}^{\mu}+\mathcal{D}_{V_8\ell u}(\bar u_R^a\gamma_\mu C{\Bar{\ell^{iT}}})V_{8ai}^\mu+\mathcal{D}_{V_9\ell Q}(\bar\ell\tau^IQ_a)V_9^{\dagger aI\mu}+{\rm h.c.}\,,
\end{align}
where the coupling constants $\mathcal{C}s$ have the mass dimension and the coupling constants $\mathcal{D}s$ are dimensionless. We define the $C_{\rm i}^{ijk}$ via the relation $C_{\rm i}^{ijk}=\frac{1}{\sqrt{2}}(C_{\rm i})^{Jk}(\epsilon\tau^J)^{ji}$ where the index i takes values $3/2,\,1/2,\,-1/2,\,3/2$, $\epsilon^{ij}$
denotes the $SU(2)$-invariant antisymmetric tensor with indices $i,j=1,2$; $\tau^J$ $(J=1,2,3)$ are the Pauli matrices and the $C_{\rm i}$ matrices are
\begin{align}
	(C_{3/2})^{Ij} = \frac{1}{\sqrt{2}}\left( \begin{array}{ccc}
		1 & 0  \\
		-i & 0  \\
		0 & 0 
	\end{array}\right),
	(C_{1/2})^{Ij} = \frac{1}{\sqrt{6}}\left( \begin{array}{ccc}
		0 & 1  \\
		0 & -i  \\
		-2 & 0 
	\end{array}\right), \nonumber \\
	(C_{-1/2})^{Ij} = -\frac{1}{\sqrt{6}}\left( \begin{array}{ccc}
		1 & 0  \\
		i & 0  \\
		0 & 2 
	\end{array}\right),
	(C_{-3/2})^{Ij} = -\frac{1}{\sqrt{2}}\left( \begin{array}{ccc}
		0 & 1  \\
		0 & i  \\
		0 & 0 
	\end{array}\right).
\end{align}

\begin{figure}[ht]
    \centering
    \begin{minipage}[t]{0.3\textwidth}
        \centering
        \includegraphics[width=\linewidth]{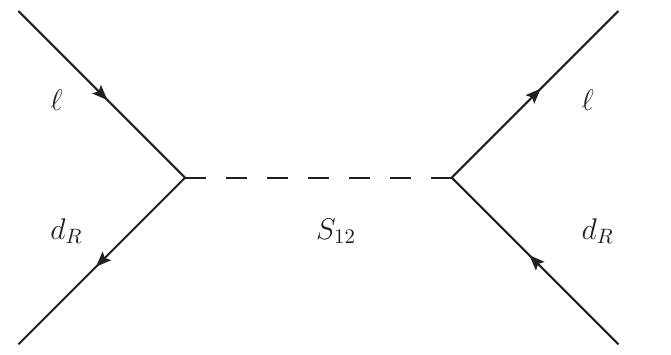}
    \end{minipage}
    \hfill
    \begin{minipage}[t]{0.3\textwidth}
        \centering
        \includegraphics[width=\linewidth]{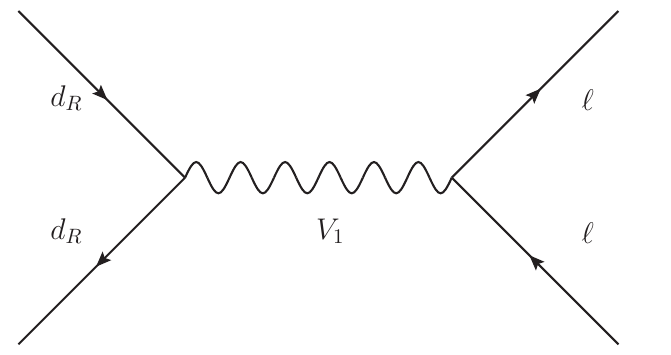}
    \end{minipage}
    \hfill
    \begin{minipage}[t]{0.3\textwidth}
        \centering
        \includegraphics[width=\linewidth]{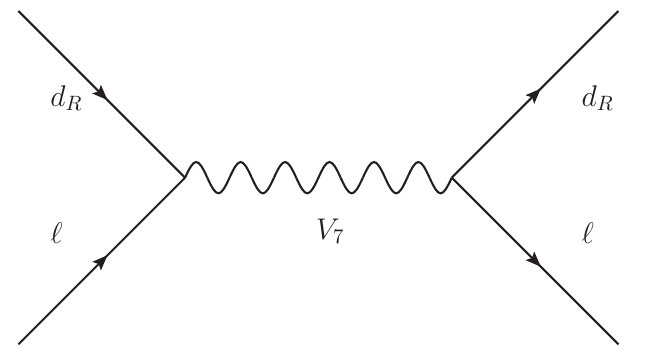}
    \end{minipage}
    \centering
    \caption{Contributions to $\mathcal{O}_{ld}$ from the UV completions with $S_{12}$\,, $V_1$ and $V_7$ particles.}
    \label{Dimension6}
\end{figure}

The SMEFT operator can be obtained through integrating out the UV particles. Here, we take the operator $\mathcal{O}_{ld}$ as the example. As is shown in Fig.~\ref{Dimension6}, these three particles can contribute to $\mathcal{O}_{ld}$. The Wilson coefficient for $\mathcal{O}_{ld}$ through the matching
\begin{align}
    \frac{C_{ld}}{\Lambda^2}=
    &\frac{\mathcal{D}_{V_{1}d^2}\mathcal{D}_{ V_{1}\ell^2}}{M_{V_1}^2}+\frac{\mathcal{D}_{d^\dagger \ell^\dagger V_{7}}\mathcal{D}_{d^\dagger \ell^\dagger V_{7}}}{M_{V_7}^2}+\frac{\mathcal{D}_{S_{12}\ell d}\mathcal{D}_{S_{12}\ell d}}{M_{S_{12}}^2}\,,
\end{align}
and the Wilson coefficients of any other SMEFT operators can be obtained through Eq.~\eqref{Inter6}. The constraints of the coupling constants and mass would be deduced from the Wilson coefficients which contribute to the low-energy experiments through the above framework.

\begin{table}[h]
    \centering
    \resizebox{0.8\textwidth}{!}{
    \begin{tabular}{c|c}
    \hline
    \hline
        $\mathcal{O}_H$ & $S_1(\mathbf{1},\mathbf{1})_0\,,S_4(\mathbf{1},\mathbf{2})_{\frac{1}{2}}\,,S_5(\mathbf{1},\mathbf{3})_0\,,S_6(\mathbf{1},\mathbf{3})_1\,,S_7(\mathbf{1},\mathbf{4})_{\frac{1}{2}}\,,S_8(\mathbf{1},\mathbf{4})_{\frac{3}{2}}\,,V_2(\mathbf{1},\mathbf{1})_1\,,V_4(\mathbf{1},\mathbf{3})_0$ \\
        \hline
        $\mathcal{O}_{HD}$ &$S_5(\mathbf{1},\mathbf{3})_0\,,S_6(\mathbf{1},\mathbf{3})_1\,,V_1(\mathbf{1},\mathbf{1})_0\,,V_2(\mathbf{1},\mathbf{1})_1\,,V_4(\mathbf{1},\mathbf{3})_0$\\ 
        \hline
        $\mathcal{O}_{H\ell}^{(1)}$ &$F_1(\mathbf{1},\mathbf{1})_0\,,F_2(\mathbf{1},\mathbf{1})_1\,,F_5(\mathbf{1},\mathbf{3})_0\,,F_6(\mathbf{1},\mathbf{3})_1\,,V_1(\mathbf{1},\mathbf{1})_0$\\ 
        \hline
        $\mathcal{O}_{H\ell}^{(3)}$ &$F_1(\mathbf{1},\mathbf{1})_0\,,F_2(\mathbf{1},\mathbf{1})_1\,,F_5(\mathbf{1},\mathbf{3})_0\,,F_6(\mathbf{1},\mathbf{3})_1\,,V_4(\mathbf{1},\mathbf{3})_0$\\ 
        \hline
        $\mathcal{O}_{Hq}^{(1)}$ &$F_8(\mathbf{3},\mathbf{1})_{-\frac{1}{3}}\,,F_9(\mathbf{3},\mathbf{1})_{\frac{2}{3}}\,,F_{13}(\mathbf{3},\mathbf{3})_{-\frac{1}{3}}\,,F_{14}(\mathbf{3},\mathbf{3})_{\frac{2}{3}}\,,V_1(\mathbf{1},\mathbf{1})_0$\\ 
        \hline
        $\mathcal{O}_{Hq}^{(3)}$ &$F_8(\mathbf{3},\mathbf{1})_{-\frac{1}{3}}\,,F_9(\mathbf{3},\mathbf{1})_{\frac{2}{3}}\,,F_{13}(\mathbf{3},\mathbf{3})_{-\frac{1}{3}}\,,F_{14}(\mathbf{3},\mathbf{3})_{\frac{2}{3}}\,,V_4(\mathbf{1},\mathbf{3})_0$\\ 
        \hline
        $\mathcal{O}_{Hu}$ &$F_{11}(\mathbf{3},\mathbf{2})_{\frac{1}{6}}\,,F_{12}(\mathbf{3},\mathbf{2})_{\frac{7}{6}}\,,V_1(\mathbf{1},\mathbf{1})_0$\\ 
        \hline
        $\mathcal{O}_{Hd}$ &$F_{10}(\mathbf{3},\mathbf{2})_{-\frac{5}{6}}\,,F_{11}(\mathbf{3},\mathbf{2})_{\frac{1}{6}}\,,V_1(\mathbf{1},\mathbf{1})_0$\\ 
        \hline
        $\mathcal{O}_{\ell q}^{(1)}$ &$S_{10}(\mathbf{3},\mathbf{1})_{-\frac{1}{3}}\,,S_{14}(\mathbf{3},\mathbf{3})_{-\frac{1}{3}}\,,V_1(\mathbf{1},\mathbf{1})_0\,,V_5(\mathbf{3},\mathbf{1})_{\frac{2}{3}}\,,V_9(\mathbf{3},\mathbf{1})_{\frac{2}{3}}$\\ 
        \hline
        $\mathcal{O}_{\ell q}^{(3)}$ &$S_{10}(\mathbf{3},\mathbf{1})_{-\frac{1}{3}}\,,S_{14}(\mathbf{3},\mathbf{3})_{-\frac{1}{3}}\,,V_4(\mathbf{1},\mathbf{3})_0\,,V_5(\mathbf{3},\mathbf{1})_{\frac{2}{3}}\,,V_9(\mathbf{3},\mathbf{3})_{\frac{2}{3}}$\\ 
        \hline
        $\mathcal{O}_{\ell u}$ &$S_{13}(\mathbf{3},\mathbf{2})_{\frac{7}{6}}\,,V_1(\mathbf{1},\mathbf{1})_0\,,V_8(\mathbf{3},\mathbf{2})_{\frac{1}{6}}$\\ 
        \hline
        $\mathcal{O}_{\ell d}$ &$S_{12}(\mathbf{3},\mathbf{2})_{\frac{1}{6}}\,,V_1(\mathbf{1},\mathbf{1})_0\,,V_7(\mathbf{3},\mathbf{2})_{-\frac{5}{6}}$\\ 
        \hline
        \hline
    \end{tabular}}
    \caption{The UV completions for dimension-6 SMEFT  operators that contribute to CE$\nu$NS.}
    \label{tab:UV6}
\end{table}

\subsection*{Dimension-7 UV Completions}

The dimension-7 SMEFT operators would violate the lepton number, and three types of the seesaw $S_{6}(\mathbf{1},\mathbf{3})_{0}$, $F_1(\mathbf{1},\mathbf{1})_0$ and $F_5(\mathbf{1},\mathbf{3})_{0}$ can contribute to the dimension-7 operators. However, these three types of the seesaw can contribute to dimension 5 and are highly suppressed. We exclude these three types of the seesaw models in the dimension-7 SMEFT operators UV completion. When considering other UV models, at least two particles should be added at tree level. We list the pairs of the UV models in Tab.~\ref{tab:UV7}.

Similarly, the gauge invariant UV Lagrangian involving these particles becomes
\begin{align}
    \mathcal{L}_{\rm UV}=&\bigg\{-S_{2}^{\dagger}(D^2+M_{S_{2}}^2)S_{2}-S_{4}^{\dagger i}(D^2+M_{S_{4}}^2)S_{4 i}-S_{12}^{\dagger a i}(D^2+M_{S_{12}}^2)S_{12 ai}\notag\\
    &\quad+\bar F_8^a(i{\slashed D}-M_{F_8})F_{8a}+\bar F_9^a(i{\slashed D}-M_{F_9})F_{9a}+\bar F_{12}^{ai}(i{\slashed D}-M_{F_{12}})F_{12ai}\notag\\
    &\quad+\bar F_{14}^{aI}(i{\slashed D}-M_{F_{14}})F_{14a}^I+V_5^{\dagger\mu}(g_{\mu\nu}D^2-D_\nu D_\mu+g_{\mu\nu} M_{V_5^2})V_5^\nu\notag\\
    &\quad+V_9^{\dagger\mu}(g_{\mu\nu}D^2-D_\nu D_\mu+g_{\mu\nu} M_{V_9}^2)V_9^\nu\bigg\}\notag\\
    &+\bigg\{\mathcal{D}_{S_2\ell^2}\epsilon^{ij}(\ell_i^T C\ell_j)S_2+\mathcal{D}_{S_4uQ}\epsilon^{ij}(\bar u_R Q_j) S_{4i}+\mathcal{D}_{S_4Qd}(\bar d_RQ_i)S_4^{\dagger i}\notag\\
    &\quad\quad+\mathcal{D}_{S_{12}\ell d}\epsilon^{ij}(\bar d_R^a\ell_j)S_{12ai}+\mathcal{D}_{F_8HQ}(\bar F^a_8Q_{ai})H^{\dagger i}+\mathcal{D}_{F_9HQ}\epsilon^{ij}(\bar F_{9}^aQ_{ai})H_j\notag\\
    &\quad\quad+\mathcal{D}_{F_{10}Hd}\epsilon^{ij}(\bar d_R^aF_{10ai})H_j+\mathcal{D}_{F_{12}Hu}(\bar u_R^aF_{12 ai})H^{\dagger i}+\mathcal{D}_{F_{14}HQ}\epsilon^{kj}(\tau^I)^i_k(\bar F_{14}^{aI}Q_{ai})H_j\,,\notag\\
    &\quad\quad+\mathcal{D}_{V_5\ell Q}(\bar\ell\gamma_\mu Q)V_5^{\dagger a\mu}-\sqrt{2}\mathcal{D}_{V_9\ell Q}(\bar\ell\tau^IQ)V_9^{\dagger aI\mu}+\rm h.c.\bigg\}\notag\\
    &+\bigg\{\mathcal{C}_{S_2S_4H}\epsilon^{ij}H_jS_{4i}S_2^\dagger+\mathcal{D}_{S_2F_8u}(\bar u_R^a F_{8a})S_2+\mathcal{D}_{S_2F_9d}(\bar d_R^aF_{9a})S_2^\dagger\notag\\
    &\quad\quad+\mathcal{D}_{S_2F_{10}Q}(F_{10}^{aiT}CQ_{ai})S_2^\dagger+\mathcal{D}_{S_2F_{12}Q}(F^T_{12ai}CQ^{ai})S_2+\mathcal{D}_{S_{12}F_9\ell}(F_{9a}^TC\ell_i)S_{12}^{\dagger ai}\notag\\
    &\quad\quad+\mathcal{D}_{S_{12}F_{14}\ell}(\tau^I)^i_j(F_{14a}^{IT}C\ell_j)S_{12}^{\dagger ai}+\mathcal{D}_{F_{12}V_5\ell}\epsilon_{ij}(\ell^j\gamma_\mu F_{12a}^i)(V_5)^{a\mu}\notag\\
    &\quad\quad+\mathcal{D}_{F_{12}V_9\ell}\epsilon_{kj}(\tau^I)^k_i(\ell^j\gamma_\mu F_{12a}^i)V_9^{aI\mu}+\rm h.c.\bigg\}\,,
\end{align}
where the first curly bracket contains the kinetic terms, the second one contains the interactions of new particles and SM particles and the last one contains the mixed term for new particles.

\begin{table}[H]
\centering
\resizebox{0.8\textwidth}{!}{  
    \begin{tabular}{c|c}
    \hline
    \hline
       \multirow{2}{*}{$\mathcal{O}_{QuLLH}$}  &  $\{S_2(\mathbf{1},\mathbf{1})_1,S_4(\mathbf{1},\mathbf{2})_{\frac{1}{2}}\}\quad\{S_2(\mathbf{1},\mathbf{1})_1,F_8(\mathbf{3},\mathbf{1})_{-\frac{1}{3}}\}\quad\{S_2(\mathbf{1},\mathbf{1})_1,F_{12}(\mathbf{3},\mathbf{2})_{\frac{7}{6}}\}$\\
       &$\{F_{12}(\mathbf{3},\mathbf{2})_{\frac{7}{6}},V_5(\mathbf{3},\mathbf{1})_{\frac{2}{3}}\}\quad\{F_{12}(\mathbf{3},\mathbf{2})_{\frac{7}{6}},V_9(\mathbf{3},\mathbf{3})_{\frac{2}{3}}\}$\\
       \hline
       $\mathcal{O}_{dLQLH1}$  & $\{S_{12}(\mathbf{3},\mathbf{2})_\frac{1}{6},F_{14}(\mathbf{3},\mathbf{3})_{\frac{2}{3}}\}$\\
       \hline
       \multirow{2}{*}{$\mathcal{O}_{dLQLH2}$} & $\{S_2(\mathbf{1},\mathbf{1})_1,S_4(\mathbf{1},\mathbf{2})_{\frac{1}{2}}\}\quad\{S_2(\mathbf{1},\mathbf{1})_1,F_9(\mathbf{3},\mathbf{1})_{\frac{2}{3}}\}\quad\{S_2(\mathbf{1},\mathbf{1})_1,F_{10}(\mathbf{3},\mathbf{2})_{-\frac{5}{6}}\}$\\
       &$\{S_{12}(\mathbf{3},\mathbf{2})_{\frac{1}{6}},F_9(\mathbf{3},\mathbf{1})_{\frac{2}{3}}\}\quad\{S_{12}(\mathbf{3},\mathbf{2})_{\frac{1}{6}},F_{14}(\mathbf{3},\mathbf{3})_{\frac{2}{3}}\}$\\
       \hline
       \hline
    \end{tabular}}
    \caption{The UV completions for dimension-7 SMEFT  operators that contribute to CE$\nu$NS.}
    \label{tab:UV7}
\end{table}

\begin{figure}[ht]
    \centering
    \begin{minipage}[t]{\textwidth}
        \centering
        \begin{subfigure}[t]{0.32\textwidth}
            \includegraphics[width=\textwidth]{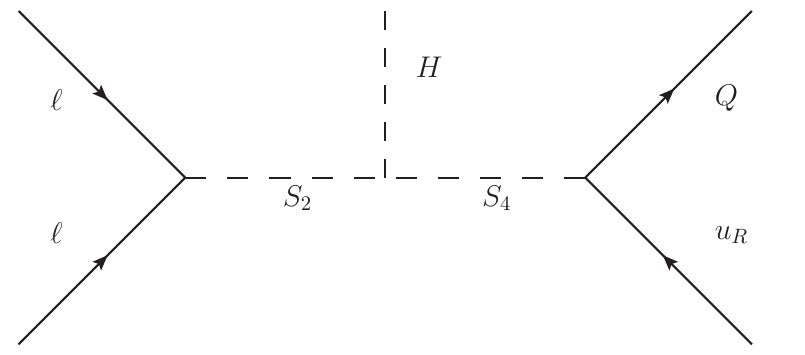}
        \end{subfigure}
        \hfill
        \begin{subfigure}[t]{0.32\textwidth}
            \includegraphics[width=\textwidth]{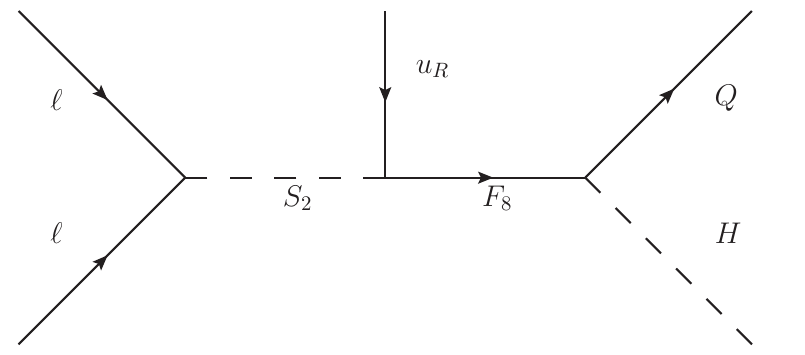}
        \end{subfigure}
        \hfill
        \begin{subfigure}[t]{0.32\textwidth}
            \includegraphics[width=\textwidth]{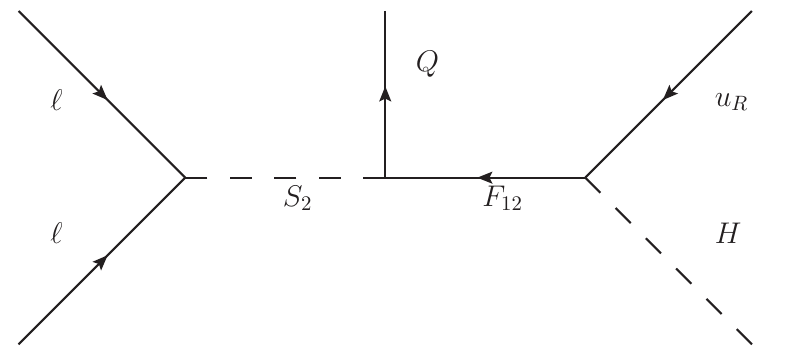}
        \end{subfigure}
    \end{minipage}
    
    \vspace{0.1cm}
    \begin{minipage}[t]{\textwidth}
        \centering
        \begin{subfigure}[t]{0.32\textwidth}
            \includegraphics[width=\textwidth]{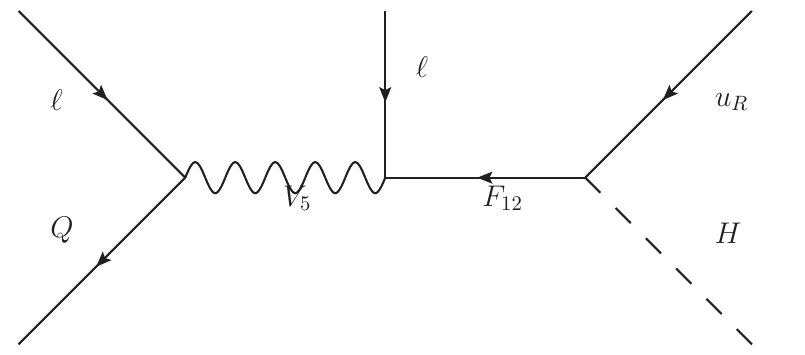}
        \end{subfigure}
        \begin{subfigure}[t]{0.32\textwidth}
            \includegraphics[width=\textwidth]{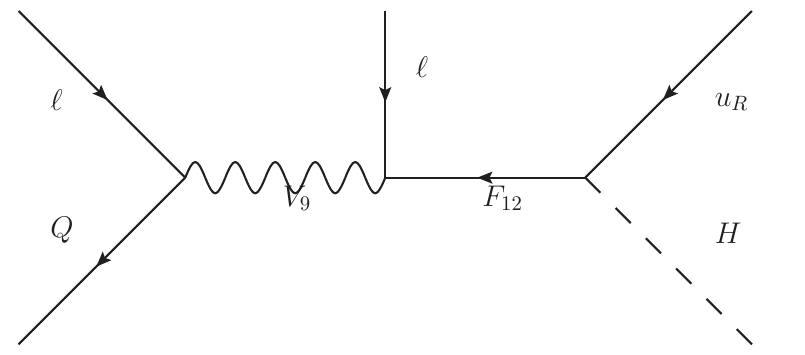}
        \end{subfigure}
    \end{minipage}
    \caption{Contributions to $\mathcal{O}_{QuLLH}$ from the UV completions with $S_{2}$\,, $S_4$\,, $F_8$\,, $F_{12}$\,, $V_5$ and $V_9$ particles.}
    \label{Dmiension7}
    \end{figure}

There are five pairs of the UV models that can contribute to $\mathcal{O}_{QuLLH}$. After integrating out the UV particles through the Fig.~\ref{Dmiension7}, the Wilson coefficient of $\mathcal{O}_{QuLLH}$ can be obtained similarly
\begin{align}
    \frac{C_{QuLLH}}{\Lambda^3}=&\frac{\mathcal{C}_{S_2S_4H}\mathcal{D}_{S_2\ell^2}D_{S_4uQ}}{M_{S_2}^2M_{S_4}^2}+\frac{\mathcal{D}_{S_2F_8u}\mathcal{D}_{S_2\ell^2}D_{F_8HQ}}{M_{S_2}^2M_{F_8}}+\frac{\mathcal{D}_{S_2F_{12}Q}\mathcal{D}_{S_2\ell^2}\mathcal{D}_{F_{12}Hu}}{M_{S_2}^2M_{F_{12}}}\notag\\
    &+\frac{\mathcal{D}_{F_{12}V_5\ell}\mathcal{D}_{V_5\ell Q}\mathcal{D}_{F_{12}Hu}}{M_{V_5}^2M_{F_{12}}}+\frac{\mathcal{D}_{F_{12}V_9\ell}\mathcal{D}_{V_9\ell Q}\mathcal{D}_{F_{12}Hu}}{M_{V_9}^2M_{F_{12}}}\,.
\end{align}
\\

\subsection*{Dimension-8 UV Completions}
\begin{figure}
    \centering
    \includegraphics[width=0.4\linewidth]{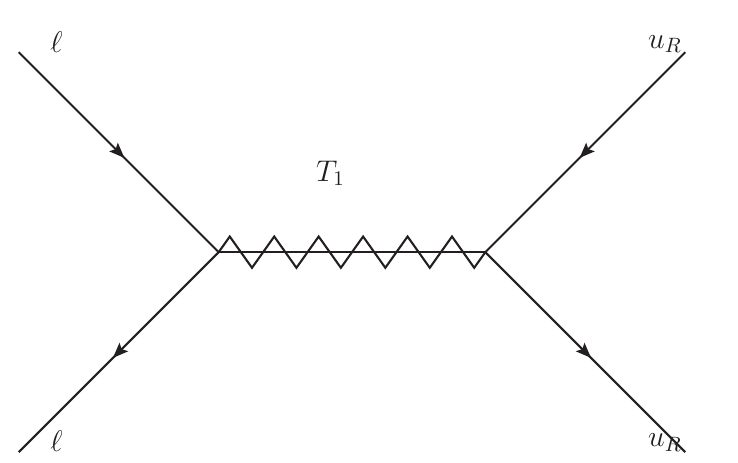}
    \caption{Contribution to $\mathcal{O}_{\ell^2u^2D^2}^{(2)}$ from the UV completions with $T_1$ particle.}
    \label{tensor}
\end{figure}

When considering the SMEFT dimension-8 operators, there are considerable numbers of the UV models. Here, we consider only the UV models that will not generate dimension-6 operators. In the dimension-8 UV completions, the spin 3/2 and spin 2 particles are included. Furthermore, the spin 3/2 and spin 2 particles are considered as Rarita-Schwinger~\cite{Rarita:1941mf} and Fierz-Pauli~\cite{Fierz:1939ix} fields. Similarly, the UV completions of the dimension-8 SMEFT operators that do not generate dimension-6 operators are listed in Tab.~\ref{tab:UV8}. Then the gauge invariant UV Lagrangian is given by 
\begin{align}
    \mathcal{L}_{UV}=& \bar F_{18}^{Ai}(i{\slashed D}-M_{F_{18}})F_{18i}^A+\mathcal{D}_{F_{18}\ell G}(\bar F_{18}^{Ai}\gamma^\mu\partial^\nu\ell_i)G_{\mu\nu}^A\notag\\
    &-\frac{1}{2}\left\{g^{\mu\lambda}g^{\nu\rho}+g^{\nu\lambda}g^{\mu\rho}-\frac{2}{3}g^{\mu\nu}g^{\rho\lambda}\right\}^{-1}T_1^{\mu\nu}(\square-M_{T_1}^{2})T_1^{\rho\lambda}\notag\\
    &-\frac{1}{2}\left\{g^{\mu\lambda}g^{\nu\rho}+g^{\nu\lambda}g^{\mu\rho}-\frac{2}{3}g^{\mu\nu}g^{\rho\lambda}\right\}^{-1}T_2^{I\mu\nu}(\square-M_{T_2}^{2})T_2^{I\rho\lambda}\notag\\
    &-\frac{1}{2}\left\{g^{\mu\lambda}g^{\nu\rho}+g^{\nu\lambda}g^{\mu\rho}-\frac{2}{3}g^{\mu\nu}g^{\rho\lambda}\right\}^{-1}T_3^{a\mu\nu}(\square-M_{T_3}^{2})T_3^{a\rho\lambda}\notag\\
    &-\frac{1}{2}\left\{g^{\mu\lambda}g^{\nu\rho}+g^{\nu\lambda}g^{\mu\rho}-\frac{2}{3}g^{\mu\nu}g^{\rho\lambda}\right\}^{-1}T_4^{aI\mu\nu}(\square-M_{T_4}^{2})T_4^{aI\rho\lambda}\notag\\
    &-\frac{1}{2}\left\{g^{\mu\lambda}g^{\nu\rho}+g^{\nu\lambda}g^{\mu\rho}-\frac{2}{3}g^{\mu\nu}g^{\rho\lambda}\right\}^{-1}T_6^{ai\mu\nu}(\square-M_{T_6}^{2})T_6^{ai\rho\lambda}\notag\\
    &-\frac{1}{2}\left\{g^{\mu\lambda}g^{\nu\rho}+g^{\nu\lambda}g^{\mu\rho}-\frac{2}{3}g^{\mu\nu}g^{\rho\lambda}\right\}^{-1}T_7^{ai\mu\nu}(\square-M_{T_7}^{2})T_7^{ai\rho\lambda}\notag\\
    &+\bar R_{18\mu}^A\left\{(\slashed p-M_{R_{18}})g^{\mu\nu}-(\gamma^\mu p^\nu+p^\mu\gamma^\nu)+\gamma^\mu \slashed p\gamma^\nu+M_{R_{18}}\gamma^\mu\gamma^\nu\right\} R_{18\nu}^A\notag\\
    &+g_{T_1\ell^2}(\bar\ell\gamma_\mu\overleftrightarrow{\partial}_\nu\ell)T_1^{\mu\nu}+g_{T_2\ell^2}(\bar\ell\gamma_\mu\overleftrightarrow{\partial}_\nu\tau^{I}\ell)T_2^{I\mu\nu}\notag\\
    &+g_{T_1Q^2}(\bar Q\gamma_\mu\overleftrightarrow{\partial}_\nu Q)T_1^{\mu\nu}+g_{T_2 Q^2}(\bar Q\gamma_\mu\overleftrightarrow{\partial}_\nu\tau^{I}Q)T_2^{I\mu\nu}\notag\\
    &+g_{T_1u^2}(\bar u_R\gamma_\mu\overleftrightarrow{\partial}_\nu u_R)T_1^{\mu\nu}+g_{T_1 d^2}(\bar d_R\gamma_\mu\overleftrightarrow{\partial}_\nu\tau^{I}d_R)T_1^{\mu\nu}\notag\\
    &+g_{T_7\ell d}( \ell^TC\gamma_\mu\overleftrightarrow{\partial}_\nu d_{Ra})T_7^{a\mu\nu}+g_{T_6\ell u}( \ell^TC\gamma_\mu\overleftrightarrow{\partial}_\nu u_{Ra})T_6^{\dagger a\mu\nu}\notag\\
    &+g_{T_3\ell Q}(\bar \ell\gamma_\mu\overleftrightarrow{\partial}_\nu Q)T_3^{a\mu\nu}+g_{T_4\ell Q}(\bar \ell\gamma_\mu\overleftrightarrow{\partial}_\nu \tau^IQ)T_3^{aI\mu\nu}+g_{T_1G^2}G_{\mu\rho}^AG_{\nu}^{\rho A}T_1^{\mu\nu}\notag\\
    &+g_{R_{18}lG}\bar R_{18\mu}^A (g^{\mu\nu}+z\gamma^\mu\gamma^\nu)\gamma^\alpha\ell G^{A}_{\nu\alpha}\,,
\end{align}
where $g_{T_1\ell^2...}$ are the coupling constants, $z$ is a constant and the dependence on $z$ will
vanish for the on-shell amplitudes. Moreover, the propagators of the Rarita-Schwinger and Fierz-Pauli fields are
\begin{align}
    \Pi^{\mu\nu}_{\rm RS}=&(\slashed p+M)\bigg[g^{\mu\nu}-\frac{2}{3M^2}p^\mu p^\nu-\frac{1}{3}\gamma^\mu\gamma^\nu-\frac{1}{3M}(p^\nu\gamma^\mu-p^\mu\gamma^\nu)\bigg]\,,\\
    \Pi_{\rm FP}^{\mu\nu\rho\lambda}=&\frac{i}{2}\frac{(\mathcal{P}^{\mu\rho}\mathcal{P}^{\nu\lambda}+\mathcal{P}^{\mu\lambda}\mathcal{P}^{\nu\rho})-\frac{2}{3}\mathcal{P}^{\mu\nu}\mathcal{P}^{\rho\lambda}}{p^2-M^2}\,,
\end{align}
where $\mathcal{P}^{\mu\nu}=(g^{\mu\nu}-\frac{p^\mu p^\nu}{M^2})$. We take the operator $\mathcal{O}_{\ell^2u^2D^2}^{(2)}$ as the example. The Feynman diagrams are shown in Fig.~\ref{tensor}, and the effective Lagrangian can be evaluated as
\begin{align}
    \mathcal{L}_{\rm EFT}=\frac{g_{T_1\ell^2}g_{T_1u^2}}{2M^2_{T_1}}\left(g^{\mu\rho}g^{\nu\lambda}+g^{\mu\lambda}g^{\nu\rho}-\frac{2}{3}g^{\mu\nu}g^{\rho\lambda}\right)(\bar\ell\gamma_{\mu}\overleftrightarrow \partial_\nu\ell)(\bar u_R\gamma_{\rho}\overleftrightarrow \partial_\lambda u_R)\,.
\end{align}
Through the matching procedure, the Wilson coefficient of the operator   $\mathcal{O}_{\ell^2u^2D^2}^{(2)}$ is obtained 
\begin{align}
 \frac{C_{\ell^2u^2D^2}}{\Lambda^4}=\frac{g_{T_1\ell^2}g_{T_1u^2}}{M_{T_1}^2}\,.   
\end{align}

\begin{table}[h]
    \centering
    \begin{tabular}{c|c}
    \hline
    \hline
        $\mathcal{O}_{\ell^2G^2D}$ &  $F_{18}(\mathbf{8},\mathbf{2})_{-\frac{1}{2}}\,,R_{18}(\mathbf{8},\mathbf{2})_{\frac{1}{2}}\,,T_{1}(\mathbf{1},\mathbf{1})_0$\\
        $\mathcal{O}_{\ell^2q^2D^2}^{(2)}$ & $T_1(\mathbf{1},\mathbf{1})_0\,,T_2(\mathbf{1},\mathbf{3})_0\,,T_3(\Bar{\mathbf{3}},\mathbf{1})_{-\frac{2}{3}}\,,T_4(\Bar{\mathbf{3}},\mathbf{3})_{-\frac{2}{3}}$ \\
        $\mathcal{O}_{\ell^2q^2D^2}^{(4)}$& $T_1(\mathbf{1},\mathbf{1})_0\,,T_2(\mathbf{1},\mathbf{3})_0\,,T_3(\Bar{\mathbf{3}},\mathbf{1})_{-\frac{2}{3}}\,,T_4(\Bar{\mathbf{3}},\mathbf{3})_{-\frac{2}{3}}$ \\
        $\mathcal{O}_{\ell^2u^2D^2}^{(2)}$ &  $T_1(\mathbf{1},\mathbf{1})_0\,,T_6(\mathbf{3},\mathbf{2})_{\frac{1}{6}}$\\
        $\mathcal{O}_{\ell^2d^2D^2}^{(2)}$ &  $T_1(\mathbf{1},\mathbf{1})_0\,,T_7(\Bar{\mathbf{3}},\mathbf{2})_{\frac{5}{6}}$\\
        \hline
        \hline
    \end{tabular}
    \caption{The UV completions for dimension-8 SMEFT  operators that contribute to CE$\nu$NS.}
    \label{tab:UV8}
\end{table}

\section{Conclusion}
\label{sec:conclusion}
In this work, we established a comprehensive end-to-end EFT framework for CE$\nu$NS studies, covering the full energy scale hierarchy from NP scale to the nuclear scale. We accomplished complete EFT matching across relevant scales with QCD RG running effects included, developed systematic power counting to clarify the contributions of one-body and two-body currents, and provided full tree-level UV completions for SMEFT operators, bridging the NP with low-energy observations. Building on a broad set of CE$\nu$NS experimental data--including the latest observation from CONUS+, multi-target-nucleus results from COHERENT, and first indications of solar $^8$B neutrinos via CE$\nu$NS from ton-scale PandaX and XENON detectors--we derived stringent and comprehensive new constraints on the scales of LEFT operators up to dimension 8 and neutrino NSI parameters.

Building on previous systematic studies of CE$\nu$NS, this work presents several methodological innovations to further advance the field, with key contributions summarized as follows:
\begin{itemize}
\item The LEFT operators up to dimension 8 and their RG running effects are incorporated, with hadronic matching achieved via the systematic spurion method.
\item A systematic power counting analysis is performed, assessing contributions of all relevant LEFT operators up to dimension 8 and nuclear response functions, which are presented in Tabs.~\ref{tab:chiral-counting} and~\ref{tab:ResponseEnhance}.
\item The contributions of SMEFT operators up to dimension 8 at high scales are incorporated, with their matching conditions onto the LEFT operators provided in App.~\ref{app:MatchingSM2L} and the full tree-level UV completions given in Tabs.~\ref{tab:UV6},~\ref{tab:UV7}, and~\ref{tab:UV8}, thereby establishing a comprehensive top-down framework.
\end{itemize}

Looking ahead to future directions, this framework can be extended to incorporate an expanded suite of experimental data, including additional measurements from ton-scale xenon detectors and results from upcoming accelerator-based neutrino experiments; such expansions will further tighten constraints on NP scenarios. 
Moreover, the present numerical treatment is restricted to one-body currents.
Since the power counting we adopt is consistent with that employed in the construction of chiral nuclear forces, the framework can be systematically extended to include two-body currents in CE$\nu$NS. In parallel, the nuclear response functions used in this work are obtained from conventional nuclear many-body calculations, while {\it ab initio} nuclear-structure inputs could be incorporated within this framework.
Furthermore, synergizing with state-of-the-art lattice QCD calculations to refine the currently unknown LECs and extending the analysis to LNV scenarios will yield deeper insights into the fundamental nature of neutrino interactions and their broader implications for NP frameworks.

\section*{Acknowledgments}
This work is supported by the National Natural Science Foundation of China under Grants No.~12347105, No.~12375099, No.~12047503, No.~12447101 and No.~12505127, and the National Key Research and Development Program of China under Grants No.~2020YFC2201501, No.~2021YFA0718304.
GL is also supported by the Guangdong Basic and Applied Basic Research Foundation (2024A1515012668).

\appendix

\section{Naive Dimension Analysis}
\label{app:nda-matching}

The matching between the LEFT operators and the chiral Lagrangian is not a one-to-one procedure, which means that every LEFT operator can be matched to several $\chi$PT operators, and each $\chi$PT operator receives contributions from several LEFT operators (as will be clarified in the subsequent examples). Thus, we need a method to assess all these contributions systematically. In detail, we want to distinguish which one is more dominant among the LEFT operators matching to the same $\chi$PT operator, and which one is more dominant among the $\chi$PT operators matched from the same LEFT operator. In this section, we use the NDA~\cite{Manohar:1983md,Gavela:2016bzc,Jenkins:2013sda,Panico:2015jxa,Buchalla:2013eza} of the effective operators to organize them, an approach consistent with that adopted in Ref.~\cite{Song:2025snz}.

The LEFT Lagrangian is organized as an expansion in powers of the electroweak scale inverse $1/\Lambda_{\text{EW}}$. The NDA master formula~\cite{Manohar:1983md,Gavela:2016bzc} in the 4-dimensional spacetime states that the operators in the Lagrangian are normalized as
\begin{equation}
\label{eq:NDA_1}
\frac{\Lambda_{\text{EW}}^4}{16\pi^2} {\left[\frac{\partial}{\Lambda_{\text{EW}}}\right]}^{N_p}{\left[\frac{4\pi\phi}{\Lambda_{\text{EW}}}\right]}^{N_\phi}{\left[\frac{4\pi A}{\Lambda_{\text{EW}}}\right]}^{N_A}{\left[\frac{4\pi\psi}{\Lambda_{\text{EW}}^{3/2}}\right]}^{N_\psi}{\left[\frac{g}{4\pi}\right]}^{N_g}{\left[\frac{y}{4\pi}\right]}^{N_y}{\left[\frac{\lambda}{16\pi^2}\right]}^{N_\lambda}\,.
\end{equation}
For convenience, we express the master formula by replacing the vector field $A$ with the field-strength tensor $F\sim \partial A$ and omitting the renormalizable coupling constants $g,y,\lambda$,
\begin{equation}
\text{LEFT:\quad }
    \frac{\Lambda_{\text{EW}}^4}{16\pi^2}\left[\frac{\partial}{\Lambda_{\text{EW}}}\right]^{N_p} \left[\frac{4\pi F}{\Lambda_{\text{EW}}^2}\right]^{N_F} \left[\frac{4\pi \psi}{\Lambda_{\text{EW}}^{3/2}}\right]^{N_\psi}\,,
\end{equation}
which implies that the 4-fermion operators of dimensions 6, 7, and 8 are normalized as
\begin{equation}
    \frac{{(4\pi)}^2}{\Lambda_{\text{EW}}^2}\psi^4\,,\quad\frac{{(4\pi)}^2}{\Lambda_{\text{EW}}^2}\frac{\partial}{\Lambda_{\text{EW}}}\psi^4\,,\quad\frac{{(4\pi)}^2}{\Lambda_{\text{EW}}^2}\frac{\partial^2}{\Lambda_{\text{EW}}^2}\psi^4\,.
\end{equation}
\iffalse
In addition, NDA is also suitable for covariant derivatives while
\begin{equation}
    \frac{1}{\Lambda_{\text{EW}}}D_\mu=\frac{\partial_\mu}{\Lambda_{\text{EW}}}+i\frac{g}{4\pi}\frac{4\pi A_\mu}{\Lambda_{\text{EW}}}\,.
\end{equation}
\fi
In particular, the spurions are dimensionless in the NDA.
On the other hand, the NDA master formula for the $\chi$PT is 
\begin{equation}
    \text{$\chi$PT:\quad }f^2\Lambda_\chi^2 \left[\frac{\partial}{\Lambda_\chi}\right]^{N_p} \left[\frac{\psi}{f\sqrt{\Lambda_\chi}}\right]^{N_\psi} \left[\frac{F}{\Lambda_\chi f}\right]^{N_A} \,,  \label{eq:nda_chiral}
\end{equation}
where $4\pi f\sim \Lambda_\chi$. To relate these two NDA formulae we first replace all occurrences of the scale $\Lambda_{\text{EW}}$ in Eq.~\eqref{eq:NDA_1} by the scale $\Lambda_\chi$,
\begin{align}
    & \frac{\Lambda_{\text{EW}}^4}{16\pi^2} \left[\frac{\Lambda_\chi}{\Lambda_{\text{EW}}}\right]^{N_p+2N_F + \frac{3}{2}N_\psi} \left[\frac{\partial}{\Lambda_\chi}\right]^{N_p} \left[\frac{F}{\Lambda_\chi^2}\right]^{N_F} \left[\frac{\psi}{\sqrt{\Lambda_\chi}f}\right]^{N_\psi} \notag \\
    =& \left[\frac{\Lambda_\chi}{\Lambda_{\text{EW}}}\right]^{\mathcal{D}} \left(f^2\Lambda_\chi^2 \left[\frac{\partial}{\Lambda_\chi}\right]^{N_p} \left[\frac{F}{\Lambda_\chi^2}\right]^{N_F} \left[\frac{\psi}{\sqrt{\Lambda_\chi}f}\right]^{N_\psi}\right)\,,
\end{align}
where $\mathcal{D}=N_p+2N_F + \frac{3}{2}N_\psi-4$. The expression inside the parentheses is similar to the NDA formula in the $\chi$PT, so we replace it with the $\chi$PT NDA formula in Eq.~\eqref{eq:nda_chiral} and obtain the NDA formula for the matching:
\begin{align}
    \text{matching:\quad }& \frac{\Lambda_{\text{EW}}^4}{16\pi^2}\left[\frac{\partial}{\Lambda_{\text{EW}}}\right]^{N_p} \left[\frac{4\pi F}{\Lambda_{\text{EW}}^2}\right]^{N_F} \left[\frac{4\pi \psi}{\Lambda_{\text{EW}}^{3/2}}\right]^{N_\psi} \notag \\
    \sim & \left[\frac{\Lambda_\chi}{\Lambda_{\text{EW}}}\right]^{\mathcal{D}} \left(f^2\Lambda_\chi^2 \left[\frac{\partial}{\Lambda_\chi}\right]^{N_p} \left[\frac{\psi}{f\sqrt{\Lambda_\chi}}\right]^{N_\psi} \left[\frac{F}{\Lambda_\chi f}\right]^{N_A} \right)\,. \label{eq:NDA_2}
\end{align}

\section{Values of Low-Energy Constants}
\label{app:LECValues}

The numerical values of the LECs $L_{1\sim 14}$ are calculated in this appendix. The LECs $L_1$ and $L_2$ are fixed by the proton and neutron magnetic moments~\cite{ParticleDataGroup:2022pth},
\begin{align}
    2L_1+2L_2
    &=2(\mu_p-1)+\mu_n
      =1.676(29)\,,\\
    2L_2
    &=2\mu_n+(\mu_p-1)
      =-2.034(29)\,,
\end{align}
and the two LECs are
\begin{align}
    L_1=1.855(14)\,,\qquad
    L_2=-1.017(15)\,.
\end{align}
The LECs $L_3$ and $L_4$ can be obtained through the axial vector currents constants
\begin{align}
    2L_3=g_A=1.2754(13)\,,\quad2L_3+4L_4=\Sigma_{ud}=0.397(40)\,,
\end{align}
where $g_A$ is determined very precisely from nuclear $\beta$ decay~\cite{ParticleDataGroup:2022pth} and the values for $\Sigma_{ud}$ follow from averages of the lattice QCD results~\cite{Liang:2018pis,Lin:2018obj}. Thus, the LECs $L_3$ and $L_4$ become
\begin{align}
   L_3=0.6377(7)\,,\quad L_4=-0.220(10)\,. 
\end{align}
The LECs $L_5$, $L_6$, $L_7$ and $L_8$ can be related to the generalized tensor form factors~\cite{Haxton:2024lyc,Gupta:2018lvp}, and the relationship can be obtained as follows
\begin{align}
    2(L_5+L_6)=&m_u\hat{F}_{T,0}^{u/p}=1.79(30){~\rm MeV}\,,\\
    2L_6=&m_d\hat{F}_{T,0}^{d/p}=-0.97(15){~\rm MeV}\,,\\  2(L_5+L_6+L_7+L_8)=&m_u\hat{F}_{T,1}^{u/p}=-6.44(54){~\rm MeV}\,,\\
2(L_6+L_8)=&m_d\hat{F}_{T,1}^{d/p}=4.32(32){~\rm MeV}\,.
\end{align}
The values of LECs $L_5$, $L_6$, $L_7$ and $L_8$ are
\begin{align}
    L_5=1.38(17){~\rm MeV}\,,\, L_6=-0.49(8){~\rm MeV}\,,\nn\\
    L_7=-6.76(37){~\rm MeV}\,,\,L_8=2.65(18){~\rm MeV}\,.
\end{align}
The LECs $L_9$ and $L_{10}$ are related to the LECs $c_1$ and $c_5$ in Ref.~\cite{Fettes:2000gb}. The following relations can be obtained~\cite{Fettes:2000xg,Hoferichter:2023ptl}
\begin{align}
    L_{10}=&B_0c_1(m_d+m_u)=24.05(48){~\rm MeV}\,,\\
    L_{9}-\frac{1}{2}L_{10}=&B_0c_5(m_d-m_u)=-0.51(8){~\rm MeV}\,.
\end{align}
Thus, the numerical values of the LECs become
\begin{align}
    L_9=11.52(25){~\rm MeV}\,,\quad L_{10}=24.05(48){~\rm MeV}\,.
\end{align}
In addition, the relations between sigma terms in Eq.~\eqref{eq:sigmaterm} and LECs are~\cite{Hoferichter:2023ptl}
\begin{align}
    \sigma^{p/n}_u=\frac{1}{2}\bigg[-4 c_1 M_\pi^2 - \frac{9 g_A^2 M_\pi^3}{64 \pi f^2}+\left( 4 c_1 + \frac{21 g_A^2 M_\pi}{128 \pi f^2} \right) \Delta_\pi\bigg](1-\xi) &\pm B c_5 (m_d - m_u)\left(1 - \frac{1}{\xi}\right), \\
\sigma^{p/n}_d=\frac{1}{2}\bigg[-4 c_1 M_\pi^2 - \frac{9 g_A^2 M_\pi^3}{64 \pi f^2}+\left( 4 c_1 + \frac{21 g_A^2 M_\pi}{128 \pi f^2} \right) \Delta_\pi\bigg](1+\xi) &\pm B c_5 (m_d - m_u)\left(1 + \frac{1}{\xi}\right),
\end{align}
where $\Delta_\pi=M_{\pi}^2-M_{\pi^0}^2$ and $\xi=(m_d-m_u)/(m_d+m_u)=0.357(6)$~\cite{ParticleDataGroup:2022pth}.

Similarly, the LECs $L_{11}$ and $L_{12}$ are related to the LECs $d_{18}$ and $d_{19}$ in Ref.~\cite{Fettes:2000gb}
\begin{align}
    L_{11}&={m_N}B_0d_{18}(m_d+m_u)=-17.7(20){~\rm MeV}\,,\\
    L_{12}&={m_N}B_0d_{19}(m_d+m_u)=-13.9(15){~\rm MeV}\,.
\end{align}
In the above, we have adopted the values $d_{18} = -0.97 \pm 0.11 \text{ GeV}^{-2}$~\cite{Fettes:2000xg} and $d_{19} = -0.76 \pm 0.08 \text{ GeV}^{-2}$, the latter of which is fixed by the axial radius $\langle r_A^2 \rangle = 0.369(30) \text{ fm}^2$~\cite{Meyer:2026kdl}.

Moreover, the LECs of gluonic current interaction have been obtained~\cite{Haxton:2024lyc,Bishara:2017pfq} as follows
\begin{align}
    L_{13}=F^N_G=-50.4(6){~\rm MeV}\,,\quad L_{14}=F^N_{\tilde{G}}=-0.306(28) {~\rm GeV}\,.
\end{align}

\section{Expansion of Chiral Building Blocks}
\label{app:chiral-expansion}

Using $u=e^{X}$ in Eq.~\eqref{eq:CCWZ}, with $X\equiv \tfrac{i}{2f}\Pi$ and $\Pi$ defined in Eq.~\eqref{eq:NGBPi}, one has $X^\dagger=-X$ and thus $u^\dagger=e^{-X}$. The chiral building blocks introduced in Eq.~\eqref{eq:buildblock} take the form
\begin{align}
  u_\mu &=
  i\left(e^{-X}\partial_\mu e^{X}-e^{X}\partial_\mu e^{-X}\right)\,, \nonumber\\
  \chi_\pm &= e^{-X}\chi e^{-X}\pm e^{X}\chi^{\dagger} e^{X}\,, \nonumber\\
  \Sigma_{\pm} &=
  e^{-X}\tau e^{-X}\pm e^{X}\tau e^{X}\,, \nonumber\\
  Q_{\pm} &=
  e^{-X}\tau e^{X}\pm e^{X}\tau e^{-X}\,, \nonumber\\
  \nabla_\mu &=
  \partial_\mu+\frac{1}{2}\left(e^{-X}\partial_\mu e^{X}+e^{X}\partial_\mu e^{-X}\right)\,.
\end{align}
Here $\tau$ denotes the spurion associated with the flavor matrices appearing in the LEFT operators. According to the Baker-Campbell-Hausdorff formula,
\begin{align}
  e^{X}A\,e^{-X}=\sum_{n=0}^\infty\frac{1}{n!}\,[A]^{(n)}_{X},
\end{align}
and
\begin{align}
  e^{-X}A\,e^{X}
  =\sum_{n=0}^\infty\frac{1}{n!}\,[A]^{(n)}_{-X}
  =\sum_{n=0}^\infty\frac{(-1)^n}{n!}\,[A]^{(n)}_{X},
\end{align}
where
\begin{align}
  [A]^{(n)}_{X} \equiv \underbrace{[X,[X,\dots,[X}_{n},\,A]\dots]] \, .
\end{align}
Therefore $Q_{\pm}$ admit the exact expansions
\begin{align}
  Q_{+} &= \sum_{n=0}^{\infty} \frac{2}{(2n)!}\,[\tau]^{(2n)}_{X}, \nonumber\\
  Q_{-} &= -\sum_{n=0}^{\infty} \frac{2}{(2n+1)!}\,[\tau]^{(2n+1)}_{X}.
\end{align}

For the derivative term, one uses
\begin{align}
  e^{-X}\,\partial_\mu e^{X}
  =\sum_{n=0}^\infty\frac{(-1)^n}{(n+1)!}\,[\partial_{\mu}X]_{X}^{(n)},
  \qquad
  [\partial_\mu X]^{(n)}_X \equiv \underbrace{[X,[X,\dots,[X}_{n},\,\partial_\mu X]\dots]] ,
\end{align}
and
\begin{align}
  e^{X}\,\partial_\mu e^{-X}
  = -\sum_{n=0}^\infty\frac{1}{(n+1)!}\,[\partial_{\mu}X]_{X}^{(n)}.
\end{align}
It follows that
\begin{align}
  u_{\mu} &= i\sum_{n=0}^\infty\frac{2}{(2n+1)!}\,[\partial_{\mu}X]_{X}^{(2n)}, \nonumber\\
  \nabla_{\mu} &= \partial_{\mu}
  -\sum_{n=0}^\infty\frac{1}{(2n+2)!}\,[\partial_{\mu}X]_{X}^{(2n+1)}.
\end{align}

For $e^{\pm X}A\,e^{\pm X}$, the Taylor expansion gives
\begin{align}
  e^{X}A\,e^{X}
  &=\sum_{n,m\ge 0}\frac{1}{n!\,m!}\,X^{n}AX^{m}
   =\sum_{k=0}^{\infty}\frac{1}{k!}\,\{A\}^{(k)}_{X}, \nonumber\\
  e^{-X}A\,e^{-X}
  &=\sum_{n,m\ge 0}\frac{(-1)^{n+m}}{n!\,m!}\,X^{n}AX^{m}
   =\sum_{k=0}^{\infty}\frac{(-1)^{k}}{k!}\,\{A\}^{(k)}_{X},
\end{align}
where
\begin{align}
  \{A\}^{(k)}_{X} \equiv \underbrace{\{X,\{X,\dots\{X}_{k},\,A\}\}\dots\}.
\end{align}
Hence,
\begin{align}
  \Sigma_{+} &= \sum_{k=0}^{\infty} \frac{2}{(2k)!}\,\{\tau\}_{X}^{(2k)}, \nonumber\\
  \Sigma_{-} &= -\sum_{k=0}^{\infty} \frac{2}{(2k+1)!}\,\{\tau\}_{X}^{(2k+1)}.
\end{align}

The expansions for $\chi_\pm$ are completely analogous. Retaining terms with at most three pion fields yields
\begin{align}
  u_\mu&=2i\,\partial_\mu X+ \frac{i}{3} [X,[X,\partial_{\mu}X]]+\cdots, \label{eq:app_umu_X}\\
  \Sigma_+&=2\tau+\{X,\{X,\tau\}\}+\cdots,\\
  \Sigma_-&= -2\{X,\tau\}- \frac{1}{3}\{X,\{X,\{X,\tau\}\}\}+\cdots,\\
  Q_+&=2\tau+[X,[X,\tau]]+\cdots,\\
  Q_-&=-2[X,\tau]-\frac{1}{3}[X,[X,[X,\tau]]]+\cdots,\\
  \chi_+&=2\chi+\{X,\{X,\chi\}\}+\cdots,\\
  \chi_-&= -2\{X,\chi\}- \frac{1}{3}\{X,\{X,\{X,\chi\}\}\}+\cdots,\\
  \nabla_\mu&=\partial_\mu -\frac{1}{2}[X, \partial_\mu X]+\cdots.
\end{align}
At leading chiral order, these expansions imply that $u_\mu$, $\Sigma_{-}$, $Q_{-}$, and $\chi_{-}$ start at $\mathcal{O}(\pi)$ and therefore contain an odd number of pion fields (i.e.\ $1,3,\ldots$). In contrast, $\Sigma_{+}$, $Q_{+}$, $\chi_{+}$, and $\nabla_\mu$ contain an even number of pion fields (i.e.\ $0,2,\ldots$).

\section{Nuclear Response}
\label{app:NuclearResponse}
In the eigenstate of nuclear angular momentum, we can expand the nuclear matrix element in Eq.~\eqref{eq:FourierMatrixElements} using a multipole expansion ~\cite{Serot:1978vj,Haxton:2008zza}. We define the operators
\begin{align}
M_{JM
}(|\vec{q}| \vec{x}) & \equiv j_J(|\vec{q}| x) Y_{JM}(\Omega_x) \,, \\
\vec{M}_{JL}^M(|\vec{q}| \vec{x}) & \equiv j_L(|\vec{q}| x) \vec{Y}_{JLM} \,,
\end{align}
where $j_{J}$ is the spherical Bessel function, $Y_{JM}$ is the spherical harmonic, and $\vec{Y}_{JLM}$ is the vector spherical harmonic.

To arrive at the nuclear response basis, we express the one-body operators in coordinate space and
Fourier transform them to momentum space. Expanding the plane wave (and its vector counterpart) in
scalar and vector spherical harmonics projects the charge and current onto definite multipoles, and
the nucleon momentum in Eq.~\eqref{eq:LDEFT} is replaced by gradients acting on the nuclear wave function,
$\vec{p}\to -i\vec{\nabla}$. For elastic ground-state matrix elements, parity and time-reversal
selection rules allow the multipole operators to be organized into six independent response
operators $\{M_{JM},\,\Delta_{JM},\,\Sigma'_{JM},\,\Sigma''_{JM},\,\tilde{\Phi}'_{JM},\,\Phi''_{JM}\}$,
where $M_{JM}$ is defined above and the remaining five are given below:
\begin{align}
\label{eq:NuclearRsponse}
\Delta_{JM}(|\vec{q}| \vec{x}) &\equiv \vec{M}_{JJ}^M(|\vec{q}| \vec{x}) \cdot {1 \over |\vec{q}|} \vec{\nabla}  \,,
\\
\Sigma^\prime_{JM}(|\vec{q}| \vec{x}) &\equiv -i \left\{ {1 \over |\vec{q}|} \vec{\nabla} \times \vec{M}_{JJ}^M (|\vec{q}| \vec{x}) \right\} \cdot \vec{\sigma} \,,
\\
\Sigma^{\prime \prime}_{JM}(|\vec{q}| \vec{x}) &\equiv \left\{ {1 \over |\vec{q}|} \vec{\nabla}  M_{JM} (|\vec{q}| \vec{x}) \right\} \cdot \vec{\sigma} \,,
\\
\tilde{\Phi}^{\prime}_{JM}(|\vec{q}| \vec{x}) &\equiv \left( {1 \over |\vec{q}|} \vec{\nabla} \times \vec{M}_{JJ}^M(|\vec{q}| \vec{x}) \right) \cdot \left(\vec{\sigma} \times {1 \over |\vec{q}|} \vec{\nabla} \right) + {1 \over 2} \vec{M}_{JJ}^M(|\vec{q}| \vec{x}) \cdot \vec{\sigma} \,,
\\
\Phi^{\prime \prime}_{JM}(|\vec{q}| \vec{x}) &\equiv i  \left( {1 \over |\vec{q}|} \vec{\nabla}  M_{JM}(|\vec{q}| \vec{x}) \right) \cdot \left(\vec{\sigma} \times {1 \over |\vec{q}|} \vec{\nabla} \right) \,,
\end{align}
which are the response operators in Eq.\eqref{eq:SixNuclearResonses}.

The many-body nuclear matrix element can be written as a sum over one-body density matrices. Using the nuclear angular momentum ($j_N$) and isospin ($T$, $M_T$) to label the nuclear state, we have
\begin{equation}
\langle j_N; T M_T ||\sum_{i=1}^A O_{J;\tau;i} || j_N; T M_T \rangle = (-1)^{T-M_T} 
\left( \begin{array}{ccc} T & \tau & T \\ -M_T & 0 & M_T \end{array} \right)  
\sum_{|\alpha|,|\beta|,i} \Psi_{|\alpha|,|\beta|} ^{J;\tau} ~\langle |\alpha|~ \vdots \vdots O_{J;\tau;i} \vdots \vdots~ |\beta| \rangle,
\end{equation}
where $|\alpha|$ and $|\beta|$ denote the quantum numbers of the single-particle basis,
$\Psi$ is the one-body density matrix for the transition between the ground states $|\alpha|$ and $|\beta|$,
and $\vdots \vdots$ indicates the reduced matrix element in spin and isospin space.
The one-body density matrix $\Psi$ can be obtained from a nuclear structure model, such as the shell model ~\cite{Haxton:2008zza}.

For the single-particle matrix elements $\langle |\alpha|~ \vdots \vdots O_{J;\tau;i} \vdots \vdots~ |\beta| \rangle$,
the harmonic oscillator model provides a convenient framework for calculation. With the harmonic oscillator
as the single-particle basis, these matrix elements can be related to the response operators
$O_{J} \rightarrow \{ M_{J},\, \Delta_{J},\, \Sigma_{J}',\, \Sigma_{J}'',\, \tilde{\Phi}_{J}',\, \Phi_{J}''\}$,
as presented in Eq.~\eqref{eq:AmplitudeSquare2}
\begin{align}
\label{eq:nucresponse}
W_{MM}^{\tau \tau^\prime}(y)&= \sum_{J=0,2,...}^\infty    \langle j_N ||~ M_{J;\tau} ~ || j_N \rangle 
\langle j_N ||~ M_{J;\tau^\prime} ~ || j_N \rangle \,, \notag\\
W_{\Sigma^{\prime \prime}\Sigma^{\prime \prime}}^{\tau \tau^\prime}(y) &=
\sum_{J=1,3,...}^\infty \langle j_N ||~ \Sigma^{\prime \prime}_{J; \tau}~ || j_N \rangle
 \langle j_N ||~ \Sigma^{\prime \prime}_{J; \tau^\prime}~ || j_N \rangle \,, \notag\\
W_{\Sigma^\prime \Sigma^\prime}^{\tau \tau^\prime}(y)  &=\sum_{J=1,3,...}^\infty  \langle j_N ||~  \Sigma_{J;\tau}^\prime ~ || j_N \rangle  \langle j_N || ~\Sigma_{J;\tau^\prime}^{\prime} ~ || j_N \rangle \,, \notag\\ 
W_{\Phi^{\prime \prime}\Phi^{\prime \prime}}^{\tau \tau^\prime}(y) &= \sum_{J=0,2,...}^\infty     ~\langle j_N ||~\Phi^{\prime \prime}_{J; \tau}~ || j_N \rangle  \langle j_N ||~\Phi^{\prime \prime}_{J; \tau^\prime}~ || j_N \rangle \,, \notag\\
W_{ \Phi^{\prime \prime}M}^{\tau \tau^\prime}(y)  &=  \sum_{J=0,2,...}^\infty  \langle j_N ||~ \Phi^{\prime \prime }_{J;\tau}~ || j_N \rangle \langle j_N ||~ M_{J;\tau^\prime}  ~|| j_N \rangle \,, \notag\\
W_{\tilde{\Phi}^\prime\tilde{\Phi}^\prime}^{\tau \tau^\prime}(y)  &=  \sum_{J=2,4,...}^\infty ~  \langle j_N || ~\tilde{\Phi}_{J; \tau}^\prime ~ || j_N \rangle  \langle j_N ||  ~ \tilde{\Phi}^{\prime}_{J; \tau^\prime}~ || j_N \rangle \,, \notag\\
W_{\Delta \Delta}^{\tau \tau^\prime}(y)  &=\sum_{J=1,3,...}^\infty  \langle j_N ||~ \Delta_{J;\tau} ~ || j_N \rangle \langle j_N ||~ \Delta_{J; \tau^\prime}~ || j_N \rangle \,, \notag\\
W_{\Delta \Sigma^\prime}^{\tau \tau^\prime}(y)  &=\sum_{J=1,3,...}^\infty 
\langle j_N || ~ \Delta_{J;\tau}  ~|| j_N \rangle  \langle j_N || ~\Sigma^{\prime}_{J; \tau^\prime}~ || j_N \rangle \,.    
\end{align}

In the harmonic oscillator basis, the reduced matrix element $\langle j_N || ~ O_{J} ~|| j_N \rangle$
can be expressed as a polynomial in the variable $y$
\begin{equation}
    \langle j_N || ~ O_{J} ~|| j_N \rangle \propto y^{(J - K)/2} e^{-y} p(y) \,,
\end{equation}
where the operators $(\Delta, \Sigma', \Sigma'')$ have abnormal parity ($K=1$), while
$(M, \tilde{\Phi}', \Phi'')$ have normal parity ($K=2$). Consequently, the nuclear responses $W(y)$
can be expressed as functions of $y$.

\section{Operator Matching: From SMEFT to LEFT}
\label{app:MatchingSM2L}
In this appendix, we give the matching between SMEFT and LEFT. We consider only the dimension-6 SMEFT operators contribution for the Higgs and electroweak gauge boson sector which was obtained in Ref.~\cite{Jenkins:2017jig}. The Higgs field can be written in unitary gauge as
\begin{align}
	\label{eq:Hvev}
	H &= \frac{1}{\sqrt 2}
		\left(\begin{array}{c}
			0 \\
			\left[ 1+ c_{H\,,\rm kin} \right]  h + v_T
		\end{array}\right) \,,
\end{align}
where
\begin{align}
	\label{eq:chkindef}
	c_{H\,,\rm kin} &\equiv \left(\hat C_{H\Box}-\frac14 \hat C_{HD}\right)v^2 \,, \qquad
	v_T \equiv \left( 1+ \frac{3 \hat C_H v^2}{8 \lambda} \right) v \,,
\end{align}
where $c_{H\,,\rm kin}$ parametrizes the Higgs kinetic term and $v_T$ is the vacuum expectation value in SMEFT. The gauge field and coupling redefinitions for $SU(2)_L$ and $U(1)_Y$ read
\begin{align}
    \bar g_2&=g_2 \left(1 + \hat C_{HW}\, v_T^2 \right)\,,&\quad\bar g_1=& g_1 \left(1 + \hat C_{HB} \, v_T^2 \right)\,,\\
    W^I_\mu  &=  \mathcal{W}^I_\mu \left(1 + \hat C_{HW} v_T^2 \right), &
B_\mu  =&  \mathcal{B}_\mu \left(1 + \hat C_{HB} v_T^2 \right)\,.
\end{align}
Thus, the weak mixing of the two neutral gauge bosons and the weak mixing angle $\overline{\theta}$ can be obtained
\begin{align}
\begin{pmatrix}
{\cal Z}^\mu \\ {\cal A}^\mu \\
\end{pmatrix} &= 
\begin{pmatrix}
\bar c- {\epsilon  \over 2} \bar s &\qquad -\bar s +  {\epsilon  \over 2} \bar c \\
\bar s+ {\epsilon  \over 2} \bar c &\qquad \bar c +  {\epsilon  \over 2} \bar s \\
\end{pmatrix}
\begin{pmatrix}
{\cal W}_3^\mu \\ {\cal B}^\mu \\
\end{pmatrix}, \quad \epsilon \equiv \hat C_{HWB} v_T^2 \,,
\end{align}
\begin{align}
\cos \overline{\theta} \equiv \bar c =& {{\bar g_2} \over \sqrt{ \bar g_1^2 + \bar g_2^2} }\left[ 1 - {\epsilon \over 2} \ {{\bar g_1} \over {\bar g_2}} \left( {{\bar g_2^2 - \bar g_1^2} \over {\bar g_1^2 + \bar g_2^2}} \right)\right]\,,\\
\sin \overline{\theta} \equiv \bar s =& {{\bar g_1} \over \sqrt{ \bar g_1^2 + \bar g_2^2} }\left[ 1 + {\epsilon \over 2} \ {{\bar g_2} \over {\bar g_1}} \left( {{\bar g_2^2 - \bar g_1^2} \over {\bar g_1^2 + \bar g_2^2}} \right)   \right]\,.
\end{align}
After the symmetry breaking, the mass of the neutral massive gauge boson ${\cal Z}^\mu$ and the eﬀective couplings to the ${\cal Z}^\mu$ boson  become 
\begin{align}
M_{\cal Z}^2 &= \frac 14 \left( \bar g_2^2 + \bar g_1^2 \right) v_T^2 \left( 1 + \frac 12 \hat C_{HD} v_T^2 \right)
+ {\epsilon \over 2} \bar g_1 \bar g_2 v_T^2 \,,\\
\bar g_Z &= \frac{\bar g_2 \,\bar s  - \frac{1}{2} \, \bar c\,\bar g_2 \, v_T^2 \, \hat C_{HWB}}{\bar c \,\bar s}  \left[1 +  \frac{\bar g_1^2+\bar g_2^2}{2 \bar g_1 \bar g_2} v_T^2\hat C_{HWB}\right]\,.
\end{align}
In addition, the fermion couplings to the massive gauge bosons ${\cal Z}^\mu$ are

\begin{align}
[Z_{\nu_L}]_{pr} &= \left[\delta_{pr}\left(\frac 12\right) - \frac12 v_T^2  \hat C^{(1)}_{\substack {Hl  pr}} + \frac12 v_T^2  \hat C^{(3)}_{\substack {Hl   pr}} \right]\,,\\
[Z_{u_L}]_{pr} &=  \left[\delta_{pr}\left(\frac 12-\frac 23 \bar s^2 \right) - \frac12 v_T^2  \hat C^{(1)}_{\substack {Hq   pr}} + \frac12 v_T^2  \hat C^{(3)}_{\substack {Hq   pr}} \right]\,,  \\
[Z_{u_R}]_{pr} &=  \left[\delta_{pr}\left(-\frac 23 \bar s^2 \right) - \frac12 v_T^2  \hat C_{\substack {Hu   pr}}  \right]\,,\\
[Z_{d_L}]_{pr} &=  \left[\delta_{pr}\left(-\frac 12+ \frac 13 \bar s^2 \right) - \frac12 v_T^2  \hat C^{(1)}_{\substack {Hq   pr}} - \frac12 v_T^2  \hat C^{(3)}_{\substack {Hq  pr}}  \right]\,, \\
[Z_{d_R}]_{pr} &=  \left[\delta_{pr}\left(+\frac13\bar s^2 \right) - \frac12 v_T^2  \hat C_{\substack {Hd   pr}}  \right]\,.
\end{align}

There are direct and indirect contributions in the matching condition. The matching conditions for dimension-5, 6 and 7 LEFT operators have been obtained in Ref.~\cite{Jenkins:2017jig,Liao:2020zyx,Hamoudou:2022tdn}. Here, we give the matching conditions for relevant LEFT operators up to dimension 8. The relevant matching results for dimension-5, 6 and 7 LEFT operators are
\begin{align}
\begin{split}
  \hat \C_1^{5}= \frac{ev_T^2}{2}\left(\hat C_{LHB}+\hat C_{LHW}\right)\,,
\end{split}
\\
\begin{split}
  \hat \C_1^{6u}= \hat C^{(1)\alpha\beta}_{lq } + \hat C^{(3)\alpha\beta}_{ lq } +\hat C_{ lu}^{\alpha\beta}
-\frac{\bar g_Z^2}{ M_Z^2}  \left( \left[Z_\nu \right]^{\alpha\beta} \left[Z_{u_L} \right]  
+   \left[Z_\nu \right]^{\alpha\beta} \left[Z_{u_R} \right]\right)\,,
\end{split}
\\
\begin{split}
  \C_1^{6d}=\hat C^{(1)\alpha\beta}_{lq} -  \hat C^{(3)\alpha\beta}_{lq}
+\hat C_{ld}^{\alpha\beta}-\frac{\bar g_Z^2}{ M_Z^2}\left( 
 \left[Z_\nu \right]^{\alpha\beta} \left[Z_{d_L} \right]   
+   \left[Z_\nu \right]^{\alpha\beta} \left[Z_{d_R} \right]\right)\,,
\end{split}
\\
\begin{split}
  \hat \C_2^{6u}= \hat C_{ lu}^{\alpha\beta}-\hat C^{(1)\alpha\beta}_{lq } - \hat C^{(3)\alpha\beta}_{ lq } 
-\frac{\bar g_Z^2}{ M_Z^2}  \left( \left[Z_\nu \right]^{\alpha\beta} \left[Z_{u_R} \right]-\left[Z_\nu \right]^{\alpha\beta} \left[Z_{u_L} \right]  
\right)\,,
\end{split}
\\
\begin{split}
  \hat \C_2^{6d}=\hat C_{ld}^{\alpha\beta}-\hat C^{(1)\alpha\beta}_{lq} +  \hat C^{(3)\alpha\beta}_{lq}
-\frac{\bar g_Z^2}{ M_Z^2}\left( 
 \left[Z_\nu \right]^{\alpha\beta} \left[Z_{d_R} \right]-\left[Z_\nu \right]^{\alpha\beta} \left[Z_{d_L} \right]   
\right)\,,
\end{split}
\\
\begin{split}
  \hat \C_1^{7u}=\frac{\sqrt{2}}{M_{\cal Z}}\delta^{\alpha\beta}\left(\bar c \hat C_{uW}-\bar s \hat C_{uB}\right)\,,\quad \hat \C_1^{7d}=-\frac{\sqrt{2}}{M_{\cal Z}}\delta^{\alpha\beta}\left(\bar c \hat C_{dW}+\bar s \hat C_{dB}\right)\,,
\end{split}
\\
\begin{split}
  \hat \C_2^{7u}=\frac{\sqrt{2}}{M_{\cal Z}}\delta^{\alpha\beta}\left(\bar s \hat C_{uB}-\bar c \hat C_{uW}\right)\,,\quad \hat \C_2^{7d}=\frac{\sqrt{2}}{M_{\cal Z}}\delta^{\alpha\beta}\left(\bar c \hat C_{dW}+\bar s \hat C_{dB}\right)\,,
\end{split}
\\
\begin{split}
  \hat \C_4^{7u}= \frac{\sqrt{2}v_T}{m_u} \hat C_{QuLLH}^{\alpha\beta}\,,\quad\quad\quad\quad\quad \hat \C_4^{7d}= \frac{\sqrt{2}v_T}{2 m_d}\hat C_{dLQLH1}^{\alpha\beta}\,,
\end{split}
\\
\begin{split}
  \hat \C_5^{7u}=-\frac{\sqrt{2}v_T}{2 m_u}\hat C_{QuLLH}^{\alpha\beta}\,,\quad\quad\quad\quad\,\hat \C_5^{7d}=-\frac{\sqrt{2}v_T}{2 m_d}\hat C_{dLQLH1}^{\alpha\beta}\,,
\end{split}
\\
\begin{split}
  \hat \C_3^{7d}=\frac{\sqrt{2}v_T}{2 m_d}\hat C_{dLQLH2}^{\alpha\beta}\,.
\end{split}
\end{align}
It should be emphasized that the operators $\mathcal{O}_{LHB}$ and $\mathcal{O}_{LHW}$~\cite{Lehman:2014jma,Liao:2016hru} are not presented in this paper. Because only three types of the seesaw can contribute to the two operators at tree level which are constrained by Weinberg operators and neglected in this work. In addition, the SMEFT operators up to dimension 8 can not contribute to $\mathcal{O}_3^{7u}$ at tree level.

Moreover, the matching conditions for LEFT dimension-8 operators become
\begin{align}
\hat C_1^{8u} &= \hat C_{l^2q^2D^2}^{(2)\alpha\beta}+\hat C_{l^2q^2D^2}^{(4)\alpha\beta}+\hat C_{l^2u^2D^2}^{(2)\alpha\beta}\,,
&&
\hat C_1^{8d} = \hat C_{l^2q^2D^2}^{(2)\alpha\beta}-\hat C_{l^2q^2D^2}^{(4)\alpha\beta}+\hat C_{l^2d^2D^2}^{(2)\alpha\beta}\,,
\\[4pt]
\hat C_2^{8u} &= \hat C_{l^2u^2D^2}^{(2)\alpha\beta}-\hat C_{l^2q^2D^2}^{(2)\alpha\beta}-\hat C_{l^2q^2D}^{(4)\alpha\beta}\,,
&&
\hat C_2^{8d} = \hat C_{l^2d^2D^2}^{(2)\alpha\beta}-\hat C_{l^2q^2D^2}^{(2)\alpha\beta}+\hat C_{l^2q^2D^2}^{(4)\alpha\beta}\,,
\\[4pt]
\hat C_3^{8u} &= \hat C_{l^2q^2G}^{(1){\alpha\beta}}+\hat C_{l^2q^2G}^{(3){\alpha\beta}}+\hat C_{l^2u^2G}^{(1){\alpha\beta}}\,,
&&
\hat C_3^{8d} = \hat C_{l^2q^2G}^{(1){\alpha\beta}}-\hat C_{l^2q^2G}^{(3){\alpha\beta}}+\hat C_{l^2d^2G}^{(1){\alpha\beta}}\,,
\\[4pt]
\hat C_4^{8u} &= \hat C_{l^2u^2G}^{(1){\alpha\beta}}-\hat C_{l^2q^2G}^{(1){\alpha\beta}}-\hat C_{l^2q^2G}^{(3){\alpha\beta}}\,,
&&
\hat C_4^{8d} = \hat C_{l^2d^2G}^{(1){\alpha\beta}}-\hat C_{l^2q^2G}^{(1){\alpha\beta}}+\hat C_{l^2q^2G}^{(3){\alpha\beta}}\,,
\\[4pt]
\hat C_5^{8u} &= \hat C_{l^2q^2D^2}^{(1)\alpha\beta}+\hat C_{l^2q^2D^2}^{(3)\alpha\beta}+\hat C_{l^2u^2D^2}^{(1)\alpha\beta}\,,
&&
\hat C_5^{8d} = \hat C_{l^2q^2D^2}^{(1)\alpha\beta}-\hat C_{l^2q^2D^2}^{(3)\alpha\beta}+\hat C_{l^2d^2D^2}^{(1)\alpha\beta}\,,
\\[4pt]
\hat C_6^{8u} &= \hat C_{l^2u^2D^2}^{(1)\alpha\beta}-\hat C_{l^2q^2D^2}^{(1)\alpha\beta}-\hat C_{l^2q^2D}^{(3)\alpha\beta}\,,
&&
\hat C_6^{8d} = \hat C_{l^2d^2D^2}^{(1)\alpha\beta}-\hat C_{l^2q^2D^2}^{(1)\alpha\beta}+\hat C_{l^2q^2D^2}^{(3)\alpha\beta}\,,
\\[4pt]
\hat C_7^{8u} &= \hat C_{l^2q^2G}^{(2)}+\hat C_{l^2q^2G}^{(4)}+\hat C_{l^2u^2G}^{(2)}\,,
&&
\hat C_7^{8d} = \hat C_{l^2q^2G}^{(2)}-\hat C_{l^2q^2G}^{(4)}+\hat C_{l^2d^2G}^{(2)}\,,
\\[4pt]
\hat C_8^{8u} &= \hat C_{l^2u^2G}^{(2)}-\hat C_{l^2q^2G}^{(2)}-\hat C_{l^2q^2G}^{(4)}\,,
&&
\hat C_8^{8d} = \hat C_{l^2d^2G}^{(2)}-\hat C_{l^2q^2G}^{(2)}+\hat C_{l^2q^2G}^{(4)}\,,
\\[4pt]
\hat C_9^{8}  &= \hat C_{l^2G^2D}\,,
&&
\end{align}
where some SMEFT operators are not presented in Tab.~\ref{tab:SMEFTop}, as the corresponding dimension-8 LEFT operators either do not contribute to CE$\nu$NS in the one-body state or are suppressed by dimension-6 LEFT operators with a similar form.

\bibliographystyle{apsrev4-1}
\bibliography{reference.bib}

\end{document}